\documentclass{aa}
\usepackage{graphicx}
\usepackage{txfonts}
\usepackage[separate-uncertainty = true]{siunitx}
\DeclareSIUnit{\dex}{dex}
\DeclareSIUnit{\Msun}{\textit{M}_\odot}
\DeclareSIUnit{\Rsun}{\textit{R}_\odot}
\DeclareSIUnit{\Lsun}{\textit{L}_\odot}
\DeclareSIUnit{\pc}{pc}
\DeclareSIUnit{\mas}{mas}
\DeclareSIUnit{\year}{yr}
\DeclareSIUnit{\mag}{mag}
\usepackage{lscape}
\usepackage{mathrsfs}
\usepackage{float}
\usepackage{natbib}
\usepackage[colorlinks,citecolor=blue]{hyperref}
\hypersetup{
  colorlinks = true,
  linkcolor = blue,
  anchorcolor = blue,
  citecolor = blue,
  filecolor = blue,
  urlcolor = blue}

\begin{document}

   \title{Quantitative spectroscopy of late O-type main-sequence stars\\ with a hybrid non-LTE method}

   \author{P. Aschenbrenner\inst{1}
          \and
          N. Przybilla\inst{1}
          \and
          K. Butler\inst{2}
          }

   \institute{Universit\"at Innsbruck, Institut f\"ur Astro- und Teilchenphysik, Technikerstr. 25/8, 6020 Innsbruck, Austria\\
              \email{patrick.aschenbrenner@student.uibk.ac.at ; norbert.przybilla@uibk.ac.at}
         \and
             LMU M\"unchen, Universit\"atssternwarte, Scheinerstr. 1, 81679 M\"unchen, Germany
             }

   \date{Received ; accepted }


  \abstract
   {Late O-type stars at luminosities $\log L/L_\odot$\,$\lesssim$\,5.2 show weak winds with mass-loss rates lower than
   10$^{-8}$\,$M_\odot$\,yr$^{-1}$. This implies that, unlike their more massive and more luminous siblings, their photospheric layers
   are not strongly affected by the stellar wind.}
   {A hybrid non-local thermodynamic
    equilibrium (non-LTE) approach -- line-blanketed hydrostatic model atmospheres computed under the assumption of LTE
    in combination with non-LTE line-formation calculations -- is tested for analyses of late O-type stars with masses up
    to $\sim$25\,$M_\odot$. A sample of 20 mostly sharp-lined Galactic O stars of spectral types O8 to O9.7 and luminosity classes V and IV,
    previously studied in the literature using full non-LTE model atmospheres, is investigated.}
   {Hydrostatic and plane-parallel atmospheric structures and synthetic spectra computed with Kurucz’s {\sc Atlas12} code together
   with the non-LTE line-formation codes {\sc Detail} and {\sc Surface}, which account for the effects of turbulent pressure on the
   atmosphere, were employed. High-resolution spectra were analysed for atmospheric parameters
   using hydrogen lines, multiple ionisation equilibria,
   and elemental abundances. Fundamental
   stellar parameters were derived by considering stellar evolution tracks and \textit{Gaia} Early Data Release 3 (EDR3) parallaxes.
   Interstellar reddening  was characterised by fitting spectral energy distributions from the UV to the mid-IR.}
   {A high precision and accuracy is achieved for all derived parameters for 16 sample stars (4 objects show composite
   spectra).
   Turbulent pressure effects turn out to be significant for the quantitative analysis.
   Effective temperatures are determined to 1-3\% uncertainty levels, surface gravities to 0.05 to 0.10\,dex, masses to better than 8\%, radii to
   better than 10\%, and luminosities to better than 20\% uncertainty typically. Abundances for C, N, O, Ne, Mg, Al, and Si are derived with
   uncertainties of 0.05 to 0.10\,dex and for helium within 0.03 to 0.05\,dex (1$\sigma$ standard deviations) in general. Overall, results
   from previous studies using unified photosphere plus wind (full) non-LTE model  atmospheres are reproduced, and with higher
   precision. The improvements are most pronounced for elemental abundances, and
   smaller microturbulent velocities are found.
   An overall good agreement is found between our spectroscopic distances and those from \textit{Gaia}. \textit{Gaia} EDR3-based distances to
   the Lac OB1b association and to the open clusters NGC\,2244, IC\,1805, NGC\,457, and IC\,1396 are determined as
   a byproduct. The derived N/C versus N/O abundance ratios tightly follow
   the predictions from stellar evolution models.
   Two ON stars show a very high degree of mixing of CNO-processed material and appear to
   stem~from~binary~evolution.
   }
  {}

   \keywords{Stars: abundances -- Stars: atmospheres -- Stars: early-type
                 -- Stars: evolution -- Stars: fundamental parameters -- Stars: massive
               }

   \maketitle
%

\section{Introduction}
The O stars span the widest range of masses of all the spectral types of the Morgan-Keenan (MK) sequence, from about 18\,$M_\odot$ at
the transition to the main-sequence B stars to upper masses beyond 100\,$M_\odot$. They and their progeny (e.g. blue supergiants, luminous blue variables,
and Wolf-Rayet stars) are the main sources of
ionising UV photons in star-forming galaxies and of momentum input to the interstellar medium (ISM) via intense stellar winds that dominate
their evolution throughout their short lifetimes. The ends of their lives give rise to some of the most extreme galactic objects, such as core-collapse
supernovae, and, in some cases, to long-duration soft-spectrum gamma-ray bursts \citep{MaMe12}. Their typical stellar remnants are black holes.
Binary O stars can form black hole binaries that can coalesce within a Hubble time, yielding measurable
gravitational wave events. O-type stars are therefore widely studied for all these aspects, and many more.

The O-star regime is a domain for hydrodynamical modelling as mass outflows are a dominant factor throughout all phases of their lives
\citep[e.g.][]{ChMa86,Pulsetal08}. However, among them exists the class of the so-called weak-wind O stars. Their existence was
finally recognised with the work of \citet{Martinsetal05} -- and subsequent work by \citet{Marcolinoetal09} and \citet{deAlmeidaetal19} --
on metal-rich Galactic O stars, despite earlier findings in the low-metallicity Small Magellanic Cloud
\citep[SMC;][]{Bouretetal03,Martinsetal04} that were less conclusive, and some first indications date back as early as the work of
\citet{ChGa91}. These weak-wind O stars have a low luminosity, with $\log L/L_\odot$\,$\lesssim$\,5.2, and show mass-loss rates lower
than 10$^{-8}$\,$M_\odot$\,yr$^{-1}$, weaker by up to two orders of magnitude than predicted by standard wind theory
(\citealt{Vinketal00,Vinketal01}, but see \citealt{ViSa21} and \citealt{Marcolinoetal22} for more recent results, and \citealt{Lucy10} and \citealt{Sundqvistetal19} for alternative approaches).
Terminal velocities of low-luminosity O stars were also
found to be low, such that the conjunction with low mass-loss rates implied a reduced line-force parameter, $\alpha$
\citep[e.g.][]{Martinsetal05}. A physical trigger mechanism for the weak-wind phenomenon may be X-ray emission from shocked gas in the
optically thin stellar wind that penetrates into the region below the critical point, thus reducing the radiative acceleration and the
mass-loss rate because of surplus ionisation (\citealt{Drewetal94}; see e.g. \citealt{Martinsetal05} and \citealt{Pulsetal08} for further
discussion). However, the issue has not yet been settled conclusively.

Low-luminosity O stars were subjected to analysis with hydrostatic hydrogen+helium model atmospheres in radiative equilibrium that accounted for
deviations from local thermodynamic equilibrium (LTE; so-called non-LTE models) as early as the seminal work by \citet{AuMi72}. A focus on
applications of such models was  given in the analysis of a few nitrogen-rich ON stars by \citet{Schoenberneretal88}, which exposed
material processed by the CNO cycle on their surfaces and on luminosity effects at spectral class O9.5 by \citet{Voelsetal89}. Several low-luminosity
O stars were investigated by \citet{Herreroetal92} within a larger sample of O stars, who identified the fundamental problem of the mass discrepancy,
the mismatch between spectroscopically derived masses and evolutionary masses, and the helium discrepancy.
Hydrostatic H+He non-LTE atmospheres were used for the last time in the works of \citet{Villamarizetal02} and \citet{ViHe05} to derive helium and CNO abundances.
These works used follow-up non-LTE line-formation calculations that considered the effects of metal line blocking and
a consistent treatment of microturbulence \citep{ViHe00}. They provided observational constraints for rotational mixing of CNO-cycled
products in late O-type stars for comparison with predictions from the first stellar evolution grids of rotating massive stars that had become available at that time \citep{HeLa00, MeMa00}.

In that period, unified model atmospheres -- self-consistent spherically extended hydrodynamic non-LTE atmospheres that
combine photospheric layers at subsonic velocities with the supersonic stellar wind -- became available. Initially, H+He compositions were realised
\citep{SantolayaReyetal97}, and applications mostly ignored low-luminosity O stars \citep{Pulsetal96}. Further
methodological progress was made by considering non-LTE metal line blanketing and including iron-group elements, first for hydrostatic
atmospheres with the codes {\sc Pam} \citep{Grigsbyetal92} and  {\sc Tlusty} \citep{HuLa95} and later for unified photosphere plus wind models
with {\sc Cmfgen} \citep{HiMi98}, {\sc PoWR} \citep{Graefeneretal02}, and {\sc Fastwind} \citep{Pulsetal05}.

First applications of these modern unified non-LTE models included the stellar parameter determination for the low-luminosity standard O star
10~Lac \citep{Herreroetal02} and work on SMC stars \citep{Bouretetal03,Martinsetal04} and Galactic objects that established
the weak-wind phenomenon \citep{Martinsetal05}. Investigations of the Orion Trapezium cluster and comparison stars followed
\citep{SimonDiazetal06}, extending the discussion to refined rotational velocities and elemental abundances.

Further progress led to more
comprehensive investigations that included a full spectral synthesis of the UV and optical spectra of O stars, including weak-wind stars
\citep[e.g.][]{Martinsetal12}. Elemental abundance investigations concentrated on nitrogen, which shows the strongest changes when mixing
with CNO-processed material occurs, seen in particular in the ON stars \citep{Walbornetal11,Martinsetal15b}. These investigations were later extended
to include all three elements, carbon, nitrogen, and oxygen \citep{Martinsetal15a,Carneiroetal19}, in order to trace the full CNO signature
in the main-sequence evolution and beyond \citep{Przybillaetal10,Maederetal14}.
At about the same time, interest in the effects of magnetic fields on the atmospheres of massive stars and their evolution grew.
Among other massive stars, a number of weak-wind O stars were investigated for the presence of magnetic fields via two large
spectropolarimetric surveys, the B fields in OB stars \citep[BOB;][]{Moreletal15,Fossatietal15,Schoelleretal17} and the Magnetism in
Massive Stars (MiMeS) surveys \citep[][]{Wadeetal16}.
Finally, the availability of larger samples of high-resolution spectra of O stars, in particular from the IACOB project -- including many
weak-wind objects -- led to mass quantitative analyses, for example with respect to rotational velocities and additional
line-broadening effects \citep{SiHe14} and stellar parameters \citep{Holgadoetal18,Holgadoetal22}.

All numerical modelling must make assumptions and simplifications in order to handle real stellar atmospheres. As a consequence, not all aspects of the observations
will be reproduced perfectly. For example, the following two issues can be improved
in the quantitative analysis of weak-wind, low-luminosity O stars. First of all, a solution to the \ion{He}{i} singlet problem is
desirable.  It is difficult to reproduce singlet and triplet lines simultaneously with non-LTE codes at effective temperatures
of about 35\,000\,K \citep{Najarroetal06}, though workarounds exist \citep{Pulsetal20}. {Secondly}, the abundance
determinations of these stars are afflicted with large uncertainties, which can reach more than 0.3\,dex \citep{Martinsetal15a}, in contrast to a mere 0.05 to 0.10\,dex for just
slightly cooler early B-type stars \citep[e.g.][]{NiPr12}. This is attributed to non-LTE effects that are much more severe in O stars
than in B stars \citep{Martinsetal15a}.

Such challenges are the motivation for the present work, which aims to improve the accuracy and, in particular, the precision of weak-wind,
low-luminosity O-star analyses in order to provide tight observational constraints on photospheric conditions {for future work on the weak-wind problem}. These analyses provide the
boundary conditions for the base of the stellar wind, the place where the (unexpectedly low) mass-loss rate is initiated.

Further motivation comes from the availability of results from the \textit{Gaia} mission \citep{Gaia16}, and in
particular from the Early Data Release 3 \citep[EDR3;][]{Gaia21}, which have a strong
impact on the determination of fundamental stellar parameters {for} all kinds of stars. These results have so far not been
considered for weak-wind late O-type stars. Parallaxes for massive stars still tend to have substantial
uncertainties in many cases, but often the stars are still located in the OB associations or open star clusters
where they were formed. This offers the possibility of verifying stellar characteristics more precisely than is
feasible from the individual distances alone -- an advantage that is exploited here.

The paper is organised as follows. Observations and data reduction are discussed in Sect.~\ref{sect:obs}. An overview of the models used
for the analysis is given in Sect.~\ref{sect:models} and of the analysis methodology in Sect.~\ref{sect:analysis}. Results from the star
sample analysis are summarised in Sect.~\ref{sect:results}. Finally, a summary of the individual objects is given
in Sect.~\ref{sect:summary} to round out their characterisation, and a general summary and discussion are presented
in Sect.~\ref{sect:conclusions}.
Appendices~\ref{appendix:A} and \ref{appendix:B} discuss the distance determination and show exemplary
spectrum fits for the standard star 10~Lac, and Appendices~\ref{appendix:C} to
\ref{appendix:F} discuss sample star distances in the context of the host star clusters \object{NGC 2244},
\object{IC 1805}, \object{NGC 457,} and \object{IC 1396}.

{\begin{table*}[ht]
\centering
\caption{Overview of the sample stars and observing log.}
\label{tab_observations}
\small
\setlength{\tabcolsep}{1.mm}
\begin{tabular}{llll
rrr
ccr}
\hline
        \hline
        ID\# & Object & Common & Sp.T.\tablefootmark{a} & $V$\tablefootmark{b} & $B-V$\tablefootmark{b} & $U-B$\tablefootmark{b} & Date & $T_{\text{exp}}$ & $S/N$\\
        & & name &  & mag & mag & mag & YYYY-MM-DD & s & \\
        \hline
        \multicolumn{3}{l}{ESPaDOnS~~~$R$\,=\,68\,000}\\[1mm]
        1\tablefootmark{f} & \object{HD 66788} & & O8\,V & 9.454(10) & $-$0.077(14) & $-$0.931(19) & 2010-02-01 & 2400 & 180\\
        2 & \object{HD 14633} & & ON8.5\,V & 7.459(12) & $-$0.204(13) & $-$1.050(28) & 2009-10-09 & 5200 & 430\\
        3\tablefootmark{f} & \object{HD 37061} & NU Ori & O9\,V\tablefootmark{c} & 6.830 & 0.23 & $-$0.59 & 2007-03-08 & $3\times 3200$ & 700\\
        4 & \object{HD 258691} & & O9\,V & 9.700(0) & 0.570(0) & $-$0.453(15) & 2008-01-23 & $2\times 3440$ & 200\\
        5 & \object{HD 201345} & & ON9.2\,IV & 7.660 & $-$0.130 & $-$0.950 & 2010-07-26 & 2800 & 220\\
        6 & \object{HD 57682} & & O9.2\,IV & 6.418(8) & $-$0.189(6) & $-$1.026(4) & 2009-05-04 & $2\times 2400$ & 700\\
        7 & \object{HD 227757} & & O9.5\,V & 9.217(13) & 0.186(13) & $-$0.744(5) & 2010-07-26 & 1160 & 90\\
        8 & \object{BD -13 4930} & & O9.5\,V & 9.435(7) & 0.273(17) & $-$0.709(32) & 2010-06-25 & 3200 & 140\\
        9 & \object{BD +60 499} & & O9.5\,V & 10.300(23) & 0.542(7) & $-$0.507(9) & 2008-08-19 & 3200 & 120\\
        10 & \object{BD +57 247} & & O9.5\,IV\tablefootmark{d} & 9.941(10) & 0.204(5) & $-$0.704(5) & 2014-02-13 & 6960 & 160\\
        11 & \object{HD 207538} & & O9.7\,IV & 7.305(14) & 0.324(8) & $-$0.640(0) & 2008-07-30 & 4000 & 400\\[1.5mm]
        \multicolumn{3}{l}{FEROS~~~$R$\,=\,48\,000}\\[1mm]
        12{\tablefootmark{f}} & \object{HD 46966} & & O8.5\,IV & 6.881(11) & $-$0.055(15) & $-$0.912(12) & 2006-01-03, 2006-01-05 & $2\times600$  & 550\\
        13 & \object{HD 46202} & & O9.2\,V & 8.186(15) & 0.178(11) & $-$0.737(13) & 2008-04-16 to 19, 2008-04-26 & $6\times720$ & 400\\
        14 & \object{HD 38666} & $\mu$ Col & O9.5\,V & 5.169(4) & $-$0.281(6) & $-$1.077(64) & 2011-11-07 & 120 & 300\\
        15{\tablefootmark{f}} & \object{HD 155889} & & O9.5\,IV & 6.551(9) & $-$0.016(8) & $-$0.881(13) & 2012-05-20 & 400 & 320\\
        16 & \object{HD 54879} & & O9.7\,V\tablefootmark{e} & 7.638(13) & $-$0.007(4) & $-$0.867(4) & 2012-05-19 & 600 & 260\\[1.5mm]
        \multicolumn{3}{l}{FOCES~~~$R$\,=\,40\,000}\\[1mm]
        17 & \object{HD 214680} & 10 Lac & O9\,V & 4.879(14) & $-$0.201(8) & $-$1.036(10) & 2005-09-21 & $6\times 600$ & 400\\
        18 & \object{HD 34078} & AE Aur & O9.5\,V & 5.956(15) & 0.225(5) & $-$0.704(15) & 2005-09-21 & $2\times 540, 3\times 450$ & 250\\
        19 & \object{HD 206183} & & O9.5\,V & 7.419(11) & 0.126(12) & $-$0.787(5) & 2005-09-27 & 2700 & 200\\
        20 & \object{HD 36512} & $\upsilon$ Ori & O9.7\,V & 4.618(13) & $-$0.264(7) & $-$1.068(8) & 2005-09-28 & $2\times 360$ & 300\\
        \hline
\end{tabular}
\tablefoot{
\tablefoottext{a}{Spectral types for most objects were adopted from \citet{Martinsetal15a}, which are mostly based on the work of \citet{Sotaetal11,Sotaetal14}}
\tablefoottext{b}{\citet{Mermilliod97}, \citet{MoMa78}}
\tablefoottext{c}{\citet{Bragancaetal12}}
\tablefoottext{d}{\citet{HoAp65}}
\tablefoottext{e}{\citet{Sotaetal11}}
{\tablefoottext{f}{These stars show spectroscopic binarity or multiplicity and are excluded from the detailed analyses.}}
}
\end{table*}
}

\section{Observations and data reduction}\label{sect:obs}
For our analysis, we obtained spectra of 20 late O-type dwarfs, ranging from O9.7 to O8 in spectral type and IV to V in luminosity class.
All have been subjected to quantitative analysis before \citep[17 objects alone were investigated by][]{Martinsetal15a}, sometimes by
several independent studies. Most of the sample stars are apparently slow rotators that show sharp lines. They were found to be
single, to be single-lined binary members (SB1), or to have only spatially resolved binary or multiple partners that were not expected to
contribute second light to the recorded spectra (see the {details} below).

Table~\ref{tab_observations} gives an overview of the sample stars, ordered
by spectral type for the observations per instrument and an observing log. An internal ID number is given, the object is identified by its
Henry-Draper (HD) or its Bonner Durchmusterung (BD) designation, where applicable a common name is given,
the $V$ magnitude and the $B-V$ and $U-B$ colours, the observing date(s), exposure time(s)
$T_\mathrm{exp}$ and the resulting signal-to-noise ratio ($S/N$) of the analysed spectrum, measured at $\sim$5000\,{\AA}.

Eleven stars were observed with the Echelle Spectro-Polari\-metric Device for the Observation of Stars \citep[{ESPaDOnS};][]{ManDon03} on
the  3.6\,m Canada-France-Hawaii Telescope  (CFHT) at Mauna Kea, Hawaii. The spectra cover a wavelength range from 3700 to 10\,500\,{\AA} at a resolving
power of $R$\,=\,$\lambda / \Delta \lambda$\,$\approx$\,68\,000. An additional five stars were observed with the Fiberfed Extended Range
Optical Spectrograph \citep[{FEROS};][]{Kauferetal99} on the Max-Planck-Gesellschaft/European Southern Observatory (ESO) 2.2\,m telescope
at La Silla in Chile, with a wavelength coverage from 3600 to 9200\,{\AA} at $R$\,$\approx$\,48\,000.
The remaining
four stars were observed with the Fibre Optics Cassegrain Echelle Spectrograph \citep[{FOCES};][]{Pfeifferetal98} on the 2.2\,m telescope at
the Calar Alto Observatory in Spain. The FOCES spectra cover a wavelength range from 3860 to 9400\,{\AA} with $R$\,$\approx$\,40\,000.

In the case of ESPaDOnS, pipeline-reduced spectra were downloaded from the CFHT Science Archive at the Canadian Astronomy
Data Centre\footnote{\url{https://www.cadc-ccda.hia-iha.nrc-cnrc.gc.ca/en/cfht/}} and in the case of FEROS, Phase 3 data
from the ESO Science Portal\footnote{\url{https://archive.eso.org/scienceportal/home}}. The spectra were  normalised by
fitting a spline function through carefully selected continuum points. For FOCES the raw data needed to be reduced. Initially, a
median filter was applied to the raw images to remove bad pixels and cosmics. The FOCES semi-automatic pipeline
\citep{Pfeifferetal98} was then used for the data reduction: subtraction of bias and darks, flat-fielding, wavelength
calibration based on Th-Ar exposures, rectification and merging  of the echelle orders. Finally, the spectra were
shifted into the laboratory rest frame via cross-correlation with appropriate synthetic spectra. If multiple spectra
-- preferably from the same night -- were available, we co-added them to increase the $S/N$.

\begin{figure*}
   \centering
   \includegraphics[width=.995\hsize]{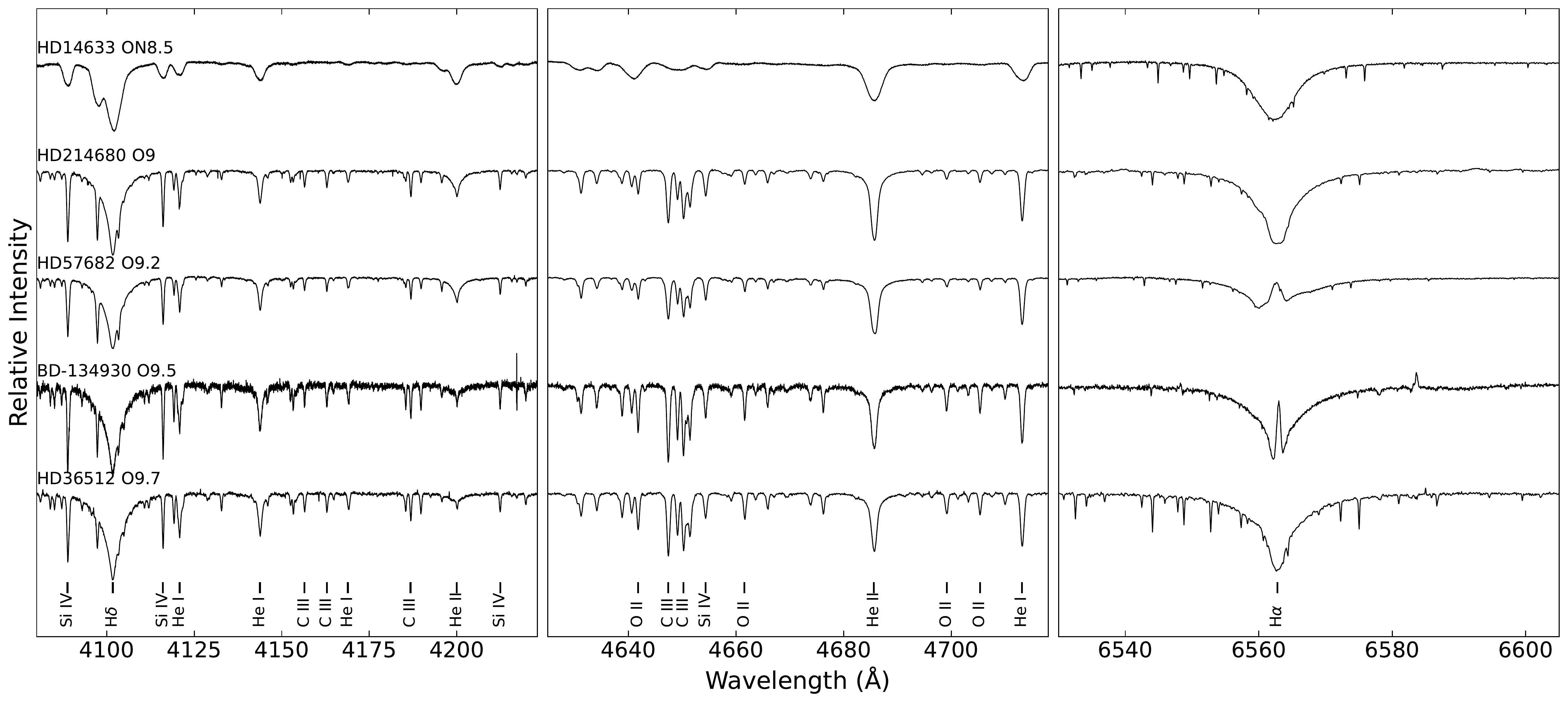}
      \caption{Examples of normalised spectra of the spectral sequence of stars investigated here, arranged from hottest (top)
      to coolest (bottom). Displayed are three spectral windows, around H$\delta$ (left), \ion{He}{ii} $\lambda$4686\,{\AA} (middle), and
      H$\alpha$ (right panel). The major spectral features are identified; the sharp absorption lines around H$\alpha$ are telluric H$_2$O
      features.}
         \label{figure:spectra_overview}
\end{figure*}

{\begin{table}[ht]
\centering
\caption{IUE spectra used for the study.}
\label{table:uvspectra}
\small
\setlength{\tabcolsep}{1.5mm}
\begin{tabular}{rcccc}
    \hline
        \hline
        ID\# & SW & Date & LW & Date\\
             &    & YYYY-MM-DD & & YYYY-MM-DD\\
        \hline
        2 & P08149 & 1980-03-03 & R07080 & 1980-03-03\\
          & P08150 & 1980-03-03 & R08633 & 1980-08-27 \\
          & P09922 & 1980-08-27 & ... & ... \\
        5 & P34159 & 1988-08-31 & P13967 & 1988-08-31\\
        6 & P11156 & 1981-01-26 & P27453 & 1994-02-20 \\
          & P50071 & 1994-02-20 & R09788 & 1981-01-26 \\
        9 & P04132 & 1979-02-03 & R03662 & 1979-02-03 \\
        10 & P20285 & 1983-06-22 & R16208 & 1983-06-22 \\
        13 & P06911 & 1979-10-19 & R05873 & 1979-10-19 \\
        14 & P14340 & 1981-06-26 & P01510 & 1982-04-05 \\
           & P21795 & 1983-12-18 & P01921 & 1983-06-28 \\
           & P22768 & 1984-04-16 & P02426 & 1983-12-18 \\
        20 & P08164 & 1980-03-04 & P11155 & 1981-01-26 \\
           & R07097 & 1980-03-04 & R09787 & 1981-01-26 \\
        \hline
\end{tabular}
\end{table}}

Examples of the final spectra are shown in Fig.~\ref{figure:spectra_overview} for three spectral windows. Optical spectra of late O-type
stars are highly valuable for elemental abundance determinations because they show the largest line density among the O stars, making the
largest number of chemical elements accessible. This line density rapidly declines towards hotter temperatures and for increasing
rotational velocity. Most of the sample stars are slow rotators with (projected) rotational velocity
$\varv \sin i$\,$\lesssim$\,30\,km\,s$^{-1}$, HD~14633 in Fig.~\ref{figure:spectra_overview} belongs to the three faster rotators investigated here,
with $\varv \sin i$\,$\approx$\,100\,km\,s$^{-1}$. It should be noted that a few of the stars show emission in H$\alpha$. This stems either
from a \ion{H}{ii} region, as in the case of BD~$-$13~4930 -- we also note the weak [\ion{N}{ii}] emission components there -- or, as is the case for HD~57682  \citep{Grunhutetal09} and HD~54879
\citep[their Fig.~9]{Castroetal15}, from magnetically confined circumstellar gas.

\begin{figure*}
\sidecaption
   \includegraphics[width=12cm]{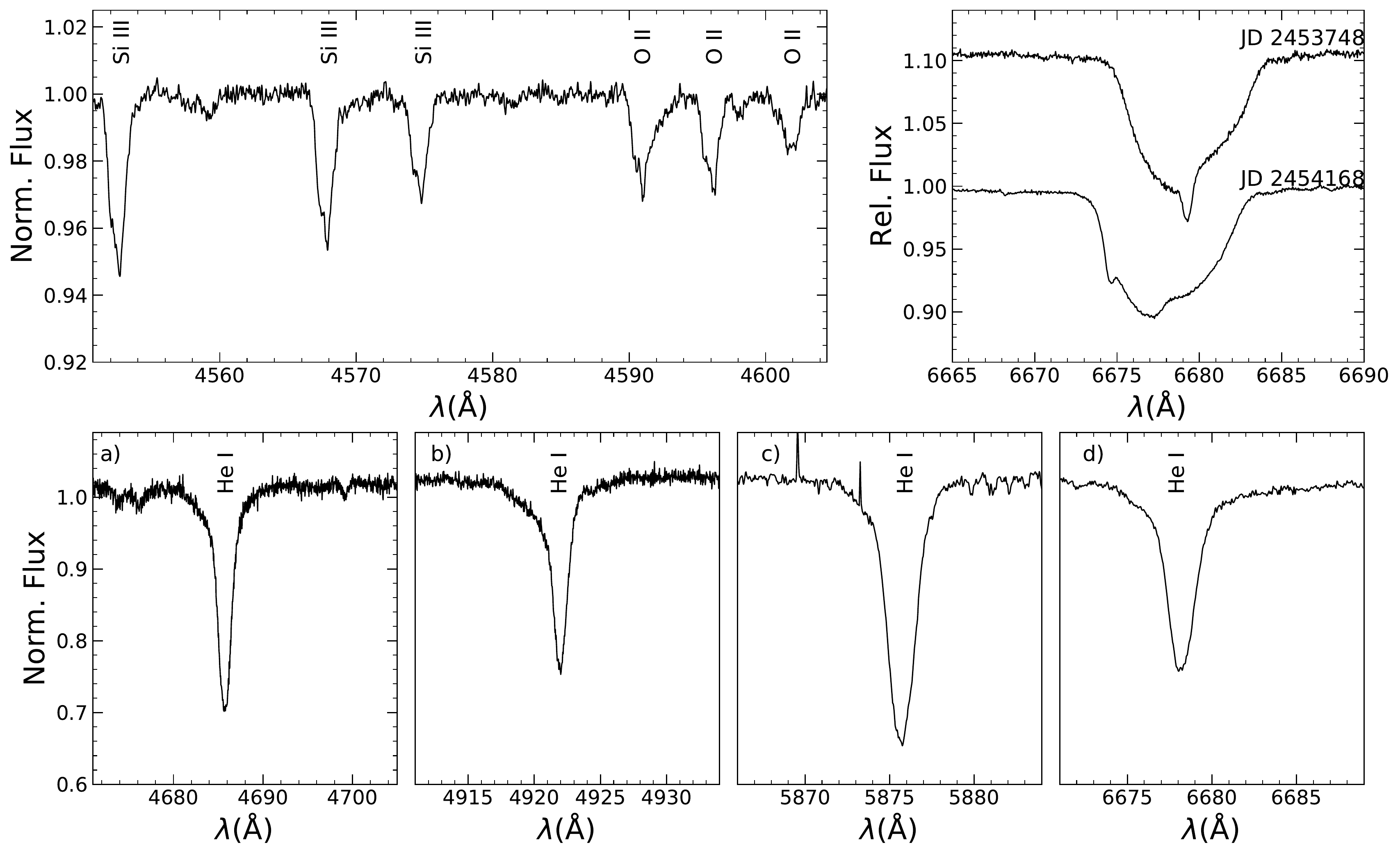}
      \caption{Examples of stars rejected for further analysis in this work, based on spectral features. The metal lines in the spectrum of
      HD\,46966 (top-left panel) show a systematic double-peak structure. The spectrum of HD\,37061 (top-right panel) has three distinct
      components, which change position in wavelength with time. The helium lines in HD\,66788 (bottom panels a and b) and HD\,155889
      (bottom panels c and d) show asymmetric line profiles, indicating a second line component in the blue wings. We note that \ion{He}{i} $\lambda$4922\,{\AA}
      is asymmetric because of a forbidden component in the blue wing, but the observed extended blue wing in panel (b) is wider than expected if the star were to be single.}
         \label{fig_rejected_spectra}
   \end{figure*}

Various (spectro-)photometric data were used in the present work in addition to the Echelle spectra. Low-dispersion, large-aperture
spectra taken with the International Ultraviolet Explorer (IUE; Table~\ref{table:uvspectra}) were downloaded from the  Mikul\-ski Archive
for Space Telescopes (MAST\footnote{\url{https://archive.stsci.edu/iue/}}). The short-wavelength (SW) data cover the range
$\lambda\lambda$1150-1978\,{\AA} and the long-wavelength (LW) data $\lambda\lambda$1851-3347\,{\AA}. For three stars (BD~$-$13\,4930, HD~34078, and HD~214680),
spectra obtained by the \textit{Hubble} Space Telescope's (HST) Space Teles\-cope Imaging Spectrograph~(STIS)~are~on hand. The
spectrophotometric data were co-added for further analysis in the cases where multiple IUE data were available for an object. Alternatively,
UV photometric data were considered, stemming either from the Astronomical Netherlands Satellite \citep[ANS;][]{Wesseliusetal82} or from the joint Belgian--UK
Ultraviolet Sky Survey Telescope \citep{Thompsonetal95} on board the European Space Research Organisation Thor-Delta satellite (TD1).
Moreover, Johnson $UBV$ magnitudes \citep{Mermilliod97, MoMa78}, $JHK$ magnitudes from the Two Micron All Sky Survey
\citep[2MASS;][]{Cutrietal03} and Wide-Field Infrared Survey Explorer (WISE) photometry \citep{Cutrietal21} were adopted.\\[-8mm]
\paragraph{Rejected stars.} Despite the fact that we had pre-selected previously analysed stars for our sample, four objects (i.e. 20\% of our
initial sample) turned out to show second or even third light. Examples for the observational evidence are presented in
Fig.~\ref{fig_rejected_spectra}. \\[1.5mm]
{\sf HD~66788} (ID\#1) shows extra broad absorption bluewards of the sharp \ion{He}{i/ii} lines of the main light source, apparently
due to a faster-rotating later O- or early B-type star. As the vast majority of massive stars are found in multiple systems
\citep{Sanaetal12,Chinietal12}, this can be viewed as the first indication that the star is a candidate physical binary and requires
further investigation. The spectral signature is more difficult to identify in the Balmer lines, but is visible in H$\alpha$ to the
trained eye, and it vanishes in the continuum noise for metal lines. A similar signature is present in a FEROS spectrum taken about 5\,yrs
earlier, which \citet{Marcolinoetal09} analysed under the assumption that HD~66788 is a single star.\\[1.5mm]
{\sf HD~37061} (NU Ori, ID\#3) is the ionising star of the \ion{H}{ii} region \object{M 43} close to the Orion nebula,
which was studied in detail on the basis that it is single by
\citet{SimonDiazetal11} based on spectra at $R$\,$\approx$\,7500-8000. It is in fact a triple system (see Fig.~\ref{fig_rejected_spectra}
for the comparison of two spectra taken at different times). A very detailed and conclusive study of the triple system by
\citet{Shultzetal19} came to our attention only later.\\[1.5mm]
{\sf HD~46966} (ID\#12) was analysed as a single star in several studies
\citep[e.g.][]{Martinsetal12,Martinsetal15a,Holgadoetal18,Carneiroetal19}
so far. However, the spectrum shows a consistent pattern of `spiky' cores best viewed in several metal lines (see
Fig.~\ref{fig_rejected_spectra}). We suggest that this may be the spectroscopic signature of a companion, with both stars being viewed
nearly pole-on, which may explain the sharp lines. This would be consistent with the finding of a constant radial velocity by
\citet{Chinietal12}. The `spikes' are also unlikely to be a signature of some low-order non-radial pulsations, as only a red noise
component was found in the variable photometry measured by the Convection, Rotation et Transits plan\'{e}taires mission \citep[CoRoT;][]{Blommeetal11}. More recently, a resolved companion
50\,mas away and fainter by $\Delta H$\,=\,1.1\,mag was found by \citet{Sanaetal14} at near-IR wavelengths and confirmed by
\citet{Aldorettaetal15} using a broad optical filter with  the Fine Guidance Sensors on board the HST, with $\Delta
m_\mathrm{F5ND}$\,$\approx$\,1.1\,mag. With such a brightness difference and little colour difference, one would indeed expect to see a
second line system in the spectrum, as suggested above.

\noindent
{\sf HD155889} (ID\#15) shows extra blue absorption, best seen in the red \ion{He}{i} lines (see Fig.~\ref{fig_rejected_spectra}). This
indicates the presence of a slightly cooler, faster-rotating companion. Indeed, \citet{Sotaetal14} note a double-lined SB2 nature for the
object, but without giving further details. The Washington Double Star Catalog lists a companion with a separation of 0.2\arcsec~at
$\Delta m$\,$\simeq$\,0.6\,mag \citep{Masonetal01}, which appears compatible with the origin of the second line system in the spectrum.

The four stars were consequently removed from further analysis in the present work, as the basic assumptions of our analysis methodology
are valid for single stars only. We note that our methodology can be extended to treat composite spectra
\citep[see e.g.][]{Irrgangetal14,Gonzalezetal17,Gonzalezetal19}, but this is beyond the scope of the present work.

In addition, HD~14633 (ID\#2) was resolved into two sources by \citet{Aldorettaetal15} using the HST, finding a distance  of $\sim$20\,mas
at $\Delta m_\mathrm{F5ND}$\,$\simeq$\,1.6$\pm$1.1\,mag. We do not see an indication for the presence of second light in the
available spectrum, but any signature could easily hide within the rotationally broadened spectral lines. However, it may be negligible,
in particular if the companion is at the fainter end of the magnitude difference range.
We kept the star in the analysed sample but note that for many of the stellar parameters and abundances we found larger uncertainties
than for most of the other sample objects. \citet{Aldorettaetal15} also found a close companion to HD~46202, at a distance of $\sim$87\,mas
and  $\Delta m_\mathrm{F583W}$\,=\,2.166$\pm$0.008\,mag. As we do not see a second line system in the spectrum, we keep the star in the
analysed sample.

\section{Model atmospheres and spectrum synthesis}\label{sect:models}

\subsection{Models and programs}
Our analysis methodology applied to the interpretation of the weak-wind O-star spectra is based on a hybrid non-LTE approach, a
non-LTE spectral synthesis is performed on LTE model atmospheres. The LTE atmospheres were computed with the {\sc Atlas9} and
{\sc Atlas12} codes \citep{Kurucz93, Kurucz05}, where the former was mostly used to compute starting models for the latter. Both codes
assume plane-parallel geometry, hydrostatic, radiative and LTE, stationarity and chemical homogeneity. Line
blanketing is realised via consideration of pre-tabulated opacity distribution functions (ODFs) in {\sc Atlas9}, whereas {\sc Atlas12} employs
opacity sampling (OS), which provides flexibility in the chemical composition. {\sc Atlas12} also allows the effects of
turbulent pressure on the atmospheric stratification to be accounted for, by considering it as an extra term in the hydrostatic equation.

The non-LTE computations were conducted with {\sc Detail} and {\sc Surface} \citep{Giddings81, BuGi85}. In a first step, the radiative
transfer equation together with the rate equations were solved with {\sc Detail} {for the LTE atmospheric structure computed with {\sc Atlas12},} using an accelerated lambda iteration scheme by
\cite{RyHu91}  and considering line blocking based on the Kurucz OS scheme \citep[with the exception of the calculations for hydrogen and helium,
where the opacity averaging via ODFs is required to avoid the \ion{He}{i} singlet problem,][]{NiPr07}. State-of-the-art model atoms were adopted according to
Table~\ref{table:model_atoms}, in which the ions considered per element, the number of explicit terms (plus superlevels) and radiative
bound-bound transitions, and references are given. All model atoms are completed by the ground term of the next higher ionisation stage.

\begin{table}
\caption{Model atoms for non-LTE calculations with {\sc Detail}.}
\label{table:model_atoms}
\centering
{\small
\begin{tabular}{llll}
\hline\hline
Ion                & Terms       & Transitions & Reference\\
\hline
H                  & 20          & 190         & [1]\\
\ion{He}{i/ii}     & 29+6/20     & 162/190     & [2]\\
\ion{C}{ii/iii/iv} & 68/70/53    & 425/373/319 & [3]\\
\ion{N}{ii}        & 77          & 462         & [4]\\
\ion{O}{ii/iii}    & 176+2/132+2 & 2559/1515   & [5]\\
\ion{Ne}{ii}       & 78          & 992         & [6]\\
\ion{Mg}{ii}       & 37          & 236         & [7]\\
\ion{Al}{iii}      & 46+1        & 272         & [8]\\
\ion{Si}{iii/iv}   & 68+4/33+2   & 572/242     & [9]\\
\hline
\end{tabular}
\tablefoot{Data for different ionisation stages are separated by a slash. If present, the number of superlevels is indicated after a plus sign.}
\tablebib{[1] \citet{PrBu04};~~[2] \citet{Przybilla05};
[3]~\citet{NiPr06,NiPr08}; [4] \citet{PrBu01};
[5] Przybilla \& Butler (in prep.); [6]~\citet{MoBu08};
[7] \citet{Przybillaetal01a}; [8] Przybilla (in prep.);
[9] Przybilla \& Butler (in prep.).
}}
\end{table}

While most of the model atoms were used previously in various studies, a new model atom for \ion{O}{ii/iii} was employed here --
replacing the previously used \ion{O}{ii} model atom of \citet{BeBu88} --, a brief
summary is given in the following. Level energies were adopted from \citet{Martinetal93} for \ion{O}{ii} and from
\citet{Moore93} for \ion{O}{iii}. These were combined into 176 LS-coupled (Russell-Saunders coupling) terms up to principal quantum number $n$\,=\,8 and the levels for
$n$\,=\,9 combined into two superlevels, one each for the doublet and quartet spin systems of \ion{O}{ii}. For \ion{O}{iii} this resulted
in 132 terms up to $n$\,=\,6 and 2 superlevels combining the levels for $n$\,=\,7 for the singlet and triplet spin systems. Oscillator
strengths and photoionisation cross-sections were for the most part adopted from the Opacity Project \citep[OP; e.g.][]{Seatonetal94}, with
several improved data taken from \citet{FFT04}. The strongest resonance lines with broad line wings were treated as fine-structure
transitions, accounting for Stark broadening using broadening coefficients from Kurucz\footnote{\url{http://kurucz.harvard.edu/atoms.html}}.
Photoionisation cross-sections missing in the OP data were assumed to be hydrogenic. Electron
impact-excitation data for a large number of transitions were available from the ab initio calculations of \citet[80 transitions]{Tayal07}
and  \citet[3748 transitions]{Maoetal20} for \ion{O}{ii} and from \citet[2346 transitions]{TaZa17} for \ion{O}{iii}. Missing data were
provided by the use of van Regemorter's formula \citep{vanRegemorter62} for radiatively permitted transitions or Allen's formula \citep{Allen73} for
forbidden transitions. All collisional ionisation data were provided via the Seaton formula \citep{Seaton62} using OP
photoionisation threshold cross-sections, or hydrogenic values.

Then, synthetic spectra were calculated with {\sc Surface}  using the non-LTE occupation numbers provided by {\sc Detail}, considering
both refined fine-structure transition data and line-broadening data. Oscillator strengths from \citet{Wieseetal96} and
recent data from Kurucz were replaced by multi-configuration Hartree-Fock data by
\citet{FFT04} for \ion{O}{ii/iii} \citep[as for other elements and ions, also accounting for data from][]{FFTI06}, which was essential to
achieve a close match between observations and model spectra. Both {\sc Detail} and {\sc Surface} were updated recently to account for
level dissolution of \ion{H}{i} and \ion{He}{ii} using the implementation of \citet{Hubenyetal94}, which yields a better modelling of the
series limits. The spectrum synthesis for hydrogen and neutral helium was further updated by Stark-broadening tables of \citet{TrBe09} and
\citet{Beauchampetal97}, respectively. For atmospheric parameters not covered by these tables, broadening data by \citet{DiSa90} were
employed instead.
\begin{figure*}
\centering
   \includegraphics[width=0.405\hsize]{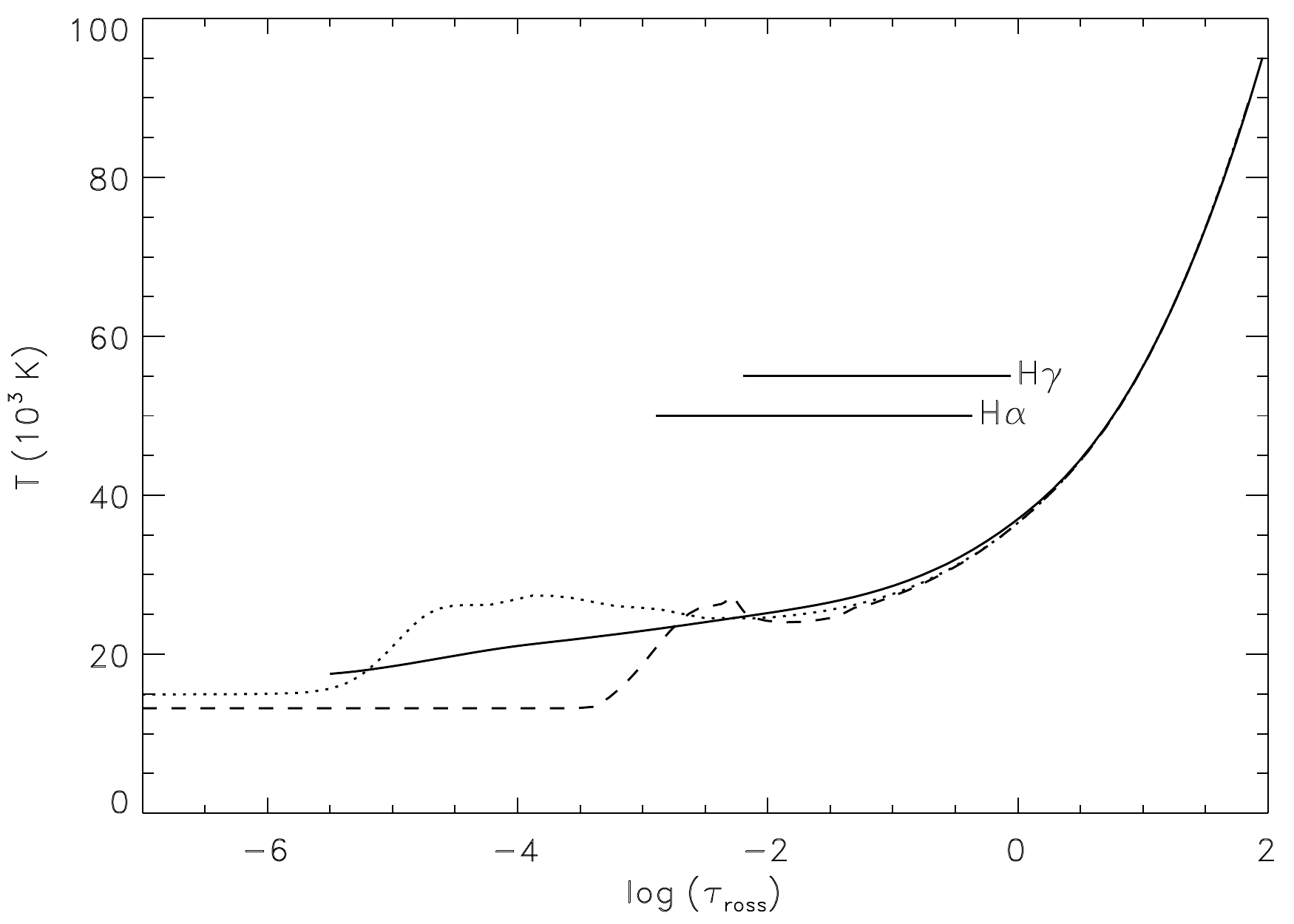}
   \includegraphics[width=0.405\hsize]{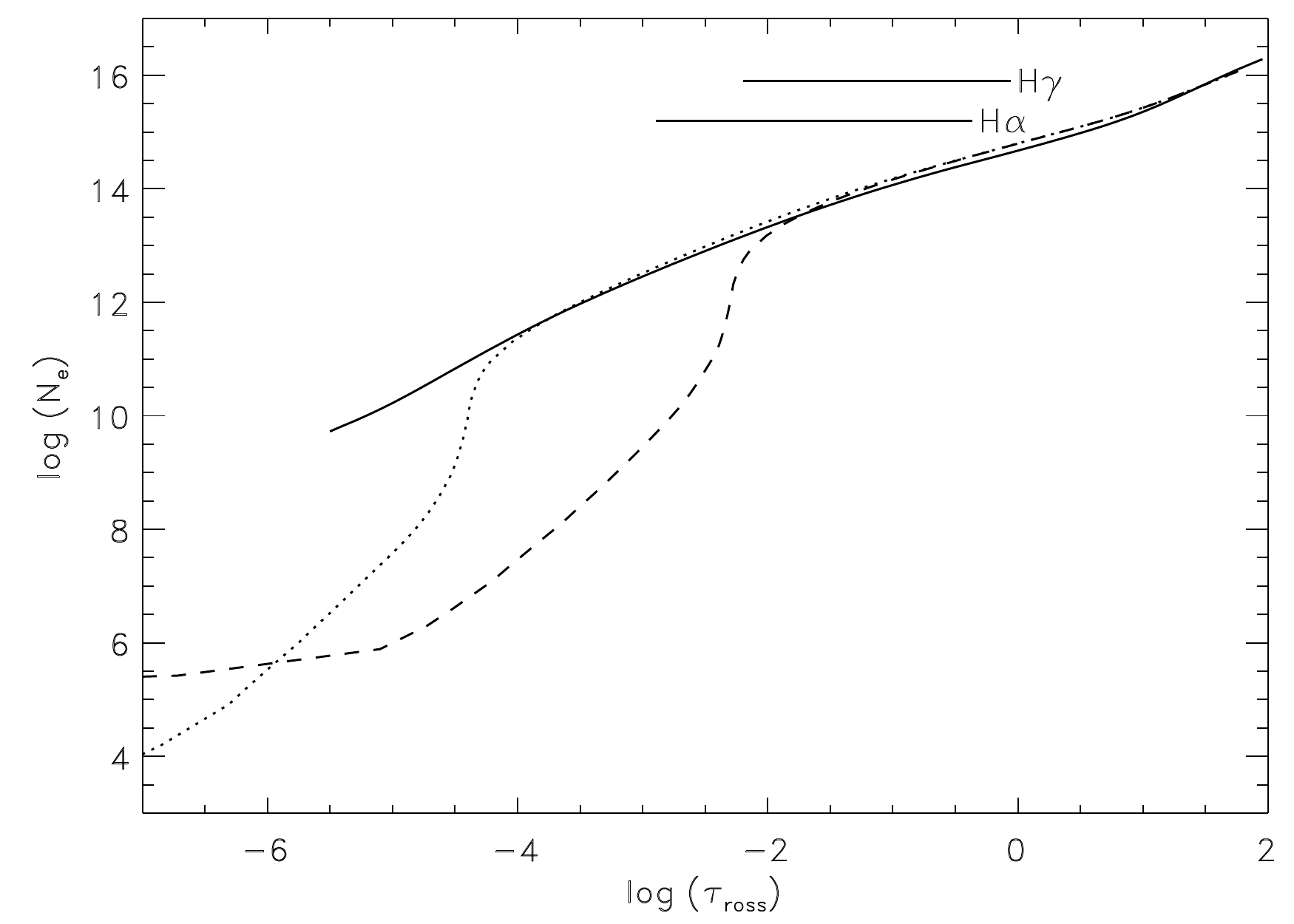}
      \caption{Comparison of {\sc Atlas12} (full line) and {\sc Fastwind} model atmospheres for an effective temperature of 33\,000\,K, surface gravity 4.0,
      microturbulent velocity 5\,km\,s$^{-1}$, and solar metallicity. {\em Left panel:} Temperature structure. {\em Right panel:} Electron density structure
      as a function of Rosseland optical depth, $\tau_\mathrm{ross}$. The two {\sc Fastwind} models were computed for the parameters stellar radius $R$\,=\,6.58\,$R_\odot$,
      wind terminal velocity $v_\infty$\,=\,1200\,km\,s$^{-1}$, $\beta$-velocity-field parameter $\beta$\,=\,1.0, and mass-loss rates
      $\log \dot{M}$\,($M_\odot$\,yr$^{-1}$)\,=\,$-$9.5 (weak wind, dotted line) and
      $\log \dot{M}$\,($M_\odot$\,yr$^{-1}$)\,=\,$-$7.41 (wind mass-loss rate according to \citealt{Vinketal00}, dashed line). Formation depths for the Balmer H$\alpha$ and H$\gamma$ lines
      are indicated.}
         \label{fig_A12FW_structures}
   \end{figure*}

\begin{figure*}
\centering
   \includegraphics[width=0.8\hsize]{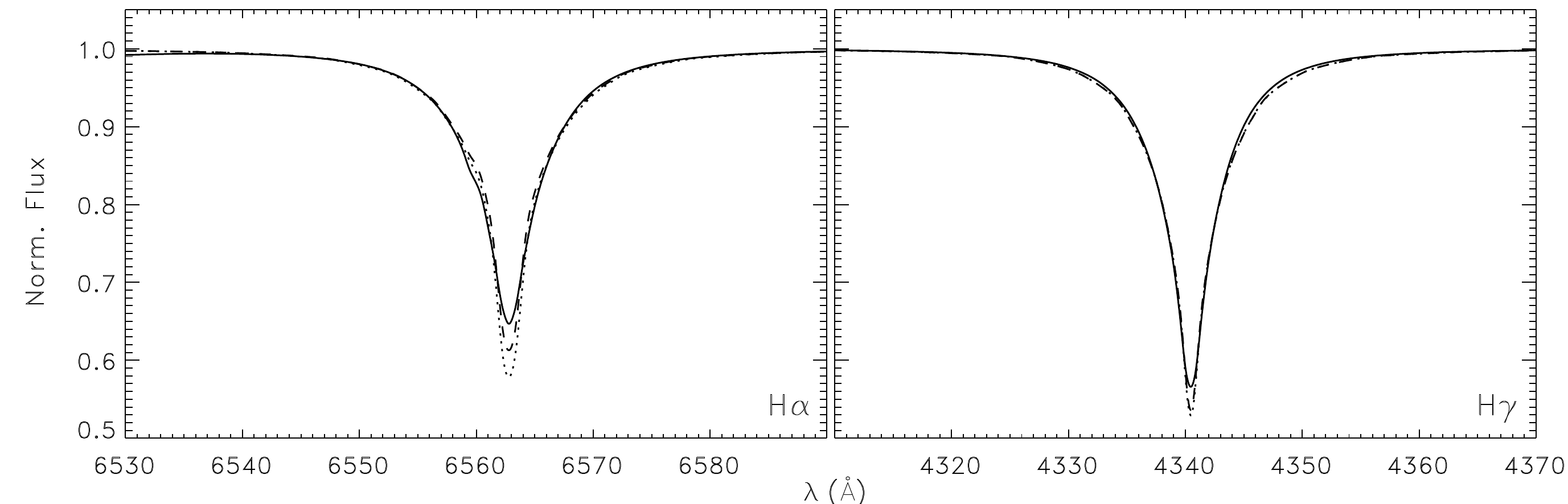}
      \caption{Comparison of Balmer line profiles computed with {\sc Atlas12-Detail-Surface} and {\sc Fastwind}. {\em Left panel:} H$\alpha$. {\em Right panel:} H$\gamma$.
      The line coding is as in Fig.~\ref{fig_A12FW_structures}. All models were convolved with a rotational profile corresponding to
      $\varv_\mathrm{rot}$\,=\,30\,km\,s$^{-1}$.}
         \label{fig_A12FW_profiles}
   \end{figure*}

For the first (coarse) step of the analysis a grid ranging from 31\,000 to 35\,000\,K in effective temperature (step size 1000\,K), 3.8 to
4.3 in logarithm of gravity (step size 0.1), 2 to 16\,km\,s$^{-1}$ in microturbulent velocity (step size 2\,km\,s$^{-1}$), and 0.07 to 0.13 in
helium abundance by number (step size 0.02) was calculated and used to determine initial solutions. The boundaries are selected to contain typical literature values for late O-type dwarfs.
Then, smaller micro-grids with half the step size and elemental abundance steps of 0.1\,dex were computed per star and further iteration steps were made to refine the analysis.

The individual line analyses were performed with the Spectral Plotting and Analysis Suite \citep[{{\sc Spas}};][]{Hirsch09}. The
program allows instrumental, (radial-tangential) macroturbulent and rotational broadening to be applied flexibly to the synthetic spectra
and can be used to fit up to three different parameters, interpolating on the pre-calculated grids
to the actual parameters, using bi-cubic spline fitting. {\sc Spas} employs the downhill simplex algorithm
\citep{NeMe65} to find minima~in~the~$\chi^2$ landscape.

\subsection{Comparison with hydrostatic full non-LTE models}
The {comparison between} our hybrid and a full non-LTE approach for the hydrostatic case was already {made} by
\citet[see their Sect.~5 for a comprehensive discussion]{NiPr07}, for parameters also relevant to the present work. In summary,
they found that atmospheric structures computed with {\sc Atlas} and {\sc Tlusty}
\citep[as adopted from the OSTAR2002 grid;][]{LaHu03} show excellent agreement in the effective temperature range from below 32\,500 to
35\,000\,K. Deviations in the temperature structures amount to less than 1\% in the inner atmosphere, which includes the formation regions
of the weaker lines and the wings of the stronger features, and increase to 2-3\% for the formation regions of the cores of most of the
H and He lines. Stronger deviations occur only in the outermost parts of the atmospheres, which, however, are modified by the hydrodynamic
outflow even in the weak wind case (this includes the formation region of the H$\alpha$ core and some metal resonance lines in the UV).
Spectral energy distributions (SEDs) computed with {\sc Atlas+Detail} and {\sc Tlusty} also agree well, with some small differences occurring
at the \ion{He}{ii} continua in the extreme UV. Hydrogen and helium lines computed with {\sc Detail+Surface} on either {\sc Atlas} or
{\sc Tlusty} atmospheric structures show excellent agreement, while the original line profiles from the OSTAR2002 grid showed the \ion{He}{i}
singlet problem.
The small deviations compared to the hydrostatic full non-LTE models come with the advantage of significantly reduced computation time.
Because of that, more complex model atoms can be employed, tailored model grids for each analysed star including variations of microturbulent velocities can be computed,
and the uncertainties introduced by the grid spacing can be reduced. Our hybrid non-LTE method is applied systematically to a sample of weak-wind late O-type main-sequence stars for the first time.

\subsection{Comparison with unified non-LTE atmospheres}\label{sect:unified_models}
A comparison of the atmospheric structures computed with {\sc Atlas12} in LTE and the unified
(photosphere + wind) non-LTE code {\sc Fastwind} \citep{SantolayaReyetal97,Pulsetal05} is shown in
Fig.~\ref{fig_A12FW_structures}. The atmospheric and wind parameters were chosen according to the
model of \citet{Martinsetal05} for HD~38666, which are representative for the sample stars. Two wind solutions
were considered, for the wind mass-loss prescription of \citet{Vinketal00} and for the observed weak wind
with a mass-loss rate reduced by about two orders of magnitude. Both solutions are practically identical for the
deepest atmospheric layers. Temperatures differ by about 1\% in the regions where the weak (metal) line
spectrum is formed, and the difference increases to $\sim$2-3\% at the formation depths of the cores of the
stronger lines (most hydrogen Balmer and the \ion{He}{i} lines), with the LTE solution showing higher
temperatures. The temperature rise of the unified models at lower optical depths is seen only by the
strongest lines in the optical, like H$\alpha$.

The effect of the wind strength is more pronounced in the
density structure, represented here by the electron density. Overall, the run of electron density with
Rosseland optical depth $\tau_\mathrm{ross}$ agrees rather well between the weak-wind {\sc Fastwind} and the
{\sc Atlas12} model, with the former showing the onset of the wind at $\log
\tau_\mathrm{ross}$\,$\approx$\,$-$4. The higher mass-loss rate on the other hand would shift the onset of the
wind to $\log \tau_\mathrm{ross}$\,$\approx$\,$-$2, well within the formation region of H$\alpha$.

Model profiles for H$\alpha$ and H$\gamma$ (including the \ion{He}{ii} blends) for the three
atmospheric stratifications are shown in Fig.~\ref{fig_A12FW_profiles}. Significant differences occur
essentially only in the line cores. The H$\alpha$ line core of the solution with a higher mass-loss rate is
shallower than in the weak-wind solution because of the density drop that sets in deeper in the atmosphere.
On the other hand, the even shallower profile of the hydrostatic hybrid non-LTE solution results from the
higher local temperature at core formation depths. The H$\gamma$ line shows close agreement.
Similar agreement between the three models is found for the \ion{He}{i} lines, while the \ion{He}{ii} lines
are stronger in the hybrid non-LTE solution, reflecting the local temperature being higher by several hundred
degrees (which can be seen by the effect on the \ion{He}{ii} blend to the blue wing of H$\alpha$ in
Fig.~\ref{fig_A12FW_profiles}). We conclude that the use of our hydrostatic hybrid non-LTE approach
overall reproduces solutions of unified non-LTE models for the parameter range investigated here.

\subsection{Turbulent pressure}
Their location close to the Eddington limit makes O-type stars
susceptible to additional factors that impact the delicate balance
between the counteracting gravitational and radiative acceleration.  One of these
may be turbulent motions (microturbulence) that give rise to `turbulent
pressure' $P_\mathrm{turb}$\,=\,$\frac{1}{2}\rho\xi^2$, with $\rho$ being the gas
density and $\xi$ the microturbulent velocity. The main effect will be on the
pressure stratification, and one may expect an increased importance of this the
closer the star is to the Eddington limit, and the larger the microturbulent
velocity. {\sc Atlas12} allows the effects of turbulent pressure on the model
atmosphere computations to be taken into account by adding the above term to the
hydrostatic equation.

\begin{figure}
   \centering
   \includegraphics[width=\hsize]{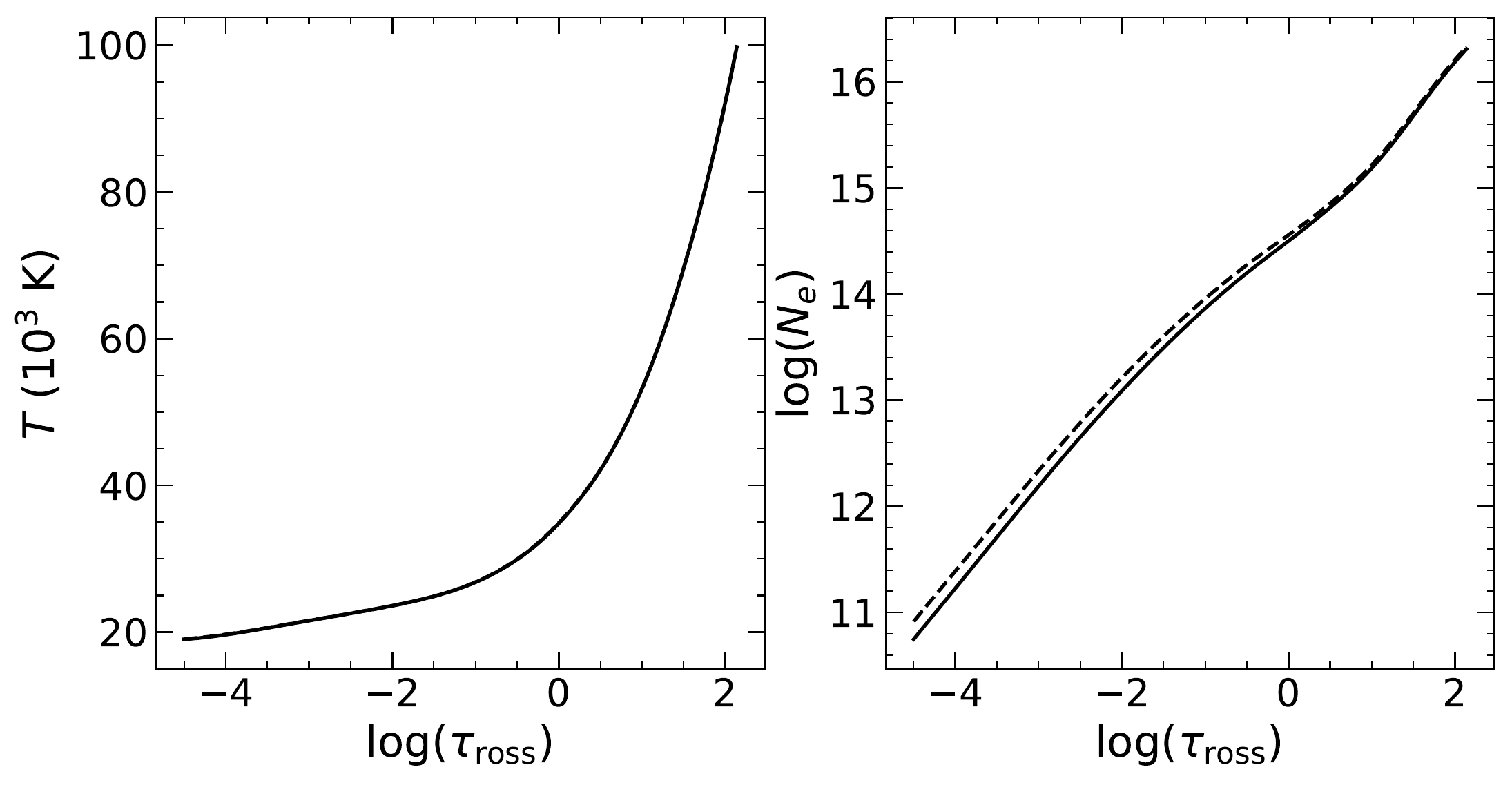}
      \caption{Effects of turbulent pressure on the atmospheric structure in terms of temperature ({\em left}) and electron number
      density ({\em right panel}) as a function of $\tau_\mathrm{Ross}$. The solid line shows the model that includes turbulent pressure,
      and the dashed line is for the same model without turbulent pressure. Atmospheric parameters for BD\,+57 247 were used for the comparison,
      with $T_\mathrm{eff}$\,=\,31\,200\,K, $\log g$\,=\,3.87, and $\xi$\,=\,16\,km\,s$^{-1}$.}
         \label{figure:turbulent_pressure}
\end{figure}

\begin{figure*}
    \centering
    \includegraphics[width=0.78\hsize]{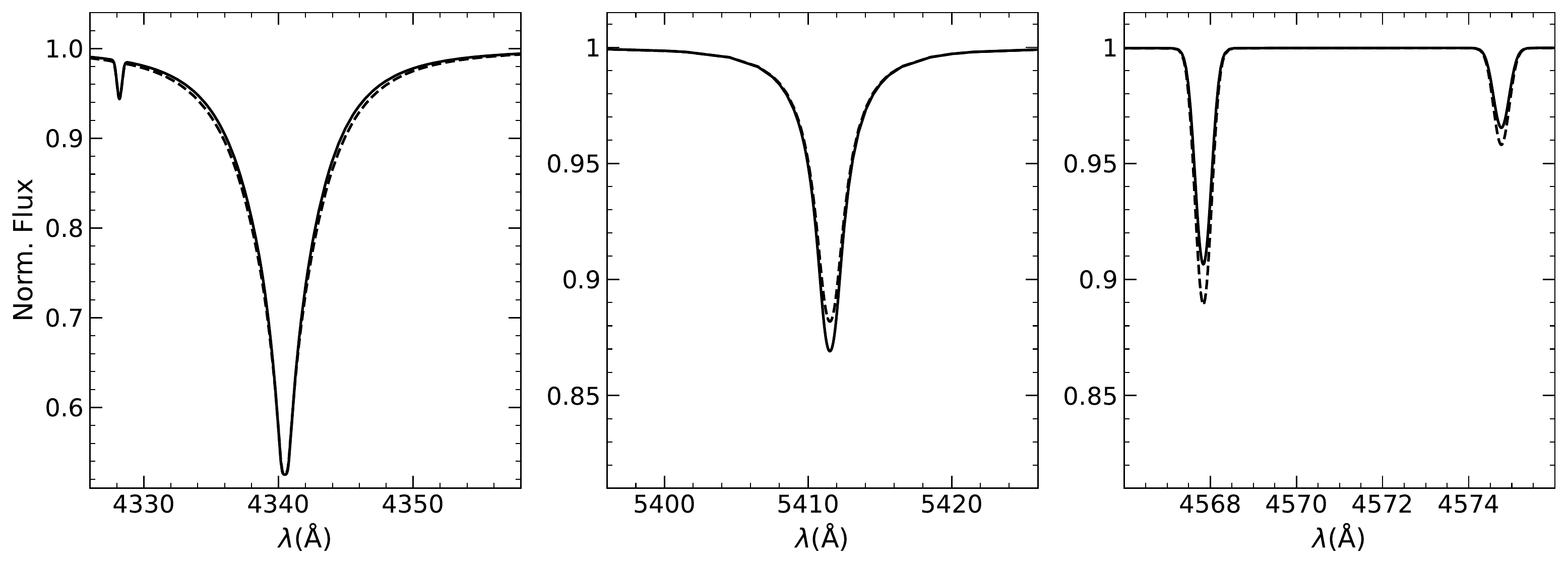}
    \caption{Effects of turbulent pressure on the line profiles of H$\gamma$ (left), \ion{He}{ii} $\lambda$5411\,{\AA} (centre), and two of the \ion{Si}{iii} triplet lines (right panel).
    The solid line shows the model that includes turbulent pressure, and the dashed line is for the same model without turbulent pressure.}
    \label{fig:turbulence_lines}
\end{figure*}

In the following we
investigate the effects of turbulent pressure on the atmospheric structure and synthetic spectra by comparing model calculations
with and without the extra term at otherwise unchanged parameters. The model for BD~+57~247 was chosen as the star shows one
of the lower gravity values and the highest microturbulence of the entire sample, so it is to be expected that the
effects will be the greatest of the sample stars.

A comparison of the resulting temperature and electron density stratifications as a function of Rosseland optical depth
$\tau_\mathrm{Ross}$ is shown in Fig.~\ref{figure:turbulent_pressure}. Differences in temperature are negligible, as they amount to less than
100\,K at line-formation depths. More pronounced are the differences in electron density, which are locally lower in the
case where turbulent pressure is accounted for. This has a small but noticeable effect on the Stark broadening of the hydrogen lines (see
Fig.~\ref{fig:turbulence_lines}). While the higher gravity values derived in the case of calculations with turbulent pressure, by
$\lesssim$0.05\,dex, lie within the uncertainties of the spectral analysis, the effect is systematic.
The cores of the helium
and some metal lines are affected, with some lines being strengthened and others being weakened. The effects may appear small in
Fig.~\ref{fig:turbulence_lines}, but, in particular, the change of equivalent width of the \ion{Si}{iii} triplet lines by 20\% for the
strongest multiplet line and gradually slightly less for the weaker multiplet members is highly significant in this case, as
the triplet lies at the heart of many quantitative analyses, like the Very Large Telescope Fibre Large Array Multi Element Spectrograph (VLT-FLAMES) massive star survey
\citep[e.g.][]{Hunteretal07} or the VLT-FLAMES Tarantula Survey \citep[e.g.][]{Garlandetal17}. Not only the ionisation balance of \ion{Si}{iii/iv} is shifted, but the
microturbulence determination may also be affected. Numerous other lines show similar effects. However, the effects of
microturbulent pressure  will be smaller for the majority of the sample stars, because of the typically smaller microturbulent velocities
(we note that often much higher values are stated in the literature), but this may change if weak-wind late O-type giants are investigated, or supergiants.

The consideration of the effects of turbulent pressure may also be important for the outer density structure of hydrodynamic models, where a depth-dependent
microturbulent velocity is sometimes considered that increases from subsonic values in the photosphere (i.e. comparable to the cases
here) up to one-tenth of the wind terminal velocity \citep{Martinsetal15a}, reaching microturbulent velocities beyond 200\,km\,s$^{-1}$
for a typical late O-type dwarf star.

\section{Spectral analysis}\label{sect:analysis}

\subsection{Stellar parameter and abundance determination}
In a first step, the atmospheric parameters  effective temperature $T_{\mathrm{eff}}$, surface gravity $\log(g)$ (in cgs units), the
microturbulent velocity $\xi$ and helium abundance by number $y$, as well as the projected rotational velocity $\varv\sin{i}$ and
macroturbulent velocity $\zeta$ were constrained by as many spectral indicators as possible. An iterative approach was employed for this
in order to overcome ambiguities because of correlations between some parameters until all parameters were constrained in a consistent way.
In a second step, the various non-LTE elemental abundances $\varepsilon\left(X\right)$\,=\,$\log\left(X/\mathrm{H}\right)$\,+\,12 for elements
$X$ were derived.

\subsubsection{Effective temperature and surface gravity}
The sample stars cover a relatively small region in the $T_\mathrm{eff}$- and $\log g$-parameter space because of the concentration on O8 to
O9.7 spectral types and luminosity classes V and IV. Overall, the sample stars can be expected to fall in the range of 31\,000 to
35\,000\,K in $T_\mathrm{eff}$ and 4.3 (about the value of the zero-age main sequence) to 3.8 in $\log g$. Ionisation equilibria of
\ion{He}{i/ii}, \ion{C}{ii/iii/iv}, \ion{O}{ii/iii,} and \ion{Si}{iii/iv} (i.e. $T_\mathrm{eff}$ was adjusted so that lines from the
different ionisation stages of an element indicate the same abundance at a given value of $\log g$) were established as temperature
indicators. The surface gravity was adjusted such that the Stark-broadened wings of the Balmer lines were reproduced at a given
$T_\mathrm{eff}$. The adopted combination of $T_\mathrm{eff}$/$\log g$ is given by the  values where all indicators match the
observations simultaneously. In principle, additional ionisation equilibria are available from the optical spectra, for example
\ion{N}{ii/iii}. However, we lack a reliable model atom for the latter ionisation stage.

\subsubsection{Microturbulence and elemental abundances}
The microturbulent velocity was adjusted so that the element abundance of individual lines in the spectral fitting became independent of the strengths of the lines
(for visualisation purposes, equivalent widths of these lines were measured). The main indicators were the carbon and silicon lines. Closely related to this was the determination of
the elemental abundances. Helium plays a particular role because of its larger mean molecular weight, which may modify the atmospheric
pressure stratification depending on the helium abundance. For a given set of atmospheric parameters, we calculated a small grid of
synthetic spectra for an element and
fitted individual line profiles employing {\sc Spas}, aiming at $\chi^2$ minimisation. The final abundance values and their uncertainties
were computed once all atmospheric parameters were fixed in the iterative process, as the statistical mean and its $1\sigma$ standard
deviation, based on all considered  lines (marked outliers were removed using $\sigma$-clipping).

\subsubsection{Projected rotational velocity and macroturbulence}
Both parameters were derived simultaneously by fitting the profiles of numerous metal lines. The synthetic lines were
broadened with rotational and a radial-tangential macroturbulence profiles
\citep{Gray05} and a Gaussian instrumental profile as appropriate for the resolving power of the three instruments. The necessity for
considering non-rotational broadening in O-type stars is well established \citep[e.g.][]{SimonDiazetal17}. Atmospheric velocity fields due to
a  sub-surface convection zone {and} stellar pulsations, among others, are the phenomena that are summarised by the macroturbulence
parameter. To account for the line depth the elemental abundance was included as a third fit parameter to achieve an optimum fit of each
line.

\subsubsection{SED fitting}\label{subsection:sed_fitting}
A comparison of the model SED with observations allows one to test whether the derived solution also
reproduces the global energy output of a star. The observed SEDs were constructed from HST or IUE spectrophotometry, or ANS or TD1
UV photometry, and Johnson, 2MASS and WISE photometry. On the model side, the starting {\sc Atlas9} model fluxes were
adopted, as they are directly usable for the comparison with the low-resolution observations without any further processing.
They are equivalent to {\sc Atlas12} models for the purpose of SED fitting because the temperature structures
are practically identical for nearly scaled-solar abundances (as is the case here). To account for interstellar extinction the reddening law
of \citet{Fitzpatrick99} was used, parameterised by the colour excess $E(B-V)$ and the ratio of total-to-selective extinction
$R_V$\,=\,$A_V/E(B-V)$. To obtain both parameters the model SEDs were matched to the observations. Synthetic magnitudes were computed by
integrating the  model flux over the normalised transmission profile of each filter, applying zero points in the Vega system, and then
reddened. The transmission profiles and zero-points were taken from the SVO Filter Profile
Service\footnote{\url{http://svo2.cab.inta-csic.es/theory/fps/}} \citep{Rodrigo12, Rodrigo20}.

The bolometric corrections $B.C.$\,= $\Bar{m}_\mathrm{bol} - \Bar{m}_V$ were obtained via synthetic magnitudes (marked
by barred variables). The zero point for the synthetic bolometric magnitude was chosen such that the absolute bolometric magnitude of the
Sun, $M_\mathrm{bol,\odot}$\,=\,4.74, was reproduced when using an {\sc Atlas9} model atmosphere for solar parameters.

\subsection{Spectroscopic distances}
The stellar flux measured at Earth $f_{\lambda}$\,=\,$F_{\lambda} \cdot (\pi R^2 / d^2)$ depends on the astrophysical flux emitted at the
surface of the star, $F_\lambda$, the distance to the star, $d$, and the stellar radius, $R$. The measurement of the magnitude $m_X$ with a
filter with normalised transmission profile $T(\lambda)$ can be expressed as
\begin{equation}
\begin{split}
m_X & = -2.5 \log \left( \int f_{\lambda} T(\lambda) \mathrm{d}\lambda \right) + zp\\
    & = -2.5 \log \left( \int F_{\lambda} T(\lambda) \mathrm{d}\lambda \right) + zp -2.5 \log(\pi R^2 / d^2)\,,
\label{equ_magnitude_observed}
\end{split}
\end{equation}
with $zp$ being the zero point of the given filter $X$.

Expressing the stellar radius in terms of the stellar mass $M$ and surface gravity $\log(g)$ and using the definition for synthetic
magnitudes one can solve {Eq.}~\eqref{equ_magnitude_observed} for the distance (in units of parsec). This distance depends on
spectroscopically derived parameters and, hence, is called the spectroscopic distance:
\begin{equation}
d_{\mathrm{spec}} = 6.63\cdot 10^{-6} \sqrt{M/M_\odot \cdot 10^{0.4(m_X-\Bar{m}_X)-\log(g)}}\,\mathrm{pc}.
\label{equ_dspec}
\end{equation}
In {Eq.}~\eqref{equ_dspec} the stellar mass -- which still needs to be determined -- is given in solar units, $m_X$ is the observed magnitude,
$\Bar{m_X}$ the synthetic magnitude of the reddened SED (we use $V$ magnitudes in the following) and $\log(g)$ is given in cgs units.

\subsection{Fundamental parameters}
In a final step the fundamental stellar parameters were determined in analogy to previous work on early B-type stars \citep{NiPr14}. The
stellar evolutionary masses $M_\mathrm{evol}$ were found by comparing the position of the stars in the spectroscopic Hertzsprung-Russell
diagram \citep[sHRD;][]{LaKu14} relative to Geneva evolutionary tracks \citep{Ekstroemetal12}. Non-rotating models were employed here,
because the sample stars are (in most cases presumably true) slow rotators. Stellar radii were then derived from the definition of the surface gravity
$\log(g)$\,=\,$G M / R^2$ and the stellar luminosity followed directly from the Stefan-Boltzmann law $L$\,=\,$4\pi\sigma R^2
T_{\mathrm{eff}}^4$, with $\sigma$ being the Stefan-Boltzmann constant. This also facilitated conversion to absolute bolometric magnitudes
$M_\mathrm{bol}$ and, considering $B.C.$, to the absolute visual magnitude $M_V$. Evolutionary ages $\tau_\mathrm{evol}$ were finally derived in the Hertzsprung-Russell
diagram (HRD) by
comparison of the sample star positions with the loci of isochrones based on the Ekstr\"om et al.~tracks for non-rotating stars.

\begin{table*}[ht]
\centering
\caption{Stellar parameters for the sample stars.}
\label{table:result_part1}
\small
\setlength{\tabcolsep}{.9mm}
    \begin{tabular}{rlr@{\hspace{0.1mm}}rcc
    rrrcc
    rrrrc
    ccrr}
    \hline
    \hline
    \# & Name & & \multicolumn{1}{c}{$T_{\mathrm{eff}}$} & \multicolumn{1}{c}{$\log(g)$} & \multicolumn{1}{c}{$y$} & \multicolumn{1}{c}{$\xi$} & \multicolumn{1}{c}{$\varv \sin(i)$} & \multicolumn{1}{c}{$\zeta$} & \multicolumn{1}{c}{$E(B-V)$} & \multicolumn{1}{c}{$R_V$} & \multicolumn{1}{c}{$B.C.$} & \multicolumn{1}{c}{$M_V$} & \multicolumn{1}{c}{$M_\mathrm{bol}$} & \multicolumn{1}{c}{$M/\si{\Msun}$} & \multicolumn{1}{c}{$R/\si{\Rsun}$} & \multicolumn{1}{c}{$\log L / \si{\Lsun}$} & \multicolumn{1}{c}{$\log\tau_{\mathrm{evol}}$} & \multicolumn{1}{c}{$d_\mathrm{Gaia}$\tablefootmark{a}} & \multicolumn{1}{c}{$d_\mathrm{spec}$}\\
     \cline{7-9}
    & & & \multicolumn{1}{c}{K} & \multicolumn{1}{c}{cgs} & & \multicolumn{3}{c}{$\si{\kilo\meter\per\second}$} & \multicolumn{1}{c}{mag} & & \multicolumn{1}{c}{mag} & \multicolumn{1}{c}{mag} & \multicolumn{1}{c}{mag} & &  &  & \multicolumn{1}{c}{$\si{\year}$} & \multicolumn{1}{c}{$\si{\pc}$} & \multicolumn{1}{c}{$\si{\pc}$}\\
    \hline
     2 & HD\,14633       &       & 34000 & 3.90 & 0.088 &  6 & 125 & 74 & 0.09 & 2.80 & $-$3.05 & $-$4.67 & $-$7.72 & 23.2 & 9.0 & 4.98 & 6.64 & 1482           & 2142\\
       &                 & $\pm$ &  1000 & 0.10 & 0.008 &  2 &   5 & 18 & 0.02 & 0.10 &    0.07 &    0.22 &    0.21 &  1.3 & 1.2 & 0.08 & 0.07 & $^{159}_{131}$ &  253\\
     2a{\tablefootmark{b}} & & & & & & & & & & & & $-$3.74 & $-$6.79 & 9.8 & 5.8 & 4.61 & ... & & ...\\
        & & $\pm$ & & & & & & & & & & 0.32 & 0.32 & 3.0 & 1.1 & 0.13 & ... & & ...\\
     4 & HD\,258691      &       & 33300 & 4.02 & 0.080 &  8 &  20 & 39 & 0.90 & 3.35 & $-$3.08 & $-$4.08 & $-$7.16 & 19.9 & 7.2 & 4.76 & 6.63 & 1563           & 1433\\
       &                 & $\pm$ &   500 & 0.05 & 0.007 &  2 &   6 &  8 & 0.04 & 0.20 &    0.04 &    0.10 &    0.09 &  0.4 & 0.5 & 0.04 & 0.07 & $^{290}_{212}$ &   84\\
     5 & HD\,201345      &       & 32000 & 3.80 & 0.089 & 12 &  91 & 61 & 0.18 & 2.70 & $-$2.97 & $-$4.66 & $-$7.63 & 21.7 & 9.7 & 4.95 & 6.72 & 1918           & 2328\\
       &                 & $\pm$ &   500 & 0.10 & 0.007 &  2 &   2 & 10 & 0.02 & 0.10 &    0.04 &    0.28 &    0.28 &  1.8 & 1.6 & 0.11 & 0.02 & $^{147}_{128}$ &  284\\
     5a{\tablefootmark{b}} & & & & & & & & & & & & $-$4.27 & $-$7.21 & 14.7 & 8.0 & 4.78 & ... & & ...\\
        & & $\pm$ & & & & & & & & & & 0.32 & 0.32 & 4.0 & 1.5 & 0.13 & ... & & ...\\
     6 & HD\,57682       &       & 33500 & 3.93 & 0.082 &  4 &  10 & 33 & 0.12 & 3.10 & $-$3.11 & $-$4.40 & $-$7.51 & 21.7 & 8.4 & 4.90 & 6.65 & 1115           & 1228\\
       &                 & $\pm$ &   800 & 0.05 & 0.012 &  2 &   1 &  4 & 0.02 & 0.10 &    0.06 &    0.11 &    0.09 &  0.8 & 0.5 & 0.04 & 0.04 & $^{100}_{85}$  &   74\\
     7 & HD\,227757      &       & 33300 & 4.04 & 0.094 &  8 &  17 & 31 & 0.51 & 3.10 & $-$3.08 & $-$4.02 & $-$7.10 & 19.6 & 7.0 & 4.73 & 6.61 & 2080           & 2141\\
       &                 & $\pm$ &   400 & 0.08 & 0.008 &  2 &   4 &  4 & 0.02 & 0.10 &    0.03 &    0.20 &    0.20 &  0.7 & 0.8 & 0.08 & 0.14 & $^{62}_{59}$   &  201\\
     8 & BD\,$-$13\,4930 &       & 32900 & 4.14 & 0.093 &  6 &   8 & 23 & 0.55 & 3.30 & $-$3.07 & $-$3.63 & $-$6.70 & 18.0 & 6.0 & 4.58 & 6.49 & 1664           & 1786\\
       &                 & $\pm$ &   300 & 0.05 & 0.007 &  2 &   2 &  4 & 0.02 & 0.10 &    0.02 &    0.10 &    0.10 &  0.2 & 0.4 & 0.04 & 0.17 & $^{52}_{49}$   &  104\\
     9 & BD\,$+$60\,499  &       & 33800 & 4.09 & 0.092 &  6 &  19 & 34 & 0.83 & 3.25 & $-$3.13 & $-$3.91 & $-$7.04 & 19.7 & 6.6 & 4.71 & 6.53 & 2052           & 2023\\
       &                 & $\pm$ &   300 & 0.05 & 0.007 &  2 &   2 &  2 & 0.02 & 0.10 &    0.02 &    0.11 &    0.11 &  0.3 & 0.4 & 0.04 & 0.11 & $^{63}_{59}$   &  118\\
    10 & BD\,$+$57\,247  &       & 31200 & 3.87 & 0.096 & 16 &  16 & 31 & 0.63 & 2.51 & $-$2.91 & $-$4.30 & $-$7.21 & 19.0 & 8.4 & 4.78 & 6.76 & 3469           & 3403\\
       &                 & $\pm$ &   500 & 0.05 & 0.010 &  2 &   1 &  5 & 0.02 & 0.10 &    0.04 &    0.11 &    0.10 &  0.5 & 0.6 & 0.04 & 0.03 & $^{234}_{206}$ &  200\\
    11 & HD\,207538      &       & 31400 & 3.93 & 0.089 &  8 &  32 & 42 & 0.62 & 2.70 & $-$2.94 & $-$4.12 & $-$7.05 & 18.4 & 7.7 & 4.72 & 6.75 &  849           &  891\\
       &                 & $\pm$ &   500 & 0.05 & 0.006 &  2 &   3 &  3 & 0.02 & 0.10 &    0.04 &    0.10 &    0.09 &  0.4 & 0.5 & 0.04 & 0.03 & $^{12}_{12}$   &   52\\
    13 & HD\,46202       &       & 33900 & 4.16 & 0.080 &  6 &  11 & 33 & 0.55 & 3.00 & $-$3.14 & $-$3.71 & $-$6.85 & 19.2 & 6.0 & 4.64 & 6.38 & 1411           & 1130\\
       &                 & $\pm$ &   500 & 0.05 & 0.008 &  2 &   1 &  4 & 0.02 & 0.10 &    0.04 &    0.09 &    0.08 &  0.4 & 0.3 & 0.03 & 0.23 & $^{273}_{197}$ &   66\\
    14 & HD\,38666       &       & 33400 & 4.12 & 0.096 &  8 & 125 & 42 & 0.04 & 3.10 & $-$3.09 & $-$3.78 & $-$6.87 & 18.9 & 6.3 & 4.64 & 6.49 &  587           &  581\\
       &                 & $\pm$ &   300 & 0.05 & 0.007 &  2 &   8 & 10 & 0.02 & 0.10 &    0.02 &    0.11 &    0.11 &  0.3 & 0.4 & 0.04 & 0.15 & $^{33}_{29}$   &   34\\
    16 & HD\,54879       &       & 32200 & 4.06 & 0.086 &  4 &   0 &  0 & 0.36 & 3.10 & $-$3.02 & $-$3.78 & $-$6.80 & 17.8 & 6.5 & 4.62 & 6.65 & 1252           & 1149\\
       &                 & $\pm$ &   700 & 0.05 & 0.008 &  2 &   1 &  1 & 0.02 & 0.10 &    0.05 &    0.10 &    0.09 &  0.6 & 0.3 & 0.03 & 0.10 & $^{57}_{52}$   &   69\\
    17 & HD\,214680      &       & 34550 & 4.04 & 0.083 &  5 &  14 & 32 & 0.11 & 3.10 & $-$3.19 & $-$4.17 & $-$7.36 & 21.6 & 7.4 & 4.84 & 6.55 &  456           &  552\\
       &                 & $\pm$ &   300 & 0.05 & 0.009 &  2 &   1 &  2 & 0.02 & 0.10 &    0.02 &    0.12 &    0.12 &  0.5 & 0.5 & 0.05 & 0.08 & $^{29}_{26}$   &   32\\
    18 & HD\,34078       &       & 33200 & 4.06 & 0.085 &  8 &   9 & 23 & 0.56 & 3.20 & $-$3.08 & $-$3.93 & $-$7.01 & 19.2 & 6.8 & 4.70 & 6.60 &  389           &  418\\
       &                 & $\pm$ &   300 & 0.05 & 0.008 &  2 &   2 &  3 & 0.02 & 0.10 &    0.02 &    0.11 &    0.11 &  0.3 & 0.5 & 0.04 & 0.08 & $^{5}_{5}$     &   24\\
    19 & HD\,206183      &       & 33100 & 4.06 & 0.085 &  4 &   4 & 21 & 0.45 & 3.10 & $-$3.08 & $-$3.91 & $-$6.99 & 19.1 & 6.8 & 4.69 & 6.60 &  921           &  970\\
       &                 & $\pm$ &   300 & 0.05 & 0.006 &  2 &   2 &  4 & 0.02 & 0.10 &    0.02 &    0.11 &    0.11 &  0.3 & 0.5 & 0.04 & 0.08 & $^{16}_{16}$   &   57\\
    20 & HD\,36512       &       & 32900 & 4.20 & 0.087 &  7 &   8 & 26 & 0.05 & 3.10 & $-$3.07 & $-$3.45 & $-$6.52 & 17.5 & 5.5 & 4.51 & 6.29 &  407           &  386\\
       &                 & $\pm$ &   600 & 0.05 & 0.006 &  2 &   4 &  4 & 0.02 & 0.10 &    0.04 &    0.09 &    0.08 &  0.5 & 0.3 & 0.03 & 0.37 & $^{25}_{22}$   &   23\\
    \hline
    \end{tabular}
    \tablefoot{Uncertainties are 1$\sigma$ values.
\tablefoottext{a}{\cite{Gaia16,Gaia21}, distances and
uncertainties correspond to inverted EDR3 parallaxes.}
{\tablefoottext{b}{Alternative solution, adopting the \textit{Gaia} distance instead of the spectroscopic distance. See Sect.~\ref{sect:results} for details.}}}
\end{table*}

\subsection{Limitations of the hybrid non-LTE approach}
As our hybrid non-LTE approach is based on hydrostatic model atmospheres it faces limitations with respect to describing late O-type
stars comprehensively. In particular, the phenomena associated with the hydrodynamical stellar wind outflow can obviously not be addressed.
Wind parameters that require modelling, such as the mass-loss rate ($\dot{M}$) or the wind velocity-law parameter ($\beta$), cannot be constrained
-- though the wind terminal velocity ($\varv_\infty$) can be empirically determined from the absorption troughs of the observed UV resonance lines. However, these parameters do not impact
the photospheric layers in a significant way in the weak-wind case, so that the analysis of the photospheric spectrum remains unaffected.

Also, the generation of X-rays in wind-embedded shocks \citep[see e.g.][]{Carneiroetal16,Pulsetal20} cannot be described within this
framework. The soft X-ray and extreme-UV back-illumination of the photosphere can affect the ionisation balance there, such that the line
formation of some
particular spectral features may be impacted. In principle, photospheric lines connected to UV-transitions are affected, but the effects are
dependent on the strength of the illumination. The impact of X-rays on the metal-line formation was studied only in the
case of carbon so far \citep{Carneiroetal19}, finding negligible effects in the $T_\mathrm{eff}$ range between 30\,000 and 35\,000\,K both for
dwarfs and supergiants, a situation that changes above a $T_\mathrm{eff}$ of 40\,000\,K. We assume that other metals may follow this
behaviour, which, however, should be verified by studies similar to that of \citet{Carneiroetal19}.

Overall, it appears that no serious limitations for meaningful analyses based on a hydrostatic hybrid non-LTE approach exist for the
photospheric spectra of weak-wind late O-type stars. This may not only apply to the dwarfs and sub-giants investigated here, but may also
be extended to giant stars later than O8 at $\log L/L_\odot$\,$\lesssim$\,5.2 \citep{deAlmeidaetal19}. {These are expected to have stronger stellar winds,
of the order 10$^{-9}$ to 10$^{-8}$\,$M_\odot$\,yr$^{-1}$, which, however, is still weaker than the `strong wind' scenario discussed
in Sect.~\ref{sect:unified_models}, which had a negligible impact on the photospheric layers.} Moreover, the hybrid non-LTE
approach may be applicable at even higher temperatures for objects below the main sequence with insignificant stellar winds, such as
for hot members of the sub-dwarf B (sdB) class \citep[e.g.][]{Przybillaetal06b} and the hotter sub-dwarf O (sdO) stars
\citep{Heber09,Heber16}, to the point where non-LTE
effects become non-negligible for the atmospheric structure calculations. This would require the provision of model atoms for more highly-ionised metal species
than employed in the present work. On the other hand, the hydrogen- or helium-only atmospheres of hot white dwarfs should be readily applicable,
again up to the point where non-LTE effects become important for the atmospheric structure modelling.


\begin{table*}[ht]
\centering
\caption{Chemical abundances for the sample stars.}
\label{table:result_part2}
\small
    \begin{tabular}{rlr@{\hspace{0.1mm}}llllclllll}
    \hline
    \hline
    \# & Name & \multicolumn{10}{c}{$\log X/{\rm H}$+12} & Z\\
    \cline{4-12}
       &                 & & \ion{He}{i/ii}  &\ion{C}{ii/iii/iv}& \ion{N}{ii}& \ion{O}{ii/iii}& $\sum$CNO    & \ion{Ne}{ii}    & \ion{Mg}{ii}    & \ion{Al}{iii}   & \ion{Si}{iii/iv} & (by mass)\\
    \hline
     2 &  HD\,14633      &       & 10.98\,(13)     & 7.64\,(7)  & 8.46\,(8)  & 8.29\,(12) & 8.72   & 8.07\,(2) & 7.55\,(1) & 6.50\,(1) & 7.41\,(7)           & 0.010\\
       &                 & $\pm$ & \phantom{1}0.04 & 0.14       & 0.11       & 0.05       & 0.15   & 0.05      & 0.05      & 0.10      & 0.11              & 0.001\\
     4 & HD\,258691      &       & 10.94\,(12)     & 8.29\,(4)  & 7.89\,(10) & 8.61\,(20) & 8.83   & 8.30\,(8) & 7.41\,(3) & 6.40\,(3) & 7.21\,(10)          & 0.013\\
       &                 & $\pm$ & \phantom{1}0.04 & 0.03       & 0.08       & 0.05       & 0.07   & 0.10      & 0.03      & 0.03      & 0.12              & 0.001\\
     5 & HD\,201345      &       & 10.99\,(14)     & 7.74\,(6)  & 8.54\,(11) & 8.56\,(17) & 8.88   & 8.32\,(2) & 7.52\,(1) & 6.50\,(1) & 7.50\,(11)          & 0.014\\
       &                 & $\pm$ & \phantom{1}0.04 & 0.09       & 0.09       & 0.12       & 0.16   & 0.05      & 0.03      & 0.10      & 0.10              & 0.001\\
     6 & HD\,57682       &       & 10.95\,(11)     & 8.19\,(13) & 8.18\,(9)  & 8.62\,(35) & 8.86   & 8.13\,(8) & 7.40\,(3) & 6.35\,(3) & 7.32\,(13)          & 0.013\\
       &                 & $\pm$ & \phantom{1}0.07 & 0.07       & 0.05       & 0.06       & 0.09   & 0.06      & 0.03      & 0.03      & 0.11              & 0.001\\
     7 & HD\,227757      &       & 11.01\,(15)     & 8.50\,(14) & 8.14\,(11) & 8.83\,(38) & 9.05   & 8.30\,(4) & 7.65\,(1) & 6.60\,(3) & 7.58\,(17)          & 0.018\\
       &                 & $\pm$ & \phantom{1}0.04 & 0.07       & 0.08       & 0.09       & 0.13   & 0.04      & 0.05      & 0.03      & 0.09              & 0.002\\
     8 & BD\,$-$13\,4930 &       & 11.01\,(15)     & 8.39\,(15) & 8.10\,(11) & 8.76\,(38) & 8.98   & 8.31\,(5) & 7.62\,(2) & 6.53\,(3) & 7.50\,(13)          & 0.016\\
       &                 & $\pm$ & \phantom{1}0.03 & 0.10       & 0.06       & 0.06       & 0.10   & 0.03      & 0.04      & 0.03      & 0.06              & 0.001\\
     9 & BD\,$+$60\,499  &       & 11.01\,(13)     & 8.32\,(16) & 8.19\,(6)  & 8.83\,(39) & 9.02   & 8.29\,(4) & 7.58\,(3) & 6.47\,(3) & 7.48\,(16)          & 0.017\\
       &                 & $\pm$ & \phantom{1}0.04 & 0.09       & 0.02       & 0.10       & 0.16   & 0.03      & 0.03      & 0.05      & 0.07              & 0.002\\
    10 & BD\,$+$57\,247  &       & 11.02\,(12)     & 8.01\,(15) & 8.35\,(12) & 8.43\,(9)  & 8.77   & 8.05\,(3) & 7.43\,(3) & 6.34\,(3) & 7.09\,(7)            & 0.010\\
       &                 & $\pm$ & \phantom{1}0.05 & 0.09       & 0.07       & 0.06       & 0.09   & 0.03      & 0.03      & 0.05      & 0.11              & 0.001\\
    11 & HD\,207538      &       & 10.99\,(13)     & 8.19\,(10) & 8.21\,(15) & 8.63\,(36) & 8.87   & 8.21\,(6) & 7.57\,(3) & 6.47\,(3) & 7.36\,(16)           & 0.013\\
       &                 & $\pm$ & \phantom{1}0.03 & 0.08       & 0.06       & 0.06       & 0.09   & 0.05      & 0.03      & 0.05      & 0.13              & 0.001\\
    13 & HD\,46202       &       & 10.94\,(15)     & 8.20\,(11) & 7.84\,(8)  & 8.62\,(34) & 8.81   & 8.10\,(7) & 7.48\,(1) & 6.36\,(3) & 7.30\,(12)           & 0.012\\
       &                 & $\pm$ & \phantom{1}0.05 & 0.10       & 0.05       & 0.05       & 0.09   & 0.06      & 0.03      & 0.03      & 0.04              & 0.001\\
    14 & HD\,38666       &       & 11.02\,(13)     & 8.50\,(6)  & 8.25\,(6)  & 8.82\,(16) & 9.06   & 8.27\,(1) & 7.58\,(1) & 6.58\,(2) & 7.45\,(11)           & 0.017\\
       &                 & $\pm$ & \phantom{1}0.04 & 0.04       & 0.15       & 0.04       & 0.08   & 0.03      & 0.03      & 0.05      & 0.10              & 0.001\\
    16 & HD\,54879       &       & 10.97\,(16)     & 8.27\,(13) & 7.67\,(10) & 8.60\,(38) & 8.80   & 8.19\,(7) & 7.50\,(3) & 6.34\,(5) & 7.26\,(15)           & 0.012\\
       &                 & $\pm$ & \phantom{1}0.04 & 0.07       & 0.04       & 0.07       & 0.11   & 0.11      & 0.04      & 0.05      & 0.16              & 0.001\\
    17 & HD\,214680      &       & 10.95\,(9)      & 8.36\,(13) & 8.11\,(13) & 8.77\,(40) & 8.98   & 8.24\,(7) & 7.59\,(1) & 6.50\,(3) & 7.48\,(15)           & 0.016\\
       &                 & $\pm$ & \phantom{1}0.05 & 0.07       & 0.08       & 0.06       & 0.10   & 0.07      & 0.03      & 0.03      & 0.10              & 0.001\\
    18 & HD\,34078       &       & 10.97\,(13)     & 8.26\,(12) & 7.78\,(12) & 8.65\,(35) & 8.84   & 8.26\,(8) & 7.50\,(3) & 6.41\,(3) & 7.33\,(16)           & 0.013\\
       &                 & $\pm$ & \phantom{1}0.05 & 0.07       & 0.11       & 0.06       & 0.10   & 0.07      & 0.03      & 0.03      & 0.12              & 0.001\\
    19 & HD\,206183      &       & 10.97\,(13)     & 8.26\,(13) & 7.89\,(12) & 8.60\,(32) & 8.82   & 8.20\,(5) & 7.41\,(1) & 6.28\,(3) & 7.29\,(13)           & 0.012\\
       &                 & $\pm$ & \phantom{1}0.04 & 0.09       & 0.08       & 0.06       & 0.10   & 0.03      & 0.03      & 0.03      & 0.08              & 0.001\\
    20 & HD\,36512       &       & 10.98\,(15)     & 8.32\,(17) & 7.85\,(12) & 8.68\,(41) & 8.88   & 8.22\,(7) & 7.47\,(1) & 6.39\,(3) & 7.32\,(15)           & 0.013\\
       &                 & $\pm$ & \phantom{1}0.03 & 0.10       & 0.05       & 0.06       & 0.11   & 0.04      & 0.03      & 0.03      & 0.08              & 0.001\\
    \hline
       & CAS~$^{a,b}$    &       & 10.99           & 8.35       & 7.79       & 8.76       & 8.93   & 8.09      & 7.56      & 6.30      & 7.50              & 0.014\\
       &                 & $\pm$ & \phantom{1}0.01 & 0.04       & 0.04       & 0.05       & 0.08   & 0.05      & 0.05      & 0.07      & 0.06              & 0.002\\
    \hline
    \end{tabular}
        \tablefoot{Numbers in brackets denote the number of analysed spectral lines. Each spectral line was given the same weight
        for calculating the average. Abundance uncertainties are 1$\sigma$ standard deviations from the
        line-to-line scatter, not standard errors of the mean.\\
        $^{(a)}$~\citet{NiPr12}~~~$^{(b)}$~\citet{Przybillaetal13}}
\end{table*}

  \section{Results}\label{sect:results}
The results of the sample star analyses are summarised in
Tables \ref{table:result_part1} and \ref{table:result_part2}. The
former lists the stellar parameters: the internal identification
number is given, the HD designation, effective temperature, surface gravity,
surface helium abundance (number fraction), microturbulent, projected rotational and
macroturbulent velocity, colour excess, total-to-selective extinction parameter,
bolometric correction, absolute visual and bolometric magnitude,
evolutionary mass, radius, luminosity, evolutionary age, the \textit{Gaia}
EDR3 distance (from the inversion of the parallax) and the spectroscopic distance.
The corresponding 1$\sigma$ uncertainties are given in the lines below the observed values.
Table~\ref{table:result_part2} concentrates on the elemental
abundances: again, the internal identification number and the HD designation
are given and abundance values for eight elements, plus the sum of the
carbon, nitrogen and oxygen abundances and the stellar metallicity (by mass). The number of spectral lines
for each ionisation stage employed for the analysis
is indicated in brackets, and 1$\sigma$ standard deviations are given
to characterise the uncertainties. For comparison, cosmic abundance
standard (CAS) values as found from early B-type stars in the solar neighbourhood
\citep{Przybillaetal08,NiPr12,Przybillaetal13} are also tabulated.

\subsection{Atmospheric and fundamental stellar parameters}
Relative uncertainties in the stars' effective
temperatures $\delta T_{\mathrm{eff}}$\,= $\Delta
T_{\mathrm{eff}}/T_{\mathrm{eff}}$\,$\approx$\,1--3\% and
uncertainties of $\Delta \log g$\,$\approx$\,0.05 to 0.10\,dex
in surface gravity are found in the present work. This is comparable to previous work
on analyses of massive early-type main-sequence to supergiant stars
using the {\sc Atlas-Detail-Surface} suite of codes
\citep{Przybillaetal06a,NiPr12,FiPr12,Wessmayeretal22}.
Relative uncertainties of the helium abundances amount to $\delta
y$\,$\approx$\,7--15\% owing to the line-to-line scatter
from the set of analysed \ion{He}{i} lines.

The uncertainty of the microturbulent velocity is limited by
the step size in the grid used for fitting, and so a conservative
estimate of $\Delta \xi$\,$\approx$\,2\,km\,s$^{-1}$ was adopted
throughout this study. The effects of projected rotational and
macroturbulent velocity on the line profiles are interconnected.
The relative uncertainties amount to from below 10\% to several 10\%.
An exception is the star HD 54879, which has an exceptionally low
rotational velocity.

Another (weaker) ambiguity is found for the pair of colour excess and
the total-to-selective extinction parameter. The fitting
described in Sect.~\ref{subsection:sed_fitting} produced uncertainties
of $\Delta E(B-V)$\,$\approx$\,0.02\,mag and $\Delta R_V$\,$\approx$\,0.1
in most cases. A $\Delta B.C.$ of 0.02 to 0.07\,mag was derived for
the sample stars and the
uncertainties in the absolute visual and bolometric magnitudes
amount to 0.1 to 0.2\,mag typically.

Values for evolved masses $M_{\mathrm{evol}}$ were derived based on
zero-age main sequence (ZAMS) mass estimates, considering the mass lost during the
main-sequence evolution by inspection of the
evolution tracks of \cite{Ekstroemetal12} (see
our Sect.~\ref{subsection:evolutionary_status}). The
uncertainties of the evolved masses are assumed to be identical to the
ZAMS mass uncertainty, $\delta M_{\mathrm{ZAMS}}$\,=\,$\delta
M_{\mathrm{evol}}$, with values ranging between 2 and 5\% typically.
Radii show relative errors $\delta R$ from 5 to slightly over 10\%
in most cases. Luminosity uncertainties amount to about 10-20\%
typically. Finally, the derived \textit{Gaia} EDR3-based and the spectroscopic distances
show a large range of relative uncertainties, from about 1 to 20\%,
with the \textit{Gaia} values typically being the more precise.

\begin{figure*}
    \centering
    \includegraphics[width=0.98\hsize]{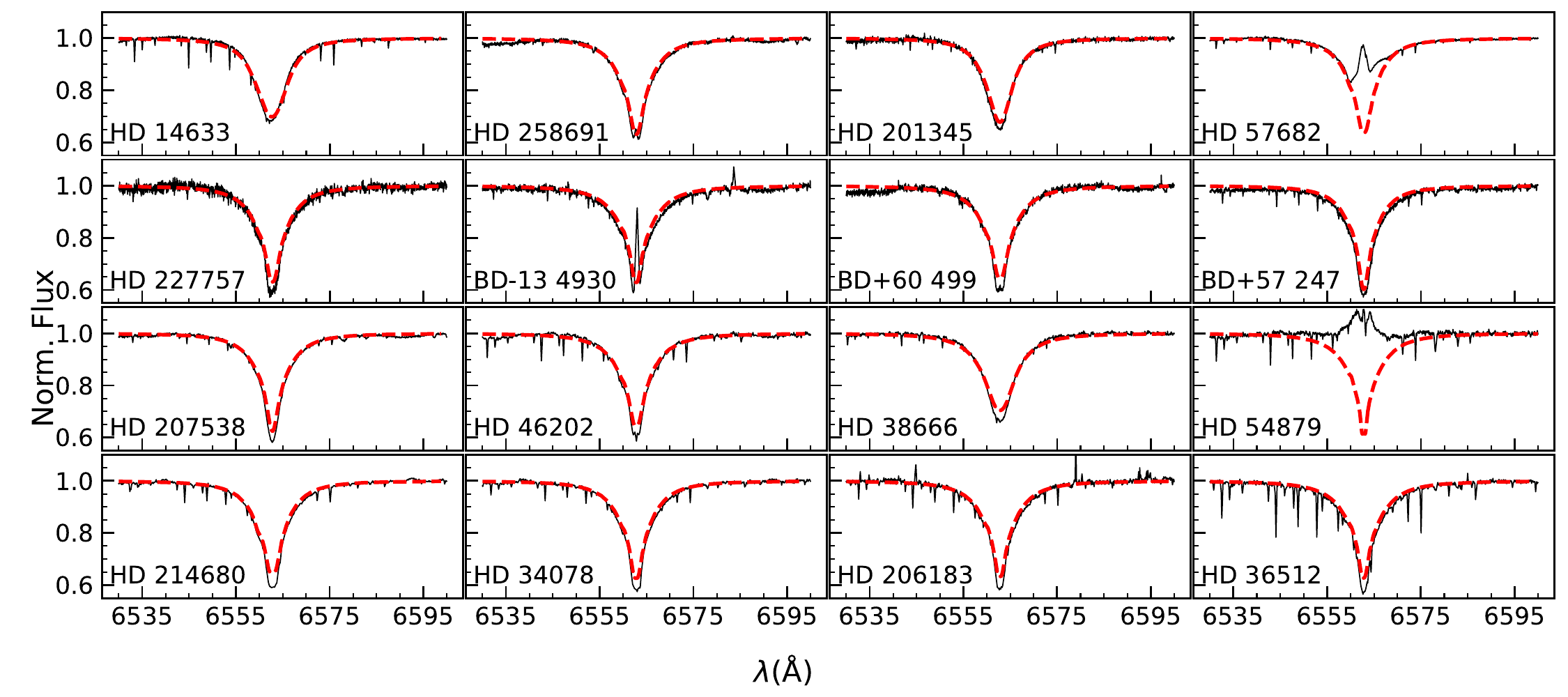}
    \caption{{Comparison of the observed spectra (black solid) and the global best fitting model (red dashed) of the H$\alpha$ line for all analysed stars. }}\label{fig:halpha}
\end{figure*}

For our two ON stars, HD~14633 and HD~201345, two solutions are given in Table~\ref{table:result_part1}. The solutions with
IDs \#2 and \#5 were derived under the standard assumptions described above. However, these two stars show notably discrepant spectroscopic
and \textit{Gaia} parallaxes. Assuming the \textit{Gaia} distances and not the spectroscopic distances to be correct, one can redetermine the stars' absolute
visual and bolometric magnitudes by adopting the \textit{Gaia}-based distance modulus (the reddening values and bolometric corrections remain
unaltered), and use {Eq.}~\ref{equ_dspec} to solve for the masses (yielding larger relative mass uncertainties) and derive the other
fundamental parameters. These are the alternative solutions \#2a and \#5a in Table~\ref{table:result_part1}. We note that all atmospheric
parameters are unchanged for these alternative solutions and the age cannot be determined, because a different evolution scenario, possibly
involving mass exchange in a close binary system, needs to be invoked, which is beyond the scope of the present paper (see Sect.~\ref{subsection:evolutionary_status} for further discussion).

{The H$\alpha$ line is the most sensitive spectral feature in the optical to effects of a stellar wind. A comparison of the global best fitting model based on the derived atmospheric parameters with the observed H$\alpha$ profiles for all analysed sample stars is shown in Fig.~\ref{fig:halpha}. Good matches are obtained typically, except for the very cores. As implied by the comparison in Sect.~\ref{sect:unified_models}, the cores are slightly too shallow, because of locally too high temperatures of the atmospheric models in the H$\alpha$ line core formation region.

The peculiar H$\alpha$ shapes of three sample stars are also understood. The star BD~$-$13~4930 shows an emission feature in the line core that stems from a \ion{H}{ii} region (we also note the weak [\ion{N}{ii}] emission components). The emission in the spectrum of HD~57682 stems from magnetically confined circumstellar gas \citep{Grunhutetal09}, and analogously in the case of HD~54879 \citep{Castroetal15}.}

\begin{figure*}
    \centering
    \includegraphics[width=0.98\hsize]{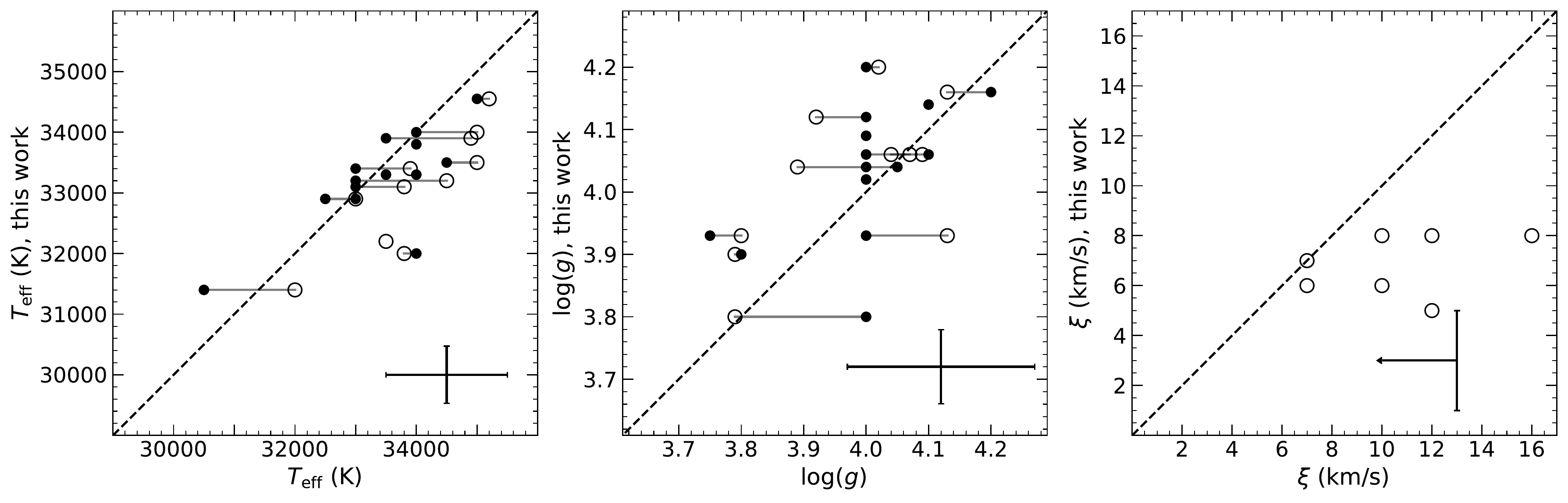}
    \caption{Comparison of our $T_\mathrm{eff}$ (left), $\log g$-values (middle), and microturbulence (right panel) with
    data from \citet[dots]{Martinsetal15a,Martinsetal15b} and \citet[circles]{Holgadoetal18,Holgadoetal22}. {A uniform value of $\xi$\,=\,10\,km\,s$^{-1}$ was adopted for the photospheric layers of the stars in the Martins et al.~work.} Different literature results for the same star are connected by black lines.
    Typical error bars from the Martins et al.~work are indicated in the left and middle panels; the uncertainties from Holgado et al.~are slightly smaller, and for the microturbulent velocities they provide only upper limits in many cases.}\label{fig:teffloggcomparison}
\end{figure*}

\subsection{Comparison with literature values}\label{subsection:comparison}
Several of our sample stars were analysed in previous studies.
In order to compare our
hybrid non-LTE approach with full non-LTE studies we
discuss two cases as examples in this section, without aiming to achieve
completeness with respect to the available literature (a few cases will be summarised in Sect.~\ref{sect:summary}).

\begin{table}[ht!]
\centering
\caption{Comparison of atmospheric parameters for 10~Lac.}
\label{table:10Lac}
\small
\begin{tabular}{llll}
\hline\hline
$T_\mathrm{eff}$\,(K) & $\log g$\,(cgs)        & $\xi$\,(km\,s$^{-1}$)    & Reference\\
\hline
35500$\pm$500         & 3.95$\pm$0.10          & 10\tablefootmark{a}      & \citet{Herreroetal02}\\
36000$^{+800}_{-900}$ & 4.03$^{+0.13}_{-0.12}$ & 15.5$^{+4.1}_{-3.8}$     & \citet{Mokiemetal05}\\
36000$\pm$1000        & 3.9$\pm$0.1            & 10\tablefootmark{a}      & \citet{SimonDiazetal06}\\
35000$\pm$1000        & 4.05$\pm$0.15          & $\ge$10\tablefootmark{b} & \citet{Martinsetal15a}\\
35200$\pm$500         & 3.89$\pm$0.05          & 12                       & \citet{Holgadoetal18}\\
34550$\pm$300         & 4.04$\pm$0.05          & 5$\pm$2                  & this work\\
\hline
\end{tabular}
\tablefoot{
\tablefoottext{a}{adopted}
\tablefoottext{b}{adopted as 10\,km\,s$^{-1}$ at the photosphere base and reaching 10\% of the
terminal velocity at the top of the atmosphere}}
\end{table}

The first case we discuss are the atmospheric parameters of the O9\,V
standard star 10~Lac (HD~214680). Modern studies with
line-blanketed hydrodynamic non-LTE atmospheres, such as {\sc Fastwind} and
{\sc Cmfgen,} have converged on consistent values for
$T_\mathrm{eff}$ and $\log g$, with values typically overlapping within the
respective uncertainties, as summarised in Table~\ref{table:10Lac}. Microturbulent velocities $\xi$
are often adopted in the literature, only sometimes determined.
Our solution is only slightly cooler, indicates a gravity on the
high side of the previous determinations and has a much smaller microturbulence, but we stress the
excellent match of the synthetic with the observed spectrum (see
Fig.~\ref{fig:10Lac_fit}), the match of the SED fit, and the
agreement of the spectroscopic with the Lac OB1b association distance (in
the absence of a consistent parallax measurement), as discussed later.
From this we can conclude that our approach produces an overall
consistent result with solutions from full non-LTE modelling for this
anchor of the MK system, though providing a significantly smaller microturbulent velocity.
The modern $T_\mathrm{eff}$ values are
about 2000\,K cooler than derived from analyses based on pure H+He
non-LTE model atmospheres in the past \citep[for a discussion, see e.g.][]{Herreroetal02}.

The second case is the comparison of the solutions for the many
sample stars in common with the work of \citet{Martinsetal15a} and \citet{Holgadoetal18,Holgadoetal22}.
This is visualised in Fig.~\ref{fig:teffloggcomparison}. The agreement
in effective temperature determinations (left panel) is very good with the Martins et al.~data,
with only one significant outlier (the ON star HD~201345), while the Holgado et al.~results are shifted overall to higher $T_\mathrm{eff}$. Our surface gravities
are in many cases consistent with the Martins et al.~results, though a small shift towards
higher values should be noted (middle panel, again with the exception of HD~201345). The same trend is more pronounced  compared with the Holgado et al.~data
\citep[in analogy to the findings from a comparison between {\sc Cmfgen} and {\sc Fastwind} by][]{Masseyetal13}.
\citet{Martinsetal15a} adopted a depth-variable microturbulent velocity starting
from 10\,km\,s$^{-1}$ at the photosphere and reaching 10\% of the terminal velocity at the top of the
atmosphere. Consequently,
the comparison of microturbulent velocities in Fig.~\ref{fig:teffloggcomparison} is restricted
to the work of \citet{Holgadoetal18}, with our values being smaller by trend
(we note that only upper limits were provided by Holgado et al. in many cases from the analysis of the \ion{He}{i/ii} lines,
which are not ideal indicators for microturbulence because of their strengths).
We conclude that our hydrostatic hybrid non-LTE approach to the quantitative
spectroscopy of weak-wind O-type dwarfs produces overall results
consistent with state-of-the-art hydrodynamic full non-LTE model atmosphere analyses, albeit at
smaller microturbulence \citep[as in B-type supergiants;][]{Wessmayeretal22}.

Finally, we briefly mention that the O9.2\,V star HD~46202 and the O9.7\,V star HD36512 were
analysed previously with a hybrid non-LTE modelling approach very similar to the one employed here,
by \citet{Briquetetal11}, and by \citet{NiSi11} and \citet{NiPr12}, respectively. The
solutions at $T_\mathrm{eff}$\,=\,34\,100$\pm$600\,K, $\log g$\,=\,4.17$\pm$0.07 and
$\xi$\,=\,6$\pm$2\,km\,s$^{-1}$ for HD~46202 and $T_\mathrm{eff}$\,=\,33\,400$\pm$200\,K,
$\log g$\,=\,4.30$\pm$0.05 and $\xi$\,=\,4$\pm$1\,km\,s$^{-1}$ for
HD~36512 are consistent with our results.

\subsection{Elemental abundances}\label{subsection:abundances}
Uncertainties for the elemental abundances in Table~\ref{table:result_part2}
were conservatively estimated as the $1\sigma$ standard deviations
from the line-to-line scatter of the individual line abundances within
a chemical species. In general, these statistical uncertainties typically
range from $\sim$0.05--0.10\,dex and rarely exceed the
upper value. The number of lines analysed per species and object is
larger or equal to two, typically. Exceptionally, only the strongest \ion{Mg}{ii}
line $\lambda$4481\,{\AA} can be analysed for the faster rotators
among the sample stars. Standard errors of the mean
amount to typically 0.01--0.03\,dex for the elemental abundances in each star.

In addition, systematic errors need to be considered for the abundances.
They depend primarily on the quality of the model atoms (on the extent, on the processes considered, and on the atomic data employed) and on the uncertainties in the atmospheric parameters (see for example \citealt{Przybillaetal00,Przybillaetal01a,Przybillaetal01b} and \citealt{PrBu01}).
Given the experience gained previously, we expect
the systematic uncertainties here to amount to $\sim$0.1\,dex as well.

Table~\ref{table:result_part2} also gives the metallicity $Z$ of the sample stars (mass fraction).
As our coverage of important elements is incomplete, missing elements were supplemented by assuming solar
abundances \citep{Asplundetal09}. Monte Carlo error propagation was used to constrain the uncertainties.

The abundance determination for many chemical species allows
global synthetic spectra to be calculated for each star. This includes
blended features that were excluded from the detailed analysis.
The global synthetic spectra can closely reproduce many details of the observed spectra, but several features, like \ion{N}{iii} lines, are still unaccounted for (see Appendix~\ref{appendix:B} for a discussion and Fig.~\ref{fig:10Lac_fit} for a comparison of synthetic and observed spectrum for the example of 10~Lac). The goodness-of-fit of our global best model for this star is representative considering the sample overall. We chose to show a detailed spectral fit for this star because it is a very important star among our sample as the MK standard for the spectral-type O9V. In addition, we chose 10~Lac because it is the hottest object of our sample, where non-LTE effects as well as wind effects can be expected to be among the strongest in the sample stars.
Many photospheric emission lines in the spectrum of 10~Lac
and the other sample stars are reliably reproduced by the global synthetic spectra as well (see Fig.~\ref{fig:emission} for some examples). The reproduction
is not as complete as reached in previous work on BA-type supergiants
\citep{Przybillaetal06a} and B-type main-sequence to supergiant stars
\citep{NiPr12,Wessmayeretal22}. This will require
additional efforts in providing the required model atoms.

\begin{figure}[ht]
   \centering
   \includegraphics[width=.995\hsize]{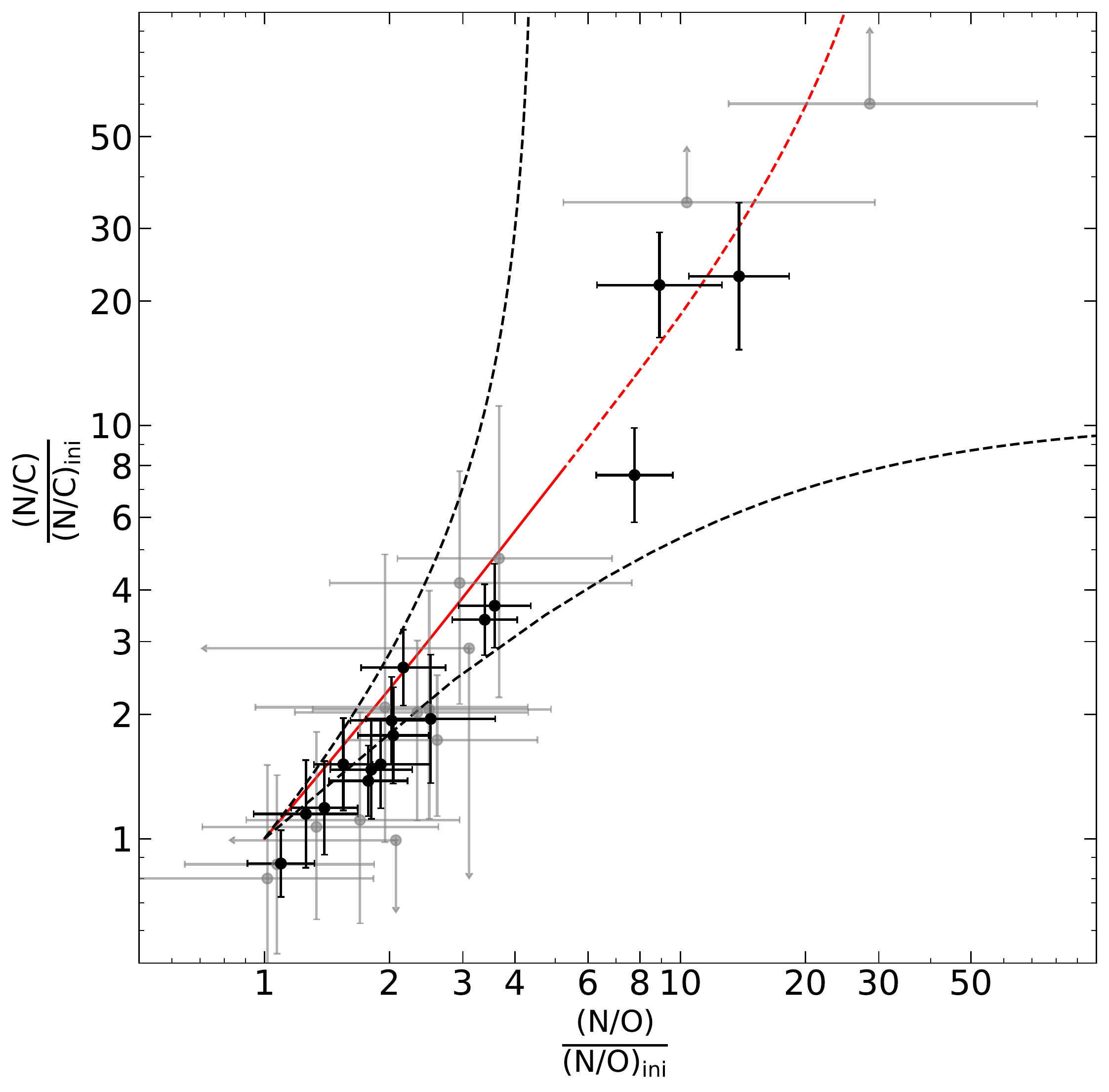}
      \caption{Ratio of the abundance number fraction (N/C) as a function of N/O. The observed abundances (black dots with error bars) are normalised
      with respect to CAS values and the theoretical data to the solar abundances used for the stellar evolution models. The red line shows a
      rotating evolutionary model of a \SI{20}{\Msun} star \citep{Ekstroemetal12}, {which is representative for the mean mass of our sample stars.
      The} transition from solid to dashed indicates the end of the main sequence. The dashed black lines show theoretical
      bounds for the CN cycle (i.e. assuming oxygen shows a constant abundance, shown with the upward-evolving curve) and for the ON cycle (i.e. assuming
      carbon to be at equilibrium abundance, shown by the horizontally evolving curve). The same stars from the analyses of \citet{Martinsetal15a,Martinsetal15b}
      are shown in grey.}
         \label{figure:CNO}
\end{figure}

Previous work on elemental abundances of early B-type stars at distances out to
about 400\,pc from the Sun has found chemical homogeneity, establishing the CAS
values (see Table~\ref{table:result_part2}). A direct comparison of the O-star abundances
is only possible for two objects, HD~34078 (ID\#18) and HD~36512 (ID\#20), with the abundances of the latter
being in good agreement with the study of \citet{NiPr12}. Overall, agreement of the abundance values
for these stars and the CAS is found within the mutual uncertainties, though the neon and
aluminium abundances appear to be slightly higher than the CAS values. This holds if the comparison
is extended to stars at distances out to $\sim$1\,kpc that are located in the Local Spiral Arm,
at similar galactocentric radii as the original CAS sample, objects HD207538 (ID\#11),
HD~38666 (ID\#14), HD~214680 (ID\#17), and HD~206183 (ID\#19). Elemental abundances in our sample stars beyond the
Local Spiral Arm are subject to Galactic abundance gradients.

\subsection{Signatures of mixing with CNO-processed material}
Several effects lead to mixing of the atmospheres of rotating stars with CNO-processed matter from the stellar core, such as
meridional circulation or shear mixing in the presence of differential rotation \citep[e.g.][]{MaMe12,Langer12}, further modified
by magnetic fields. The most sensitive indicators for mixing with nuclear-processed matter are the nitrogen-to-carbon (N/C)
and nitrogen-to-oxygen (N/O) ratios, which can be combined into a N/C versus N/O plot \citep{Przybillaetal10}. This plot shows little dependence
on the initial stellar masses, rotation velocities, and nature of the mixing processes up to relative enrichment of N/O by a factor of $\sim$4, and thus it constitutes an ideal test for the quality of observational results \citep{Maederetal14}.

Ratios of the surface carbon, nitrogen
and oxygen fractions are expected to appear in a relatively tight locus in the N/C versus N/O diagram, as displayed in Fig.~\ref{figure:CNO}.
The abundance ratios were normalised to the initial values so as to make the comparison with the evolution tracks easier -- the
observations were normalised relative to CAS abundances and the models to their respective (solar) initial values. Two analytical boundary
solutions can be defined. Either the CN cycle can be assumed to be operational, with oxygen to remain constant at the initial high value, which
corresponds to the curve that moves almost vertically away from the pristine values; or the ON-cycle is assumed to be at work, with carbon to remain
at a low equilibrium value, which corresponds to the horizontal boundary curve. The predictions of stellar evolution models fall between the two
extremes, as with the track for a rotating 20\,$M_\odot$ star of \citet{Ekstroemetal12} shown in Fig.~\ref{figure:CNO}.

The CNO signatures of the sample stars indeed follow the predicted theoretical locus in Fig.~\ref{figure:CNO} tightly, with most objects being
compatible with the predicted amount of mixing on the main sequence and falling below a value of four times the initial value.
However, we note that the prediction is for a star with an initial rotation $\varv_\mathrm{ini}$ of 40\% of the critical velocity
$\varv_\mathrm{crit}$, while the true rotational
velocities of the sample stars are not known, only their $\varv \sin i$ values. The most mixed object that may still be compatible with
a main-sequence scenario is BD\,$+$57\,247 (ID\#10) at N/C and N/O values around eight times the initial values, probably stemming from an
initially fast rotator, as the star is still well within its core hydrogen-burning phase (see Sect.~\ref{subsection:evolutionary_status}).
The two other objects with a much higher mixing signature are the two ON stars, HD~14633 (ID\#2) and HD~201345 (ID\#5), which are likely to have
reached their extremely high amount of mixing not by rotationally induced processes in a single-star scenario, but from a binary channel
involving mass overflow (see the discussion in Sect.~\ref{subsection:evolutionary_status}).

As for the atmospheric parameters, our CNO results are also compared to the data from the full non-LTE model atmosphere analyses by
\citet{Martinsetal15a,Martinsetal15b}, which
cover almost all of the sample stars (grey symbols in Fig.~\ref{figure:CNO}). Noticeable are the much smaller error bars in our analysis (we note the
logarithmic axes of the figure), and some systematic shifts, which are easily recognised for the two ON stars. Moreover, several upper or lower
limits from the work of Martins et al.~could be replaced by actual determinations in our work. We consider the significant
reduction of uncertainties in abundance determinations with respect to previous work on weak-wind late O-type stars as one of the most
important achievements of the present work.

Finally, we want to draw attention to the sum of CNO abundances, $\sum$CNO\,=\,$\log$\,($\sum$CNO/H)\,+\,12, which can be used to verify
the catalytic nature of the CNO-process, where individual nuclei of C, N and O are only transmuted, the sum being conserved.
Again, a meaningful comparison with the CAS values can only be made for stars IDs \#18 and \#20 (see Sect.~\ref{subsection:abundances}),
which leads to compatible values. Also, for the extended range of stars \#11, \#14, \#17 and \#19 compatibility is found, with the exception
of slightly high values for HD~38666 (ID\#14), which, however, are in agreement when systematic uncertainties of the abundances
are considered. For the other objects Galactic abundance gradients have to be considered once more.
Only the ON star, HD~14633 (ID\#2), is noticeable for its low $\sum$CNO value. {We note that this might be explained by contamination from an additional source (see Sect.~\ref{sect:obs}).}
A second continuum reduces the contrast between line to continuum opacity,
weakening the original line depths with respect to a single-star scenario and consequently reducing the derived abundances, which may be a factor
here.

\begin{figure}
   \centering
   \includegraphics[width=\hsize]{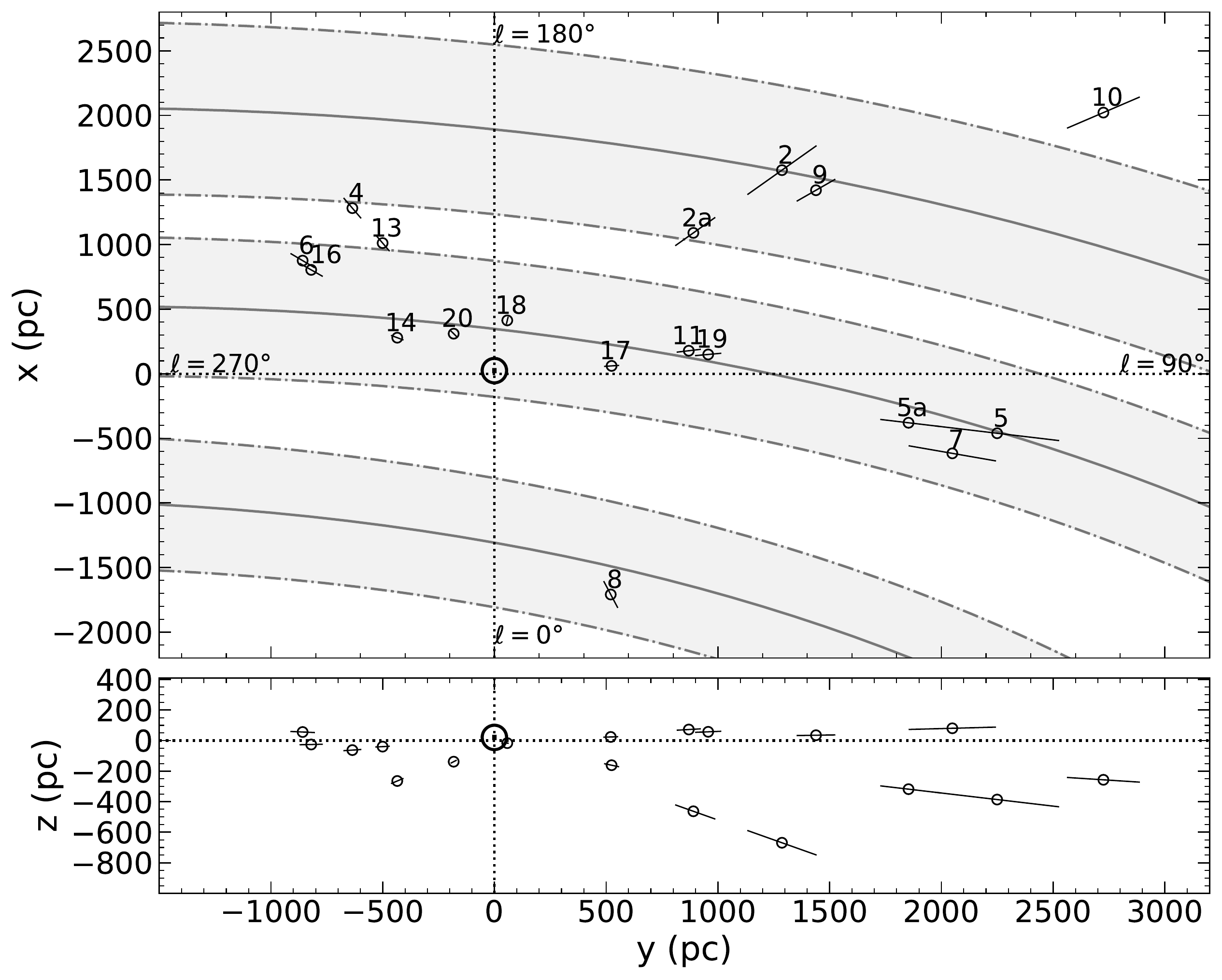}
      \caption{Distribution of the analysed stars in the Milky Way, relative to the position of the Sun at the origin of the Cartesian
      coordinate system. {\em Upper panel}: Distribution of the stars in the Galactic plane; the Galactic centre lies in
      the direction $\ell$\,=\,0\degr. The sketched spiral arm locations are according to \citet{Xuetal21}, based on the spiral structure model of \citet{Reidetal19}, from top to bottom: the Perseus, Local, and Sagittarius spiral arms.
      {\em Lower panel}: Elevation of the stars above or below the plane. The error bars indicate
      distance uncertainties from the spectroscopic distances and for IDs\#2a and 5a from the inverted \textit{Gaia} EDR3 parallaxes. Numbers correspond to the internal ID numbers.}
      \label{figure:position_distribution}
\end{figure}

\subsection{Spectroscopic distances}\label{subsection:spec_distances}
Spectroscopic distances for the sample stars as calculated using {Eq.}~\ref{equ_dspec} are quoted in Table~\ref{table:result_part1}. Based
on these the distribution of the sample stars in the Milky Way in a Cartesian coordinate system centred on the Sun is shown in
Fig.~\ref{figure:position_distribution}. This shows the projection on the Galactic plane in the upper panel and perpendicular to it in the
lower panel. The positions of the stars can be compared to the local spiral structure based on \textit{Gaia} EDR3 data by \citet{Xuetal21},
implying that stars ID\#5 to 7, 11, 14, and 16 to 20 are located in the Local spiral arm, star ID\#8 is located in the Sagittarius spiral arm
and the remainder of the analysed stars in the Perseus spiral arm. {It is reassuring to find these young stars overall within spiral arms, as expected, and it is a first qualitative indicator that the spectroscopic distances are not too far off from \textit{Gaia} EDR3 constraints on the star-forming regions of the Milky Way.}
With regard to the $z$ distance, most stars are located within one scale
height of the thin disc, {again as can be expected}. Exceptions are the two runaway and ON stars HD~14633 (ID\#2) and HD~201345 (ID\#5), with BD\,+57~247 (ID\#10),
appearing to be an intermediate case \citep[we note that the scale height is in principle increasing towards the outer Galactic
disc,][]{Yuetal21}. The classical runaway star HD~38666 ($\mu$\,Col, ID\#14) has also reached some distance from the Galactic plane, about two
scale heights. The other classical runaway in the star sample, HD34078 (AE Aur, ID\#18) is located well within the thin disc.

\begin{figure}
   \centering
   \includegraphics[width=.9\hsize]{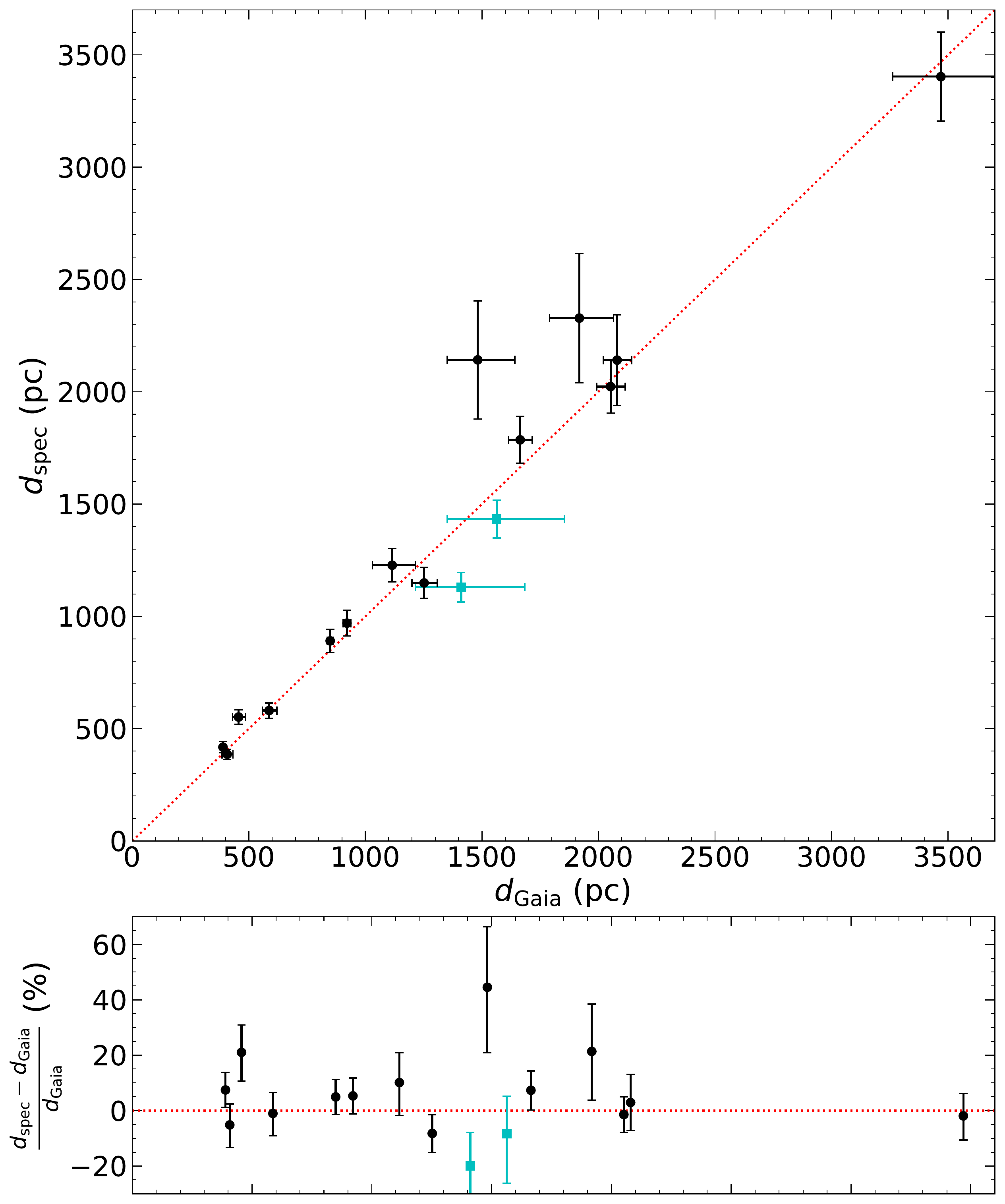}
      \caption{Comparison of \textit{Gaia} EDR3 parallax-based distances with our spectroscopic values (upper panel) and the relative differences between them (lower panel). Stars with a RUWE factor $>$4 are marked by cyan squares.}
         \label{figure:d_spec}
\end{figure}

A comparison of our spectroscopic distances with the independent distances based on inverted \textit{Gaia} EDR3 parallaxes is shown in
Fig.~\ref{figure:d_spec}. Overall, good agreement of the two independent distances is found, with the relative differences amounting to
typically less than $\sim$10\%. Significantly larger spectroscopic distances
are found for HD~214680 (10~Lac) and the two runaway ON stars HD~14633 and HD~201345, a significantly lower spectroscopic distance is found
for HD~46202. The case of 10~Lac is discussed in detail in Appendix~\ref{appendix:A}. In brief, the \textit{Gaia} EDR3 parallax measurement seems to
be systematically biased, but an average \textit{Gaia} EDR3-based distance to the Lac OB1b association -- if adopted for 10~Lac -- is in good
agreement with the spectroscopic distance. The differences for the two ON stars are likely related to their particular evolutionary history
(see below), and the spectroscopic distance to HD~46202 may be affected by unrecognised systematic effects of the companion on the analysed spectrum. We note that the \textit{Gaia} EDR3 distance to HD~46202 fits the parent NGC~2244 cluster distance better (see Appendix~\ref{appendix:C}) despite the close companion not being accounted for in the five-parameter solution of EDR3, probably leading to the large re-normalised unit weight error (RUWE) $>$4 for this star.

\begin{figure}
    \centering
    \includegraphics[width=0.97\hsize]{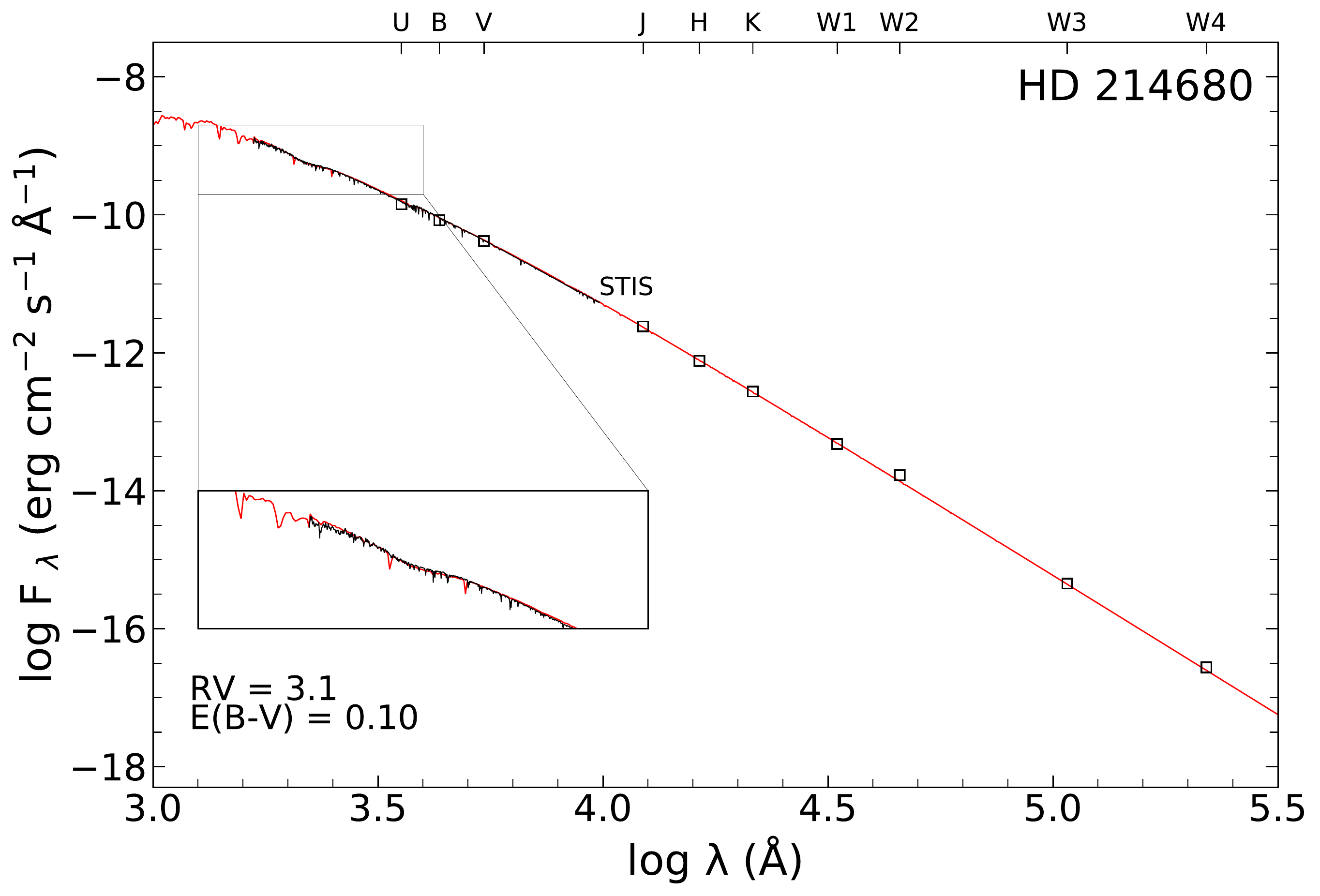}\\
    \includegraphics[width=0.97\hsize]{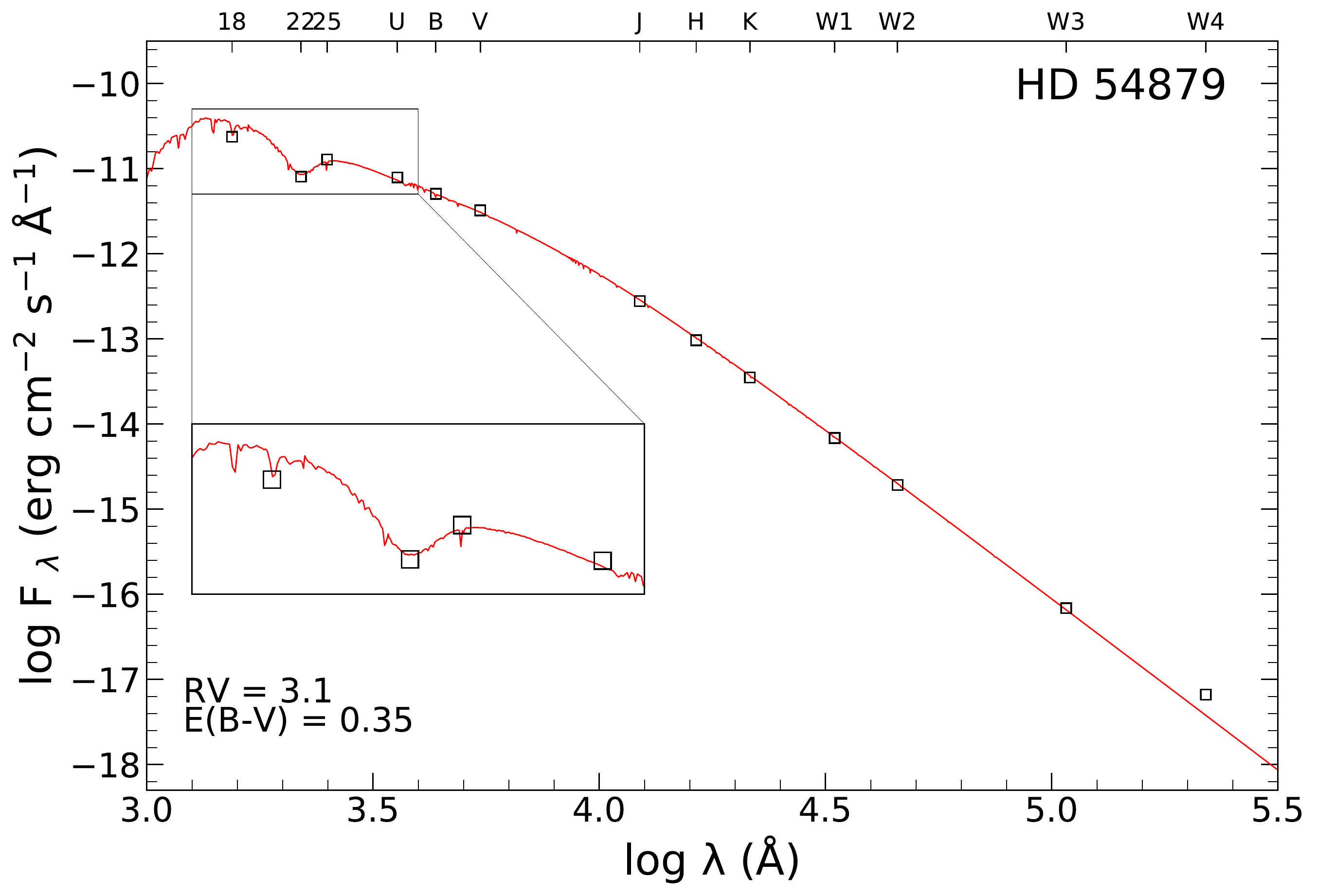}
    \caption{Example SED fits of reddened {\sc Atlas9} fluxes (red) to observed UV-spectra (black), if available, and photometric
    measurements for two sample stars: Johnson $UBV$ \citep{Mermilliod97}, 2MASS $JHK$ \citep{Cutrietal03}, WISE $W1$ to $W4$
    \citep{Cutrietal21}, and ANS magnitudes \citep{Wesseliusetal82}. The insets zoom in on the UV range.}
    \label{figure:SED_fit}
\end{figure}

\begin{figure}
    \centering
    \includegraphics[width=0.95\hsize]{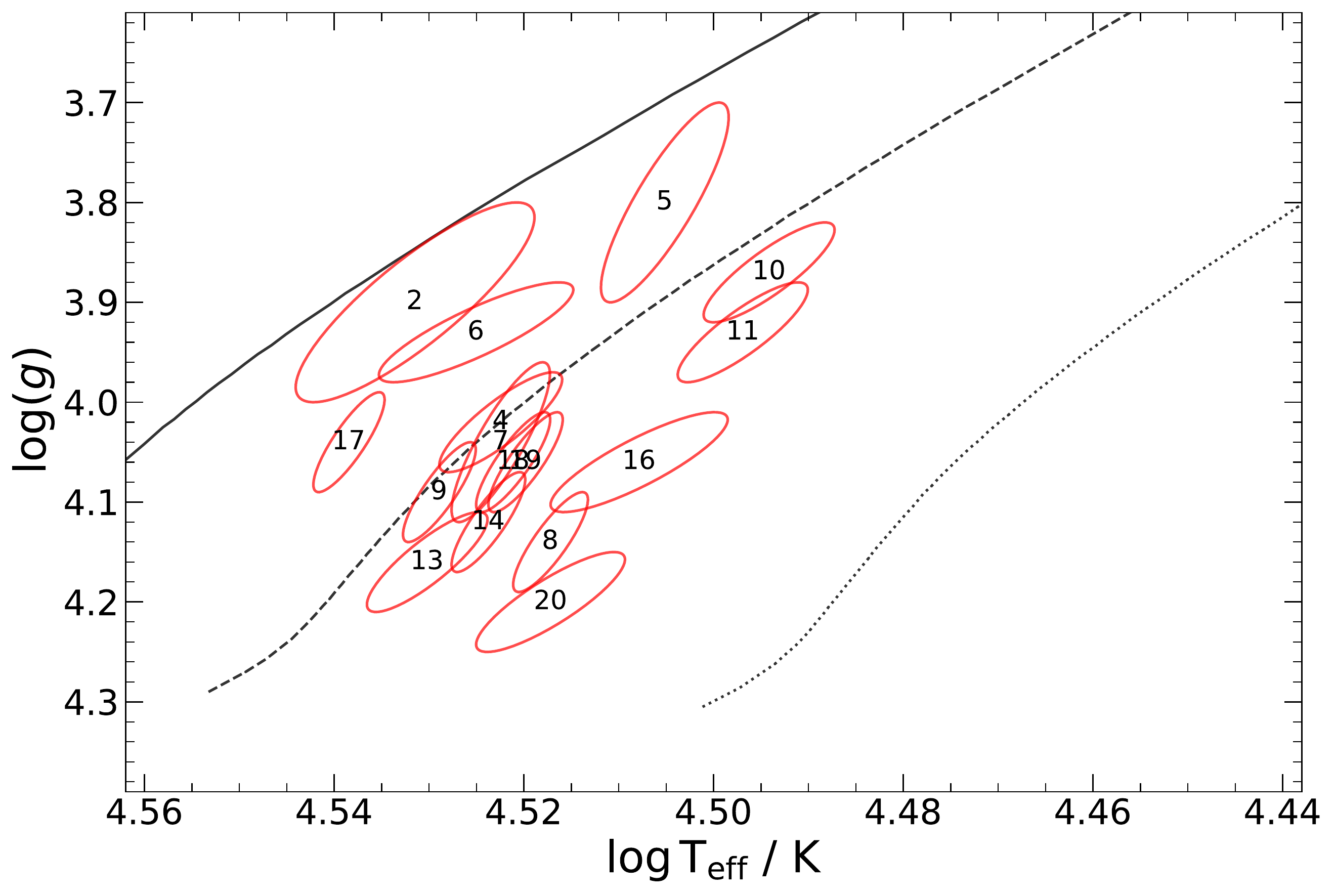}\\
    \includegraphics[width=0.95\hsize]{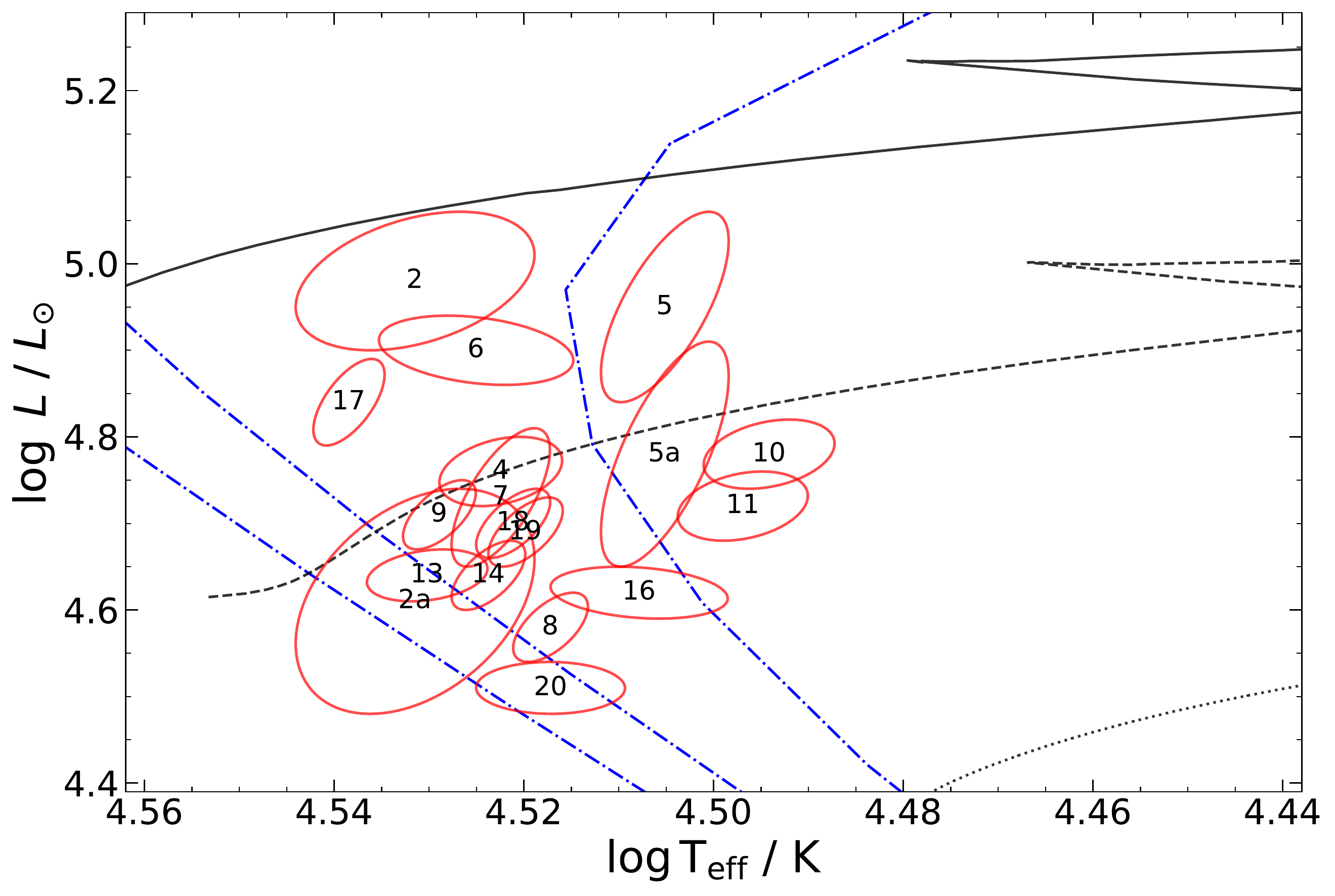}\\
    \includegraphics[width=0.95\hsize]{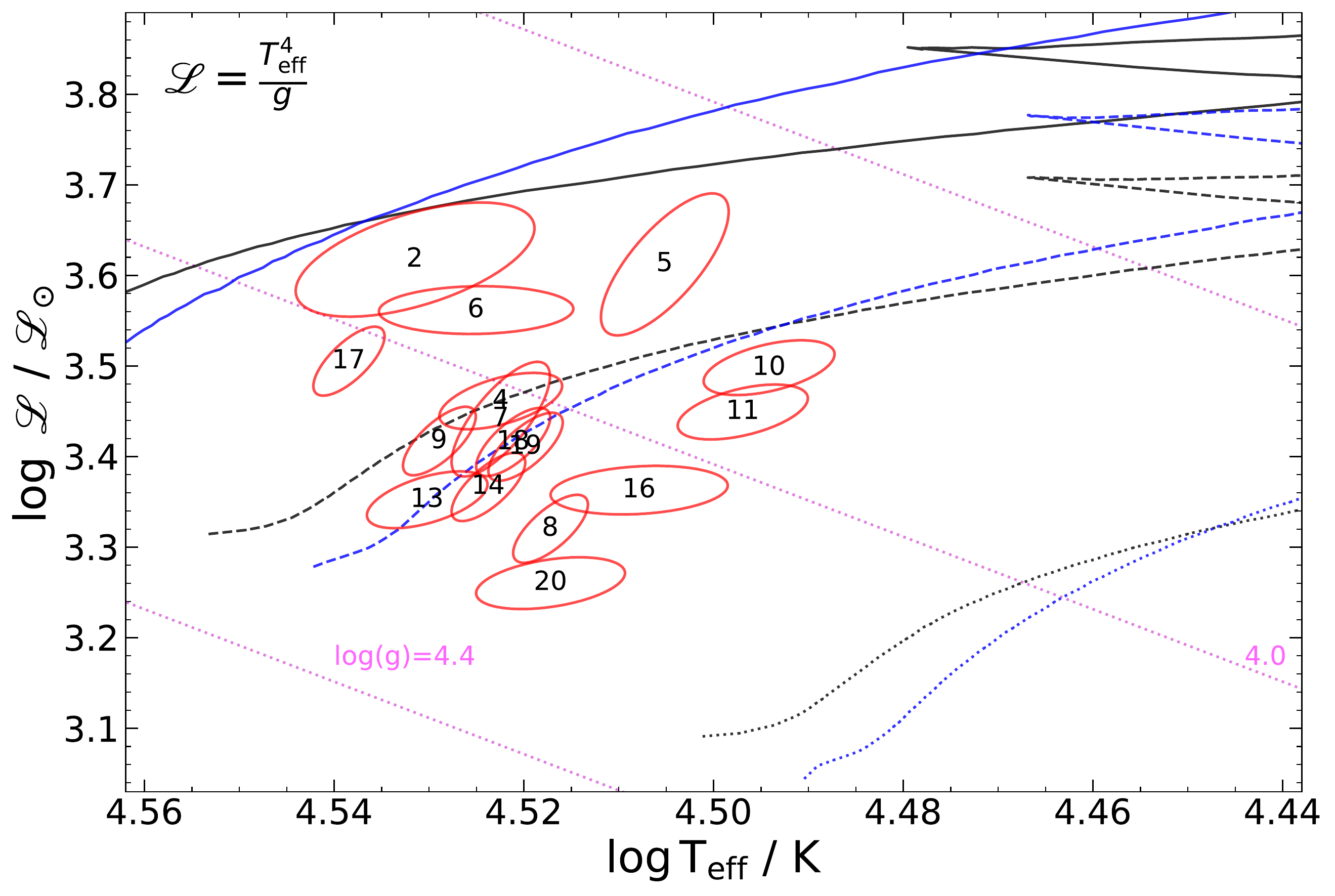}
    \caption{Position of the analysed stars (red ellipses) in the $T_\mathrm{eff}$-$\log(g)$ diagram ({\em upper panel}), HRD ({\em middle panel}) and
    sHRD ({\em lower panel}). The error ellipses outline 1$\sigma$-uncertainties, the stars are identified by their ID number according to
    Table~\ref{tab_observations}.
    Additionally, evolutionary tracks of \citet{Ekstroemetal12} for non-rotating stars at solar metallicity
    $Z$\,=\,0.014 are indicated (\SI{25}{\Msun}: solid, \SI{20}{\Msun}: dashed, \SI{15}{\Msun}: dotted line). The HRD includes isochrones from \citet{Ekstroemetal12} for $\log\tau_{\mathrm{evol}}(\si{\year})=6.0$, $6.4$ and $6.7$ (blue dash-dotted lines, left to right). In the sHRD lines of constant $\log g$ are indicated by dotted magenta lines. {In addition the sHRD shows evolutionary tracks for rotating stars  ($v/v_{\text{crit}}=0.4$) in blue for comparison.}}
    \label{figure:HRD}
\end{figure}

\subsection{Interstellar sight lines -- Reddening law}
Two examples of {\sc Atlas9} model flux fits to observed SEDs are shown in Fig.~\ref{figure:SED_fit}. One example is for 10~Lac
(HD~214680), which has a wide HST spectrophotometric coverage throughout the UV and optical, as extracted from CALSPEC \citep{Bohlinetal14} in
addition to our adopted standard photometric coverage. The second example is for the more reddened star HD~54879, which has only photometric
coverage of the observed SED. A similar fit quality was achieved for the other sample stars. Both parameters, $E(B-V)$ and $R_V$
could be determined to high precision and accuracy facilitating further studies of ISM properties that depend on such high-quality
characterisations of the sight lines towards the background stars \citep[e.g.][]{Ebenbichleretal22}.
The data for $R_V$ range between 2.5 and 3.35 and concentrate mostly around the typical ISM value of 3.1, and
the reddening values vary between 0.05 and 0.90 (see Table~\ref{table:result_part1} for a summary).

We emphasise that UV coverage is highly important for a tight ISM sight line characterisation via SED fitting.
While the HST data have the highest quality, it is a rare resource. The extensive coverage of bright hot stars
in the IUE archive is therefore an asset. However, many valuable complementary data are found in the less well-known
archives of the ANS \citep{Wesseliusetal82} and TD1 \citep{Thompsonetal95} missions.

\begin{figure*}
    \sidecaption
    \includegraphics[width=12cm]{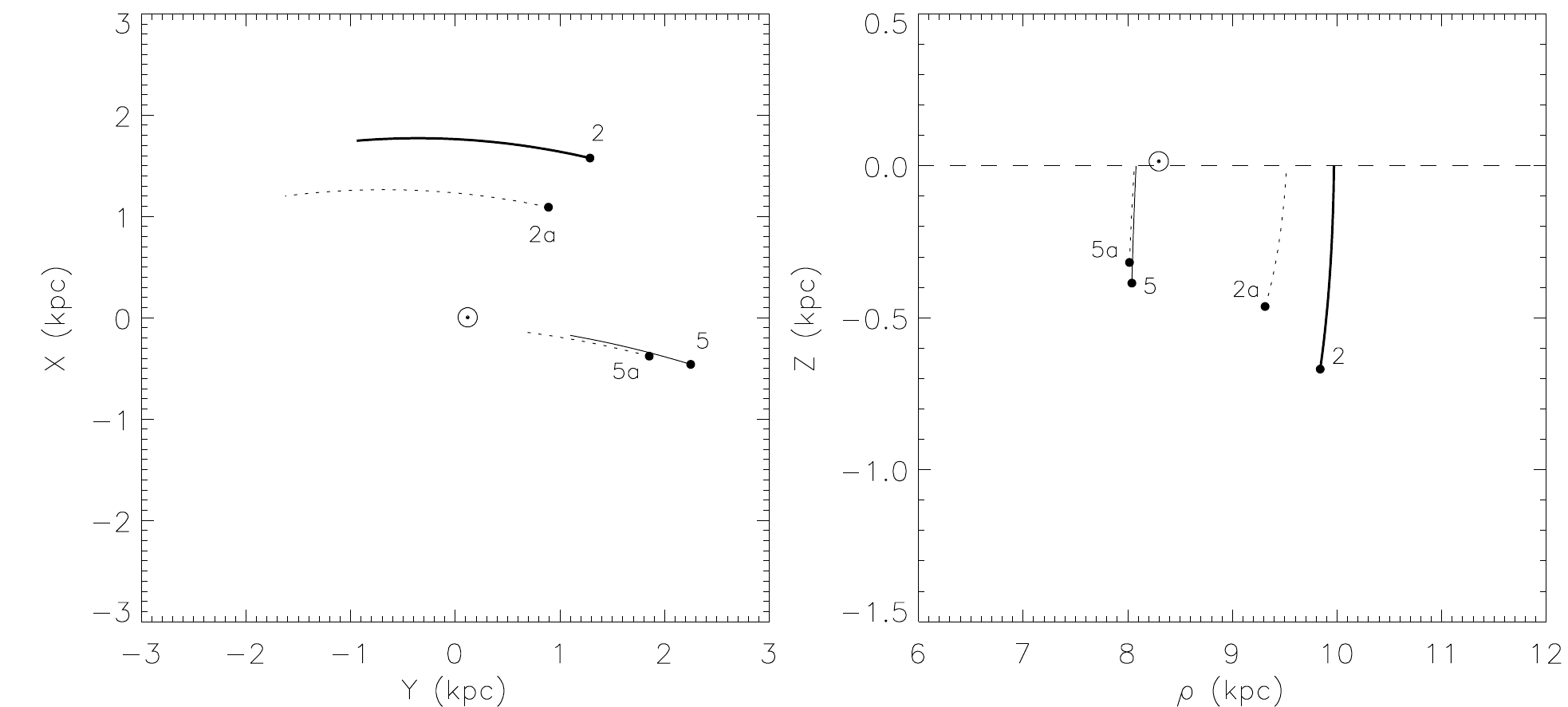}
    \caption{Sketch of the orbits of the two sample runaway ON stars HD~14633 (IDs\#2, 2a) and HD~201345 (IDs\#5, 5a) in the Galactic potential. Galactic Cartesian coordinates $XYZ$ are employed,
    with the origin shifted to the position of the Sun for better comparison with Fig.~\ref{figure:position_distribution}, adopting
    the galactocentric distance from the \citet{GravityCollaboration19}. {\em Left panel:} Galactic plane projection. {\em Right panel:} Meridional projection, $\rho$, is the galactocentric distance, with the Galactic mid-plane indicated by the dashed line. The current positions of the stars are indicated by black dots, the curves show the trajectory calculated backwards in time until the Galactic mid-plane was crossed.}
    \label{fig:kinematics}
\end{figure*}

\subsection{Evolutionary status}\label{subsection:evolutionary_status}
The evolutionary status of the sample stars on the main sequence can be constrained by comparison with stellar evolution tracks.
Three complementary diagnostic diagrams may be employed for this, the Kiel diagram ($\log g$ vs $\log T_\mathrm{eff}$), the
physical HRD ($\log L/L_\odot$ vs $\log T_\mathrm{eff}$) and the spectroscopic HRD \citep[sHRD, $\log(\mathscr{L}/\mathscr{L}_{\odot})$
versus $\log T_\mathrm{eff}$, introduced by][with $\mathscr{L}$\,=\,$T_\mathrm{eff}^4/g$]{LaKu14}. The Kiel diagram and the sHRD are based
solely on observed atmospheric parameters. On the other hand, the physical HRD requires knowledge of the distance and interstellar
extinction. The extinction is derived from $A_V$\,=\,$R_V E(B-V)$, and we prefer spectroscopic distances.
The positions of the sample stars in all
three diagrams with respect to evolutionary tracks for non-rotating stars by \citet{Ekstroemetal12} are given in Fig.~\ref{figure:HRD} and we
stress the very similar positions relative to the evolution tracks in all three diagrams. Ages are then determined from the comparison
with the isochrones. The picture is consistent with the evolution of core hydrogen-burning single
stars with masses of 17.5\,$M_\odot$ to about 23\,$M_\odot$  distributed from close to the ZAMS to an age of $\log \tau_\mathrm{evol}$(yr)\,$\simeq\,$6.8, that is, quite advanced in the main-sequence
evolution.

The comparison with evolution tracks for non-rotating stars is mostly motivated by the relatively small
$\varv \sin i$ of the stars. However, the values of $i$ are not known, and therefore one can expect the stars to rotate faster than is apparent. {A comparison with rotating evolutionary tracks shows mean deviations below $3\%$ in $M/\si{\Msun}$ and below $1\%$ in $\log\tau_{\mathrm{evol}}$.}
This view is supported by the CNO mixing signature (Fig.~\ref{figure:CNO}), but the mixing is compatible with initial rotation velocities
$\varv_\mathrm{ini}/\varv_\mathrm{crit}$\,$<$\,0.4, that is, excluding faster rotation with the possible exception of BD~+57~247 (and the different
evolution scenarios for the two ON stars; see below).
Systematic effects on the derived parameters are covered well by the error ellipses, as the evolution of non-rotating and initially
40\% critically rotating stars is very similar in this part of the HRD (see Fig.~5 of \citealt{Ekstroemetal12}).

The presence of large-scale magnetic fields is known to potentially  affect the evolution of massive stars, including magnetic
braking \citep{Meynetetal11}. Two stars of the sample were found to show magnetic fields: HD~54879 \citep{Castroetal15} and HD~57682
\citep{Grunhutetal09}, both with field strengths below 1\,kG. A spurious detection of a magnetic field in HD~34078 (AE~Aur) was reported by
\citet{Grunhutetal17}. For all other objects at least one spectropolarimetric non-detection of magnetic fields has been reported in the literature
\citep{Fossatietal15,Grunhutetal17,Schoelleretal17}, except for BD~+57~247. As the spectrum of this star adopted for the present
work was taken in spectropolarimetric mode, we briefly inspected the pipeline-reduced Stokes $V$ spectrum, but no obvious
spectropolarimetric signature is present (a more thorough analysis is desirable, but beyond the scope of the present work).
The presence of magnetic fields leads to rotational modulation because of the formation of spots on the surface, which in
turn allows rotational periods to be determined. Indeed, both magnetic stars turn out to be (extremely) slow rotators, with HD~57682 having a
rotational period of $\sim$63.5\,d \citep{Grunhutetal12} and HD~54879 of possibly 7.2\,yr \citep{Jaervinenetal22}.
In terms of CNO mixing
signatures, HD~57682 shows the fifth strongest signature of our star sample, while HD~54879 is the least-mixed star.
Rotational spin-down for stars with radiation-driven winds in the presence of a magnetic field can be fast,
of the order of $\sim$1\,Myr \citep{udDoulaetal09}, and so one may speculate that HD~57682 has been spun down from an
initially larger rotation rate that allowed mixing to occur, whereas HD~54879 appears to have always been a slow rotator,
now spun down to almost zero rotation.

Finally, we wish to address the evolutionary status of the two ON stars in our sample, HD~14633 (ID\#2) and HD~201345 (ID\#5).
Their mixing signature can only be reached with difficulty on the main sequence in the framework of rotational mixing in single star evolution.
Instead, they may have resulted from the evolution of close binaries
\citep[e.g.][]{Vanbeverenetal98,Langer12}. As the evolution of close binaries involves additional (continuous) parameters, such as the
initial masses of the stars and the initial period of the binary \citep[e.g.][]{Wellsteinetal01}, properties of the resulting
evolving binary systems are best studied via population synthesis \citep[e.g.][]{Senetal22}. Since the calculation of tailored evolution
scenarios is beyond the scope of the present paper, we can only sketch a likely scenario.

The mass donors in close binaries typically evolve into helium stars that only briefly pass the O-type range in the HRD
\citep{Wellsteinetal01}, so our two sample ON stars may be the long-lived mass gainers, that is, the initially less-massive secondaries of
such systems. We stress two findings from the work of \citet{Wellsteinetal01}: {\sc i)} mass gainers can reach the region of
effective temperatures and luminosities of ordinary late O-type stars as investigated here at significantly lower stellar mass and
{\sc ii)} they can spend millions of years there.

Both sample ON stars are located untypically far from the mid-plane of the Galactic disc, where the higher gas densities favour star
formation to occur, at more than two to three scale heights \citep{Yuetal21} (see our Sect.~\ref{subsection:spec_distances}).
HD~14633 is a runaway in a SB1 system with a $\sim$15.4\,d orbital period, high eccentricity ($e$\,$\approx$\,0.7) and small
mass function \citep{BoRo78,Boyajianetal05,Mahyetal22}, in which the low-mass companion is suggested to  possibly be a neutron star in a  `quiet'
massive X-ray binary by \citet{Boyajianetal05}. On the other hand, HD~201345 shows no signs of radial-velocity variability \citep{Martinsetal15b} -- it may
therefore be single --, but like HD~14633 it is a runaway star. The question arises as to which mechanism led the stars to become runaway
stars and whether this may be related to them showing an extremely pronounced CNO mixing signature.

We employed the Galactic potential as described by \citet{AlSa91} together with the code of \citet{OdBr92} to calculate the Galactic orbit
and kinematic parameters of the two stars for two solutions, depending on whether $d_\mathrm{spec}$ (solution for ID\#2 and 5) or
alternatively $d_\mathrm{Gaia}$ (solution ID\#2a and 5a)
is used to test the single-star or binary hypothesis. Otherwise, \textit{Gaia} EDR3 proper motions were used and radial velocities of
$-$38.17$\pm$0.08\,km\,s$^{-1}$ \citep{Pourbaixetal04} for HD~14633 and of 19.2$\pm$2.5\,km\,s$^{-1}$ \citep{Gontcharov06} for HD~201345.
The results are sketched in Fig.~\ref{fig:kinematics}, in the Galactic plane projection and for a meridional cut. For reasons of clarity,
we do not show error ranges (from Monte Carlo calculations), as they are comparatively large and dominated by the uncertainties in the
distances. The space velocities of HD~14633 relative to the standard of rest are 75 and 58\,km\,s$^{-1}$ for cases 2 and 2a, and  102 and
86\,km\,s$^{-1}$ for HD~201345 for cases 5 and 5a, respectively. The velocity components pointing away from the Galactic mid-plane are 47 and
26\,km\,s$^{-1}$ for cases 2 and 2a, and 73 and 59\,km\,s$^{-1}$ for cases 5 and 5a, respectively. The flight times from the Galactic
mid-plane to the current position amount to 12 and 13\,Myr for cases 2 and 2a, and to $\sim$5\,Myr for~both~cases~5~and~5a.

A possible scenario for case 5a for HD~201345 may be that it had experienced mass overflow as the gainer in a binary system, now exposing
highly  CNO-processed material at its surface that was close to the core of its initial binary companion. This companion may have subsequently
exploded in a core-collapse supernova, breaking up the system and thus explaining the current single-star nature of HD~201345.
The flight time necessary to reach the current position is likely compatible with its life time, in particular if the breakup happened from
a position within one scale height below, instead of the mid-plane itself. We note that the flight time is still compatible with the stellar
lifetime for case 5, with a dynamical ejection scenario \citep{Povedaetal67} as an alternative to the supernova ejection scenario
\citep{Blaauw61}. However, the CNO mixing signature remains difficult to explain in that case. We therefore favour the binary scenario for HD~201345, with
current fundamental stellar parameters according to case 5a.

The travel time for HD~14633 to its current position rules out a scenario where it has evolved as a single star without interaction with
its binary partner (case 2), as this is about 50\% longer than the main-sequence lifetime of such a late O-type star. Even if it started from within one scale height below the disc, case 2 is unlikely. Consequently case 2a is more likely, implying that HD14633 accreted
highly processed matter from its companion during the evolution of the binary. The apparently more massive companion exploded in a
core-collapse supernova \citep[if the neutron star nature of the companion is correct;][]{Boyajianetal05}, without breaking up the system,
but the supernova kick sufficed to make the system a runaway. A central question is whether accretion onto a main-sequence star can produce
a $\sim$10\,$M_\odot$ object with the observed luminosity of $\log L/L_\odot$\,$\simeq$\,4.6 at the derived $T_\mathrm{eff}$ of sufficient
lifetime (an ordinary 10\,$M_\odot$ star produces $\log L/L_\odot$\,$\simeq$\,3.0 for a $\sim$25\,Myr lifetime). Perhaps accretion onto a
helium star needs to be invoked, implying a more complex mass exchange history for the binary. As already mentioned, without
detailed binary evolution modelling -- which is beyond the scope of the present paper -- no final conclusions on the
past evolution of HD~14633 can be drawn here.

\section{Summary of individual objects}\label{sect:summary}
Important details for the individual sample stars are summarised in the following, both from the literature and the present work. The comparison with literature data from Sect.~\ref{subsection:comparison} is not repeated here in detail.\\[2mm]
{\sf HD~14633 (ID\#2).} This ON star with markedly enriched nitrogen and both depleted carbon and oxygen is probably the most
enigmatic object of our star sample. It is the visible component of a runaway SB1 system with $\sim$15.4\,d orbital period, high
eccentricity and small mass function \citep{BoRo78,Boyajianetal05,Mahyetal22}, located about 3 scale heights below the Galactic mid-plane, a position
that requires $\sim$13\,Myrs to reach it from the mid-plane within the Perseus spiral arm.  This is longer than the lifetime of a normal late O-type star.
The system is further complicated by the presence of a second optical source, resolved by the Fine Guidance Sensor on board the HST,
at $\sim$20\,mas distance with $\Delta m_\mathrm{F5ND}$\,$\simeq$\,1.6$\pm$1.1\,mag \citep{Aldorettaetal15}, which corresponds to a linear
distance of $\sim$30\,AU at $d_\mathrm{Gaia}$\,$\simeq$\,1.5\,kpc (i.e. it is not related to the close companion). As a runaway triple
system appears to be unlikely, this is probably a chance alignment along the line of sight \citep[it should be on the faint end of the
magnitude difference found, as we see no indication of features from this second light source, neither in our spectrum nor in the SED; see also][]{Mahyetal22}. HD~14633 has a RUWE of 1.090 in \textit{Gaia} EDR3, apparently the second source is not interfering with the astrometric
solution either (in view of the apparent zero impact of the second light source on all observables an independent confirmation of its existence is desirable). Our derived parameters agree very well with those from the most recent {\sc Cmfgen}-based analysis by \citet{Mahyetal22},
in particular with respect to the $d_\mathrm{Gaia}$-based (spectroscopic) mass. The O-star luminosity
of such a $\sim$10\,$M_\odot$ star may require consideration of a more complex evolution history, possibly with accretion of highly
CNO-processed material onto a helium star. Certainly, the star deserves further investigations in particular regarding population synthesis
in the context of binary evolution in order to pinpoint its origin.\\[1mm]
{\sf HD~258691 (ID\#4).} The star is one of the seven O-stars members of the young open cluster \object{NGC 2244} that excite the
Rosette Nebula \citep{RoLa08}. The probably most reliable distance to NGC~2244 (1.39$\pm$0.1\,kpc) and age (2.3$\pm$0.2\,Myr) determination
available in the literature stems from the analysis of the detached eclipsing massive binary \object{V578 Mon} \citep{Hensbergeetal00}. The
distance is shorter than photometric estimates of typically 1.6 to 1.7\,kpc \citep[see][]{RoLa08}, but it is in excellent agreement with our spectroscopic distance of
HD~258691 and the \textit{Gaia} EDR3-based distance to NGC~2244 of 1464$_{-84}^{+94}$\,pc (1$\sigma$ standard deviation;  see Appendix~\ref{appendix:C}); the \textit{Gaia} EDR3 astrometric solution for HD~258691 is unreliable because of a RUWE of $\sim$7. Our analysis complements the work of
\citet{Martinsetal12} who have analysed all the O stars of NGC~2244 using full non-LTE techniques with {\sc Cmfgen}, except for
HD~258691. They find an age of about 2 to 4\,Myrs for the O-star population in NGC~2244, which is in {good} agreement with our derived age
of HD~258691. Finally, we note the apparent visual faintness of HD~258691 when compared to the other NGC~2244 O stars, it is about
1.5\,mag fainter in $V$ than the next brightest O star, HD~46202. This comes from a much higher reddening, as the star is located
close to a dark lane in the Rosette Nebula. In the $K$-band it is only 0.2\,mag fainter than HD~46202.\\[1mm]
{\sf HD~201345 (ID\#5).} The second runaway ON star of our star sample that has travelled untypically far from the Galactic mid-plane, its flight time being compatible with its age.
It appears to be single and has likely accreted highly processed matter from a companion, which apparently exploded in a core-collapse
supernova that broke apart the binary and accelerated HD~201345 out from the Galactic plane, with a footpoint of the orbit in the Local
spiral arm. Its mass is less than expected for a normal late O-type star, but in the range of mass-gaining secondaries according to the
binary evolution models of \citet{Wellsteinetal01}.\\[1.5mm]
{\sf HD~57682 (ID\#6).} This is a runaway star from the CMa OB1 association, ejected about 1\,Myr ago possibly in a binary supernova event, showing a bow shock in the WISE
W3 and W4 bands \citep{Fernandesetal19}. The star is magnetic, with a field strength of below 1\,kG \citep{Grunhutetal09}, and it is a slow
rotator with a period of $\sim$63.5\,d \citep{Grunhutetal12}. It shows one of the stronger CNO-mixing signatures of the sample
stars.\\[1mm]
{\sf HD~227757 (ID\#7).} The star is a member of the Cyg OB3 association (e.g. \citealt[][]{GaSt92}; see \citealt{QuWr21} for an \textit{Gaia} EDR3 view on this). Our $T_\mathrm{eff}$ and $\log g$ values are significantly higher
than those derived by \citet{Mahyetal15} based on an analysis with {\sc Cmfgen}, but agreement is found for the fundamental stellar
parameters. The authors estimate an age between 3 and 5\,Myr for the Cyg OB2 association, which is in excellent agreement with our value
for HD~227757. The star shows an intermediate level of nitrogen enrichment.\\[1.5mm]
{\sf BD~$-$13~4930 (ID\#8).} The star is a member of the open cluster \object{NGC 6611} in the Eagle Nebula (M16). BD~$-$13~4930 is the
brightest star in the original HST image of the `Pillars of Creation', in between the two easternmost pillars
\citep[also in 3D; see the discussion by][]{McLeodetal15}. It was previously observed
within the VLT-FLAMES survey of massive stars \citep{Evansetal05} under the ID NGC~6611-006 and analysed using {\sc Tlusty} models
\citep{Hunteretal09}. \citet{Maederetal14} indicated that the model found using the atmospheric parameters of \citet{Hunteretal09} does not
provide a convincing fit to the observed spectrum, which is bolstered here by finding significantly different atmospheric parameters and
elemental abundances. In particular, we find a higher metallicity than for similar objects in the solar neighbourhood, which is expected
for a star in the Sagittarius spiral arm. \citet{Hunteretal09} on the other hand found the star to be metal-poor. It is mildly
nitrogen-enhanced. NGC~6611 was subject to a thorough investigation based on \textit{Gaia} EDR3 data by \citet{Stoopetal22}. BD~$-$13~4930 is fully
compatible with their derived cluster distance of 1706$\pm$7\,pc (standard error of the mean), proper motion and radial velocity and a member of the `young' population of massive stars
according to its age. Its extinction $A_V$\,=\,1.82 is smaller than the mean of the young population of stars
\citep[$A_V$\,=3.6$\pm$0.1,][]{Stoopetal22}, but this may be explained by a location in front of the bulk of the cluster as the region shows a marked radial increase in extinction \citep{McLeodetal15} and the next
brightest star of the same spectral type in NGC~6611, BD~$-$13~4928, is 0.65\,mag fainter in $V$, while they are equally bright in $K$.\\[1.5mm]
{\sf BD~+60~499 (ID\#9).} The star is located in the open cluster IC~1805, near the centre of the Heart
Nebula. We find excellent agreement between the spectroscopic and the \textit{Gaia} EDR3 distance, which is consistent with
the \textit{Gaia} EDR3 cluster distance of 2108$_{-115}^{+129}$\,pc (1$\sigma$ standard deviation; see
Appendix~\ref{appendix:D}). The stellar age is also consistent with the cluster age of 1--3\,Myr
\citep[e.g.][]{Masseyetal95}. BD~+60~499 was suggested to be a radial velocity
variable \citep{HuGa06a,RaNa16}; however, there is no sign of significant second light from the available data. It shows one
of the stronger CNO-mixing signatures.\\[1.5mm]
{\sf BD~+57~247 (ID\#10).} The star is a probable member of the open cluster NGC~457 according to the WEBDA
database\footnote{\url{https://webda.physics.muni.cz/}}. However, its derived distance puts it at the far side of
the cluster distance of 3026$_{-223}^{+262}$\,pc (1$\sigma$ standard deviation; see Appendix~\ref{appendix:E}). BD~+57~247 is also younger than the cluster age of about 24\,Myr found in the literature \citep{Kharchenkoetal13,Diasetal21} and also younger than our revised cluster age of $\sim$9--10\,Myr in Appendix~\ref{appendix:E}.
The star was considered a blue straggler by \citet{Mermilliod82}, but the
present results imply it to be a background object to NGC~457, lying about 0.4\,kpc farther away. It shows one of the larger reddening values among the sample stars despite a Galactic latitude of $-$4\fdg3.
Our analysis is the first using non-LTE techniques.
BD\,+57~247 has the largest CNO mixing signature among the sample stars that may be compatible with rotational mixing on the main sequence,
assuming it to be an initially faster rotator than average.\\[1.5mm]
{\sf HD~207538 (ID\#11).} The star is a member of IC~1396, lying outside the Elephant's Trunk Nebula. Both its
spectroscopic and \textit{Gaia} EDR3 distance are compatible with the \textit{Gaia} EDR3 cluster distance of 954$_{-95}^{+119}$\,pc
(1$\sigma$ standard deviation; see Appendix~\ref{appendix:F}), though the \textit{Gaia} distance is on the short side. It appears to
be at the older limit of the cluster age of 3 to 5\,Myr \citep{Errmannetal13}.
HD~207538 shows one of the stronger nitrogen enrichments.\\[1.5mm]
{\sf HD~46202 (ID\#13).} Like HD~258691 the star is a member of the NGC~2244 cluster inside the Rosetta Nebula
\citep{RoLa08}. Observations with the CoRoT satellite detected the presence of $\beta$ Cep-like pulsations of very
small amplitude \citep[$\sim$0.1\,mmag,][]{Briquetetal11}. The stellar parameters derived in that study are in
excellent agreement with the present findings based on two independent data reductions. The spectroscopic distance
to HD~46202 is shorter than the \textit{Gaia} EDR3-based cluster distance of 1464$_{-84}^{+94}$\,pc (1$\sigma$ standard deviation; see Appendix~\ref{appendix:C}), while
the \textit{Gaia} EDR3 astrometric solution for HD~46202 is unreliable because of a RUWE\,$\simeq$\,5. The poor astrometric
solution may be related to the presence of a close companion to HD~46202, at a distance of $\sim$87\,mas and  $\Delta m_\mathrm{F583W}$\,=\,2.166$\pm$0.008\,mag \citep{Aldorettaetal15}.
This would correspond to a spectral type of B1.5 or B2 for the companion, which is in the middle of the range of spectral types where $\beta$ Cephei pulsations are strongest. Possibly, the observed sub-millimagnitude pulsational amplitudes do not stem from the O star, but from the companion, diluted in magnitude in the combined light. This may have the potential to explain why the excitation of the observed pulsation modes
is inconsistent with the observed atmospheric parameters of the O star \citep{Briquetetal11}. The presence of a binary companion of such a magnitude usually leads to an observable second line system. In order to hide this within the observed narrow lines, one would need to see the system from near-pole on, such that radial-velocity variations are absent \citep{Chinietal12} -- and also implying that HD~46202 is actually a faster rotator. Further investigations should be undertaken to study whether second light and possibly fast (differential) rotation and gravity darkening \citep{Zorecetal17} can resolve these remaining discrepancies.\\[1mm]
{\sf HD~38666 ($\mu$\,Col, ID\#14), HD~34078 (AE Aur, ID\#18).} Both stars are classical late O-type runaway stars originating in the
Local Spiral arm. HD~34078 shows a bow shock \citep{Franceetal07}. They were suggested to stem from a common ejection event in the nascent
Orion Trapezium cluster based on Hipparcos data
\citep{Hoogerwerfetal00}. However, more recent investigations based on \textit{Gaia} EDR3 measurements dismiss this scenario \citep{Bhatetal22}.
Our analysis finds very similar atmospheric and fundamental stellar parameters for both stars. Within the mutual uncertainties they have
the same age. HD~38666 shows enriched nitrogen, while HD~34078 is compatible with pristine CAS abundances.\\[1.5mm]
{\sf HD~54879 (ID\#16).} Like HD~57682, this star is a member of the CMa OB1 association, located close to the association centre
\citep{Fernandesetal19}. It also shows a magnetic field of strength below 1\,kG \citep{Castroetal15} and is an even slower rotator
with a period of possibly longer than 7\,yr \citep{Jaervinenetal22}. HD~54879 is the least CNO-mixed star of the sample.\\[1.5mm]
{\sf HD~214680 (10 Lac, ID\#17).} 10~Lac is the ionising source of the \ion{H}{ii} region Sh2-126 in the Lac OB1b
association and the anchor point of the MK system for spectral type O9\,V. While it was too distant to derive
a meaningful distance by the original Hipparcos data release \citep{ESA97} it is on the  bright side for the \textit{Gaia}
mission. Our spectroscopic distance is in agreement with the re-reduction of the Hipparcos mission by
\citet{vanLeeuwen07a} and our determination of the Lac OB1b association distance of
$d_{\mathrm{Lac OB1b}}$\,=\,542$_{-52}^{+65}$\,pc (1$\sigma$ standard deviation; see Appendix~\ref{appendix:A} for
a discussion). Several of the more recent analyses with full non-LTE model atmosphere techniques and our work have
converged on consistent stellar parameters for 10~Lac (see Sect.~\ref{subsection:comparison}). It
shows an intermediate level of nitrogen enrichment.\\[1.5mm]
{\sf HD~206183 (ID\#19).} The star is a member of IC~1396, lying inside the Elephant's Trunk Nebula. As for
HD~207538, both the spectroscopic and the \textit{Gaia} EDR3 distance agree with the cluster distance (see
Appendix~\ref{appendix:F}), and the stellar age fits with the cluster age. HD~206183 shows near-pristine CNO
abundances.\\[1.5mm]
{\sf HD~36512 ($\upsilon$ Ori, ID\#20).} The star is the intrinsically bluest member of the Ic subgroup of the Ori
OB1 association \citep[for an overview, see e.g.][]{Bally08}. Traditionally, it was used as an anchor point of the
spectral type B0\,V in the MK system until a fine adjustment of the classification scheme at the transition from the
B to the O stars by \citet{Sotaetal11} such that it now defines the O9.7\,V spectral type. It was reanalysed here as
a connection point to previous work using hybrid non-LTE modelling \citep{NiSi11,NiPr12,NiPr14}, finding agreement in
the derived stellar parameters. The star shows CNO abundances consistent with pristine values.

\section{Summary and discussion}\label{sect:conclusions}
We implemented and tested a hybrid non-LTE modelling technique for the quantitative analysis of weak-wind late
O-type stars. The technique provides accurate and precise atmospheric parameters and elemental abundances and, when coupled with
stellar evolution models and \textit{Gaia} parallaxes, fundamental stellar parameters as well. This allows a full
characterisation of this class of stars to be achieved, except for the stellar wind parameters. Overall, results from
previous full non-LTE modelling were reproduced. We went beyond the current state-of-the-art approaches by considering the effects of
turbulent pressure on the atmospheric stratification and line formation, which produce small but systematic effects.

Of the initial sample of 20 carefully selected stars from the literature, four had to
be excluded from further analysis because of clear contamination of the spectra by second light, which highlights the
usefulness of high-S/N and high-resolution spectra in this context. The remaining 16 stars were comprehensively
analysed (one for the first time in non-LTE), including interstellar sight line characterisation,
and they were individually put in the context of recent findings reported in the literature. Fourteen are `ordinary'
main-sequence stars, though some of them would benefit from further detailed investigations, in particular
the binary status of HD~46202. For the two ON
stars of the sample, we argue that they accreted strongly CNO-processed matter from their (former) binary
companion stars, and that the supernova explosions of these companions led to their expulsion from the Galactic plane
as runaway stars. They certainly need to be investigated further in the context of population synthesis models for
binary evolution to pinpoint their exact evolution channels.

In the future, extensions of the model atom database, focusing on \ion{N}{iii},
\ion{S}{iii/iv,} and \ion{Fe}{iv}, would be desirable. Once the observed spectra are comprehensively covered by the models,
faster-rotating weak-wind late O-type stars will be fully accessible for analysis. Moreover, it should be possible to analyse lower-resolution
spectra of such stars, observed with existing 10-metre-class telescopes even in galaxies beyond the
Local Group, such as in the dwarf irregular galaxy Sextans A \citep{Lorenzoetal22}.
Furthermore, automation of the analysis is required for the analysis of larger spectroscopic datasets \citep[e.g.][]{Blommeetal22,Xiangetal22,Moreletal22,Holgadoetal22}, with machine-learning
techniques~being~the~most~promising avenues.

The most important aspect of the present work, besides the atmospheric and spectral modelling, is the impact of
\textit{Gaia} data on the characterisation of the weak-wind late O-type stars. Here parallaxes to infer distances,
and therefore luminosities and proper motions, are crucial since they allow cluster membership or runaway status to be constrained.
They were made available most recently in \textit{Gaia} EDR3. A reassuring result from the present work is the
overall good agreement of spectroscopic and EDR3-based distances. However, the uncertainties of both can be
considerable. As several of our objects are located in young star clusters, we estimated cluster distances
on the basis of EDR3 parallaxes of massive cluster member stars (see the appendices). Noticeable is the large
scatter of the member stars in parallaxes, which is also apparent in other works that consider extended cluster star
samples (see, for example, the detailed study of NGC~6611 by \citealt{Stoopetal22}). Often, the parallaxes of individual cluster
member stars disagree, with the
differences being (much) larger than the uncertainties of the individual parallaxes. These differences cannot be
overcome by considering zero-point corrections for the parallaxes \citep[e.g.][]{Lindegrenetal21}. Because of the
relative brightness of the cluster members and their relative proximity, a high significance of the parallaxes,
$\varpi$/$\Delta\varpi$\,$>$\,20, is typical, so we ignored zero-point corrections altogether.
We conclude that EDR3 parallaxes apparently still contain significant systematics, for example unaccounted
binarity, in particular for early-type stars, which require identification and minimisation in future data
releases.

The full \textit{Gaia} DR3 \citep{Gaia22} also provides the atmospheric parameters $T_\mathrm{eff}$ and $\log g$ for 12 of
the 16 stars analysed here. Among them, two-thirds are (mis)classified as white dwarfs with $T_\mathrm{eff}$ between
about 21\,000 and 30\,000\,K and $\log g$ values between 3.5 and 4.0. The other objects are classified as stars
with $T_\mathrm{eff}$ between about 6600 and 24\,000\,K and $\log g$ values between 2.8 and 3.7. Obviously, the \textit{Gaia}
astrophysical parameter inference system (Apsis) needs to be improved in order to enable it to correctly characterise
late O-type stars. For the very limited parameter space investigated here, we can confirm that $\log g$ values
are underestimated by $\sim$0.4\,dex by Apsis for O stars \citep{Fouesneauetal22} -- though the maximum deviation
for the present stars is $\sim$1.2\,dex in one case. The $T_\mathrm{eff}$ underestimation is, however, much larger
than the 1000 to 5000\,K claimed by Fouesneau et al.~and amounts to $\sim$4000 to 26\,000\,K, with half of the stars showing deviations of
more than 10\,000\,K.

Finally, one may ask whether the analyses conducted in the present work can contribute to the development of a better
understanding of the weak-wind phenomenon. A trivial answer would be no, as the wind lies outside the physical
framework established by our hydrostatic approach. However, our work allows the
photospheric layers to be looked at in detail. These layers form  the base of the stellar wind or, in other words, the lower boundary condition for the wind solution.
{As we were able to reproduce the observed optical spectra with our global solutions (see Appendix~\ref{appendix:B} for an example), without consideration of stellar winds, this indicates that
the sample stars have winds that are weak enough not to affect the photospheric layers. Much larger mass-loss rates than predicted by \citet{Vinketal00} would be required for this.}

A difference from {most} previous works is the smaller
microturbulent velocities (by trend) derived here.
Using the hybrid non-LTE approach, we were able to derive individual microturbulence velocities for our sample stars from minimising abundance trends as a function of the strengths of metal lines, in contrast to the majority of literature studies, where the microturbulent velocities at photospheric layers were assumed to have a particular value. The few studies where photospheric microturbulence values for late O-type dwarfs were in fact determined (for SMC stars) found similar values \citep{Bouretetal03,Martinsetal04}, or they gave compatible upper limits \citep[e.g.][]{Holgadoetal18}.
Microturbulence has an impact on the broadening of
spectral lines and therefore on the line acceleration. The effect is such that
larger microturbulent velocities give broader lines and therefore provide a higher $g_\mathrm{L}$ in the subsonic region. On the other hand, the shadowing of the continuum flux by the broader photospheric lines leads to a lower line acceleration beyond the sonic point. The combined effect gives a reduced mass-loss rate (\citealt{Lucy07}, see also \citealt{Babel96} for some earlier considerations). This behaviour of reduced mass-loss rates in conjunction with larger terminal velocities due to increased turbulent velocities was independently derived by \citet{Sundqvistetal19}. As a result, the lower microturbulent velocities found here are unlikely to contribute to the solution of the weak-wind problem.

\begin{acknowledgements}
We are grateful to A.~Irrgang for updates and extensions of the {\sc Detail/Surface} codes. We thank J.~Puls for valuable comments on the manuscript,
and our referee E.~S.~G. de~Almeida for suggestions that helped to improve the manuscript.
This research has made use of the services of the ESO Science Archive Facility. Based on data obtained from the ESO Science Archive Facility
with DOI(s): {\url{https://doi.org/10.18727/archive/24}}.
Based on observations collected at the Centro Astron\'omico Hispano Alem\'an
at Calar Alto (CAHA), operated jointly by the Max-Planck Institut f\"ur  Astronomie and the Instituto de Astrof\'isica de Andaluc\'ia
(CSIC), proposal H2005-2.2-016. This research used the facilities of the Canadian Astronomy Data Centre operated by the National Research
Council of Canada with the support of the Canadian Space Agency. It is also based on observations made with the NASA/ESA Hubble Space
Telescope obtained from the Space Telescope Science Institute, which is operated by the Association of Universities for Research in
Astronomy, Inc., under NASA contract NAS 5–26555.
This work has made use of data from the European Space Agency (ESA) mission
{\it Gaia} (\url{https://www.cosmos.esa.int/gaia}), processed by the {\it Gaia}
Data Processing and Analysis Consortium (DPAC,
\url{https://www.cosmos.esa.int/web/gaia/dpac/consortium}). Funding for the DPAC
has been provided by national institutions, in particular the institutions
participating in the {\it Gaia} Multilateral Agreement. This publication makes use
of data products from the Two Micron All Sky Survey, which is a joint project
of the University of Massachusetts and the Infrared Processing and Analysis
Center/California Institute of Technology, funded by the National Aeronautics
and Space Administration and the National Science Foundation.
It also  makes use of data products from the Wide-field Infrared Survey Explorer,
which is a joint project of the University of California, Los Angeles, and the Jet
Propulsion Laboratory/California Institute of Technology, funded by the National
Aeronautics and Space Administration. We have made use of the WEBDA database, operated
at the Department of Theoretical Physics and Astrophysics of the Masaryk University.
\end{acknowledgements}

%
%

\bibliography{refs}

\begin{thebibliography}{210}
\expandafter\ifx\csname natexlab\endcsname\relax\def\natexlab#1{#1}\fi

\bibitem[{{Aldoretta} {et~al.}(2015){Aldoretta}, {Caballero-Nieves}, {Gies},
  {Nelan}, {Wallace}, {Hartkopf}, {Henry}, {Jao}, {Ma{\'\i}z Apell{\'a}niz},
  {Mason}, {Moffat}, {Norris}, {Richardson}, \& {Williams}}]{Aldorettaetal15}
{Aldoretta}, E.~J., {Caballero-Nieves}, S.~M., {Gies}, D.~R., {et~al.} 2015,
  \aj, 149, 26

\bibitem[{{Allen} \& {Santillan}(1991)}]{AlSa91}
{Allen}, C. \& {Santillan}, A. 1991, \rmxaa, 22, 255

\bibitem[{{Allen}(1973)}]{Allen73}
{Allen}, C.~W. 1973, {Astrophysical quantities, 3rd ed.} ({London: Athlone
  Press})

\bibitem[{{Asplund} {et~al.}(2009){Asplund}, {Grevesse}, {Sauval}, \&
  {Scott}}]{Asplundetal09}
{Asplund}, M., {Grevesse}, N., {Sauval}, A.~J., \& {Scott}, P. 2009, \araa, 47,
  481

\bibitem[{{Auer} \& {Mihalas}(1972)}]{AuMi72}
{Auer}, L.~H. \& {Mihalas}, D. 1972, \apjs, 24, 193

\bibitem[{{Babel}(1996)}]{Babel96}
{Babel}, J. 1996, \aap, 309, 867

\bibitem[{{Bally}(2008)}]{Bally08}
{Bally}, J. 2008, in Handbook of Star Forming Regions, Volume I, ed.
  B.~{Reipurth} ({San Francisco: ASP}), 459

\bibitem[{{Beauchamp} {et~al.}(1997){Beauchamp}, {Wesemael}, \&
  {Bergeron}}]{Beauchampetal97}
{Beauchamp}, A., {Wesemael}, F., \& {Bergeron}, P. 1997, \apjs, 108, 559

\bibitem[{{Becker} \& {Butler}(1988)}]{BeBu88}
{Becker}, S.~R. \& {Butler}, K. 1988, \aap, 201, 232

\bibitem[{{Bhat} {et~al.}(2022){Bhat}, {Irrgang}, \& {Heber}}]{Bhatetal22}
{Bhat}, A., {Irrgang}, A., \& {Heber}, U. 2022, \aap, 663, A39

\bibitem[{{Blaauw}(1961)}]{Blaauw61}
{Blaauw}, A. 1961, \bain, 15, 265

\bibitem[{{Blomme} {et~al.}(2022){Blomme}, {Daflon}, {Gebran}, {Herrero},
  {Lobel}, {Mahy}, {Martins}, {Morel}, {Berlanas}, {Blaz{\`e}re}, {Fr{\'e}mat},
  {Gosset}, {Ma{\'\i}z Apell{\'a}niz}, {Santos}, {Semaan},
  {Sim{\'o}n-D{\'\i}az}, {Volpi}, {Holgado}, {Jim{\'e}nez-Esteban}, {Nieva},
  {Przybilla}, {Gilmore}, {Randich}, {Negueruela}, {Prusti}, {Vallenari},
  {Alfaro}, {Bensby}, {Bragaglia}, {Flaccomio}, {Francois}, {Korn},
  {Lanzafame}, {Pancino}, {Smiljanic}, {Bergemann}, {Carraro}, {Franciosini},
  {Gonneau}, {Heiter}, {Hourihane}, {Jofr{\'e}}, {Magrini}, {Morbidelli},
  {Sacco}, {Worley}, \& {Zaggia}}]{Blommeetal22}
{Blomme}, R., {Daflon}, S., {Gebran}, M., {et~al.} 2022, \aap, 661, A120

\bibitem[{{Blomme} {et~al.}(2011){Blomme}, {Mahy}, {Catala}, {Cuypers},
  {Gosset}, {Godart}, {Montalban}, {Ventura}, {Rauw}, {Morel}, {Degroote},
  {Aerts}, {Noels}, {Michel}, {Baudin}, {Baglin}, {Auvergne}, \&
  {Samadi}}]{Blommeetal11}
{Blomme}, R., {Mahy}, L., {Catala}, C., {et~al.} 2011, \aap, 533, A4

\bibitem[{{Bohlin} {et~al.}(2014){Bohlin}, {Gordon}, \&
  {Tremblay}}]{Bohlinetal14}
{Bohlin}, R.~C., {Gordon}, K.~D., \& {Tremblay}, P.~E. 2014, \pasp, 126, 711

\bibitem[{{Bolton} \& {Rogers}(1978)}]{BoRo78}
{Bolton}, C.~T. \& {Rogers}, G.~L. 1978, \apj, 222, 234

\bibitem[{{Bouret} {et~al.}(2003){Bouret}, {Lanz}, {Hillier}, {Heap}, {Hubeny},
  {Lennon}, {Smith}, \& {Evans}}]{Bouretetal03}
{Bouret}, J.~C., {Lanz}, T., {Hillier}, D.~J., {et~al.} 2003, \apj, 595, 1182

\bibitem[{{Boyajian} {et~al.}(2005){Boyajian}, {Beaulieu}, {Gies},
  {Grundstrom}, {Huang}, {McSwain}, {Riddle}, {Wingert}, \& {De
  Becker}}]{Boyajianetal05}
{Boyajian}, T.~S., {Beaulieu}, T.~D., {Gies}, D.~R., {et~al.} 2005, \apj, 621,
  978

\bibitem[{{Bragan{\c{c}}a} {et~al.}(2012){Bragan{\c{c}}a}, {Daflon}, {Cunha},
  {Bensby}, {Oey}, \& {Walth}}]{Bragancaetal12}
{Bragan{\c{c}}a}, G.~A., {Daflon}, S., {Cunha}, K., {et~al.} 2012, \aj, 144,
  130

\bibitem[{{Briquet} {et~al.}(2011){Briquet}, {Aerts}, {Baglin}, {Nieva},
  {Degroote}, {Przybilla}, {Noels}, {Schiller}, {Vu{\v{c}}kovi{\'c}}, {Oreiro},
  {Smolders}, {Auvergne}, {Baudin}, {Catala}, {Michel}, \&
  {Samadi}}]{Briquetetal11}
{Briquet}, M., {Aerts}, C., {Baglin}, A., {et~al.} 2011, \aap, 527, A112

\bibitem[{{Butler} \& {Giddings}(1985)}]{BuGi85}
{Butler}, K. \& {Giddings}, J.~R. 1985, Newsletter of Analysis of Astronomical
  Spectra, 9 (Univ. London)

\bibitem[{{Carneiro} {et~al.}(2019){Carneiro}, {Puls}, {Hoffmann}, {Holgado},
  \& {Sim{\'o}n-D{\'\i}az}}]{Carneiroetal19}
{Carneiro}, L.~P., {Puls}, J., {Hoffmann}, T.~L., {Holgado}, G., \&
  {Sim{\'o}n-D{\'\i}az}, S. 2019, \aap, 623, A3

\bibitem[{{Carneiro} {et~al.}(2016){Carneiro}, {Puls}, {Sundqvist}, \&
  {Hoffmann}}]{Carneiroetal16}
{Carneiro}, L.~P., {Puls}, J., {Sundqvist}, J.~O., \& {Hoffmann}, T.~L. 2016,
  \aap, 590, A88

\bibitem[{{Castro} {et~al.}(2015){Castro}, {Fossati}, {Hubrig},
  {Sim{\'o}n-D{\'\i}az}, {Sch{\"o}ller}, {Ilyin}, {Carrol}, {Langer}, {Morel},
  {Schneider}, {Przybilla}, {Herrero}, {de Koter}, {Oskinova}, {Reisenegger},
  {Sana}, \& {BOB Collaboration}}]{Castroetal15}
{Castro}, N., {Fossati}, L., {Hubrig}, S., {et~al.} 2015, \aap, 581, A81

\bibitem[{{Chen} \& {Lee}(2008)}]{ChLe08}
{Chen}, W.~P. \& {Lee}, H.~T. 2008, in Handbook of Star Forming Regions, Volume
  I, ed. B.~{Reipurth} ({San Francisco: ASP}), 124

\bibitem[{{Chini} {et~al.}(2012){Chini}, {Hoffmeister}, {Nasseri}, {Stahl}, \&
  {Zinnecker}}]{Chinietal12}
{Chini}, R., {Hoffmeister}, V.~H., {Nasseri}, A., {Stahl}, O., \& {Zinnecker},
  H. 2012, \mnras, 424, 1925

\bibitem[{{Chiosi} \& {Maeder}(1986)}]{ChMa86}
{Chiosi}, C. \& {Maeder}, A. 1986, \araa, 24, 329

\bibitem[{{Chlebowski} \& {Garmany}(1991)}]{ChGa91}
{Chlebowski}, T. \& {Garmany}, C.~D. 1991, \apj, 368, 241

\bibitem[{{Cutri} {et~al.}(2003){Cutri}, {Skrutskie}, {van Dyk}, {Beichman},
  {Carpenter}, {Chester}, {Cambresy}, {Evans}, {Fowler}, {Gizis}, {Howard},
  {Huchra}, {Jarrett}, {Kopan}, {Kirkpatrick}, {Light}, {Marsh}, {McCallon},
  {Schneider}, {Stiening}, {Sykes}, {Weinberg}, {Wheaton}, {Wheelock}, \&
  {Zacarias}}]{Cutrietal03}
{Cutri}, R.~M., {Skrutskie}, M.~F., {van Dyk}, S., {et~al.} 2003, VizieR Online
  Data Catalog, II/246

\bibitem[{{Cutri} {et~al.}(2021){Cutri}, {Wright}, {Conrow}, {Fowler},
  {Eisenhardt}, {Grillmair}, {Kirkpatrick}, {Masci}, {McCallon}, {Wheelock},
  {Fajardo-Acosta}, {Yan}, {Benford}, {Harbut}, {Jarrett}, {Lake}, {Leisawitz},
  {Ressler}, {Stanford}, {Tsai}, {Liu}, {Helou}, {Mainzer}, {Gettngs},
  {Gonzalez}, {Hoffman}, {Marsh}, {Padgett}, {Skrutskie}, {Beck}, {Papin}, \&
  {Wittman}}]{Cutrietal21}
{Cutri}, R.~M., {Wright}, E.~L., {Conrow}, T., {et~al.} 2021, VizieR Online
  Data Catalog, II/328

\bibitem[{{de Almeida} {et~al.}(2019){de Almeida}, {Marcolino}, {Bouret}, \&
  {Pereira}}]{deAlmeidaetal19}
{de Almeida}, E.~S.~G., {Marcolino}, W.~L.~F., {Bouret}, J.~C., \& {Pereira},
  C.~B. 2019, \aap, 628, A36

\bibitem[{{De Becker} {et~al.}(2006){De Becker}, {Rauw}, {Manfroid}, \&
  {Eenens}}]{DeBeckeretal06}
{De Becker}, M., {Rauw}, G., {Manfroid}, J., \& {Eenens}, P. 2006, \aap, 456,
  1121

\bibitem[{{Dias} {et~al.}(2021){Dias}, {Monteiro}, {Moitinho}, {L{\'e}pine},
  {Carraro}, {Paunzen}, {Alessi}, \& {Villela}}]{Diasetal21}
{Dias}, W.~S., {Monteiro}, H., {Moitinho}, A., {et~al.} 2021, \mnras, 504, 356

\bibitem[{{Dimitrijevic} \& {Sahal-Brechot}(1990)}]{DiSa90}
{Dimitrijevic}, M.~S. \& {Sahal-Brechot}, S. 1990, \aaps, 82, 519

\bibitem[{{Drew} {et~al.}(1994){Drew}, {Denby}, \& {Hoare}}]{Drewetal94}
{Drew}, J.~E., {Denby}, M., \& {Hoare}, M.~G. 1994, \mnras, 266, 917

\bibitem[{{Ebenbichler} {et~al.}(2022){Ebenbichler}, {Postel}, {Przybilla},
  {Seifahrt}, {We{\ss}mayer}, {Kausch}, {Firnstein}, {Butler}, {Kaufer}, \&
  {Linnartz}}]{Ebenbichleretal22}
{Ebenbichler}, A., {Postel}, A., {Przybilla}, N., {et~al.} 2022, \aap, 662, A81

\bibitem[{{Ekstr{\"o}m} {et~al.}(2012){Ekstr{\"o}m}, {Georgy}, {Eggenberger},
  {Meynet}, {Mowlavi}, {Wyttenbach}, {Granada}, {Decressin}, {Hirschi},
  {Frischknecht}, {Charbonnel}, \& {Maeder}}]{Ekstroemetal12}
{Ekstr{\"o}m}, S., {Georgy}, C., {Eggenberger}, P., {et~al.} 2012, \aap, 537,
  A146

\bibitem[{{Errmann} {et~al.}(2013){Errmann}, {Neuh{\"a}user}, {Marschall},
  {Torres}, {Mugrauer}, {Chen}, {Hu}, {Briceno}, {Chini}, {Bukowiecki},
  {Dimitrov}, {Kjurkchieva}, {Jensen}, {Cohen}, {Wu}, {Pribulla},
  {Va{\v{n}}ko}, {Krushevska}, {Budaj}, {Oasa}, {Pandey}, {Fernandez},
  {Kellerer}, \& {Marka}}]{Errmannetal13}
{Errmann}, R., {Neuh{\"a}user}, R., {Marschall}, L., {et~al.} 2013, Astron.
  Nachr., 334, 673

\bibitem[{{ESA}(1997)}]{ESA97}
{ESA}. 1997, ESA Spec. Publ., Vol. 1200 ({Noordwijk: ESA Publications
  Division})

\bibitem[{{Evans} {et~al.}(2005){Evans}, {Smartt}, {Lee}, {Lennon}, {Kaufer},
  {Dufton}, {Trundle}, {Herrero}, {Sim{\'o}n-D{\'\i}az}, {de Koter}, {Hamann},
  {Hendry}, {Hunter}, {Irwin}, {Korn}, {Kudritzki}, {Langer}, {Mokiem},
  {Najarro}, {Pauldrach}, {Przybilla}, {Puls}, {Ryans}, {Urbaneja}, {Venn}, \&
  {Villamariz}}]{Evansetal05}
{Evans}, C.~J., {Smartt}, S.~J., {Lee}, J.~K., {et~al.} 2005, \aap, 437, 467

\bibitem[{{Fernandes} {et~al.}(2019){Fernandes}, {Montmerle}, {Santos-Silva},
  \& {Gregorio-Hetem}}]{Fernandesetal19}
{Fernandes}, B., {Montmerle}, T., {Santos-Silva}, T., \& {Gregorio-Hetem}, J.
  2019, \aap, 628, A44

\bibitem[{{Firnstein} \& {Przybilla}(2012)}]{FiPr12}
{Firnstein}, M. \& {Przybilla}, N. 2012, \aap, 543, A80

\bibitem[{{Fitzpatrick}(1999)}]{Fitzpatrick99}
{Fitzpatrick}, E.~L. 1999, \pasp, 111, 63

\bibitem[{{Flynn} {et~al.}(2022){Flynn}, {Sekhri}, {Venville}, {Dixon},
  {Duffy}, {Mould}, \& {Taylor}}]{Flynnetal22}
{Flynn}, C., {Sekhri}, R., {Venville}, T., {et~al.} 2022, \mnras, 509, 4276

\bibitem[{{Fossati} {et~al.}(2015){Fossati}, {Castro}, {Sch{\"o}ller},
  {Hubrig}, {Langer}, {Morel}, {Briquet}, {Herrero}, {Przybilla}, {Sana},
  {Schneider}, {de Koter}, \& {BOB Collaboration}}]{Fossatietal15}
{Fossati}, L., {Castro}, N., {Sch{\"o}ller}, M., {et~al.} 2015, \aap, 582, A45

\bibitem[{{Fouesneau} {et~al.}(2022){Fouesneau}, {Fr{\'e}mat}, {Andrae},
  {Korn}, {Soubiran}, {Kordopatis}, {Vallenari}, {Heiter}, {Creevey}, {Sarro},
  {de Laverny}, {Lanzafame}, {Lobel}, {Sordo}, {Rybizki}, {Slezak},
  {{\'A}lvarez}, {Drimmel}, {Garabato}, {Delchambre}, {Bailer-Jones},
  {Hatzidimitriou}, {Lorca}, {Le Fustec}, {Pailler}, {Mary}, {Robin},
  {Utrilla}, {Abreu Aramburu}, {Bakker}, {Bellas-Velidis}, {Bijaoui}, {Blomme},
  {Bouret}, {Brouillet}, {Brugaletta}, {Burlacu}, {Carballo}, {Casamiquela},
  {Chaoul}, {Chiavassa}, {Contursi}, {Cooper}, {Dafonte}, {Demouchy},
  {Dharmawardena}, {Garc{\'\i}a-Lario}, {Garc{\'\i}a-Torres}, {Gomez},
  {Gonz{\'a}lez-Santamar{\'\i}a}, {Jean-Antoine Piccolo}, {Kontizas},
  {Lebreton}, {Licata}, {Lindstr{\o}m}, {Livanou}, {Magdaleno Romeo},
  {Manteiga}, {Marocco}, {Martayan}, {Marshall}, {Nicolas}, {Ordenovic},
  {Palicio}, {Pallas-Quintela}, {Pichon}, {Poggio}, {Recio-Blanco}, {Riclet},
  {Santove{\~n}a}, {Schultheis}, {Segol}, {Silvelo}, {Smart}, {S{\"u}veges},
  {Th{\'e}venin}, {Torralba Elipe}, {Ulla}, {van Dillen}, {Zhao}, \&
  {Zorec}}]{Fouesneauetal22}
{Fouesneau}, M., {Fr{\'e}mat}, Y., {Andrae}, R., {et~al.} 2022, arXiv e-prints,
  arXiv:2206.05992

\bibitem[{{France} {et~al.}(2007){France}, {McCandliss}, \&
  {Lupu}}]{Franceetal07}
{France}, K., {McCandliss}, S.~R., \& {Lupu}, R.~E. 2007, \apj, 655, 920

\bibitem[{{Froese Fischer} \& {Tachiev}(2004)}]{FFT04}
{Froese Fischer}, C. \& {Tachiev}, G. 2004, At. Data Nucl. Data Tables, 87, 1

\bibitem[{{Froese Fischer} {et~al.}(2006){Froese Fischer}, {Tachiev}, \&
  {Irimia}}]{FFTI06}
{Froese Fischer}, C., {Tachiev}, G., \& {Irimia}, A. 2006, At. Data Nucl. Data
  Tables, 92, 607

\bibitem[{{Gaia Collaboration}(2022)}]{Gaia22}
{Gaia Collaboration}. 2022, VizieR Online Data Catalog, I/355

\bibitem[{{Gaia Collaboration} {et~al.}(2018){Gaia Collaboration}, {Brown},
  {Vallenari}, {Prusti}, {de Bruijne}, {Babusiaux}, {Bailer-Jones}, {Biermann},
  {Evans}, {Eyer}, \& et~al.}]{Gaia18}
{Gaia Collaboration}, {Brown}, A.~G.~A., {Vallenari}, A., {et~al.} 2018, \aap,
  616, A1

\bibitem[{{Gaia Collaboration} {et~al.}(2021){Gaia Collaboration}, {Brown},
  {Vallenari}, {Prusti}, {de Bruijne}, {Babusiaux}, {Biermann}, {Creevey},
  {Evans}, {Eyer}, {Hutton}, {Jansen}, {Jordi}, {Klioner}, {Lammers},
  {Lindegren}, {Luri}, {Mignard}, {Panem}, {Pourbaix}, {Randich}, {Sartoretti},
  {Soubiran}, {Walton}, {Arenou}, {Bailer-Jones}, {Bastian}, {Cropper},
  {Drimmel}, {Katz}, {Lattanzi}, {van Leeuwen}, {Bakker}, {Cacciari},
  {Casta{\~n}eda}, {De Angeli}, {Ducourant}, {Fabricius}, {Fouesneau},
  {Fr{\'e}mat}, {Guerra}, {Guerrier}, {Guiraud}, {Jean-Antoine Piccolo},
  {Masana}, {Messineo}, {Mowlavi}, {Nicolas}, {Nienartowicz}, {Pailler},
  {Panuzzo}, {Riclet}, {Roux}, {Seabroke}, {Sordo}, {Tanga}, {Th{\'e}venin},
  {Gracia-Abril}, {Portell}, {Teyssier}, {Altmann}, {Andrae}, {Bellas-Velidis},
  {Benson}, {Berthier}, {Blomme}, {Brugaletta}, {Burgess}, {Busso}, {Carry},
  {Cellino}, {Cheek}, {Clementini}, {Damerdji}, {Davidson}, {Delchambre},
  {Dell'Oro}, {Fern{\'a}ndez-Hern{\'a}ndez}, {Galluccio}, {Garc{\'\i}a-Lario},
  {Garcia-Reinaldos}, {Gonz{\'a}lez-N{\'u}{\~n}ez}, {Gosset}, {Haigron},
  {Halbwachs}, {Hambly}, {Harrison}, {Hatzidimitriou}, {Heiter},
  {Hern{\'a}ndez}, {Hestroffer}, {Hodgkin}, {Holl}, {Jan{\ss}en}, {Jevardat de
  Fombelle}, {Jordan}, {Krone-Martins}, {Lanzafame}, {L{\"o}ffler}, {Lorca},
  {Manteiga}, {Marchal}, {Marrese}, {Moitinho}, {Mora}, {Muinonen}, {Osborne},
  {Pancino}, {Pauwels}, {Petit}, {Recio-Blanco}, {Richards}, {Riello},
  {Rimoldini}, {Robin}, {Roegiers}, {Rybizki}, {Sarro}, {Siopis}, {Smith},
  {Sozzetti}, {Ulla}, {Utrilla}, {van Leeuwen}, {van Reeven}, {Abbas}, {Abreu
  Aramburu}, {Accart}, {Aerts}, {Aguado}, {Ajaj}, {Altavilla}, {{\'A}lvarez},
  {{\'A}lvarez Cid-Fuentes}, {Alves}, {Anderson}, {Anglada Varela}, {Antoja},
  {Audard}, {Baines}, {Baker}, {Balaguer-N{\'u}{\~n}ez}, {Balbinot}, {Balog},
  {Barache}, {Barbato}, {Barros}, {Barstow}, {Bartolom{\'e}}, {Bassilana},
  {Bauchet}, {Baudesson-Stella}, {Becciani}, {Bellazzini}, {Bernet}, {Bertone},
  {Bianchi}, {Blanco-Cuaresma}, {Boch}, {Bombrun}, {Bossini}, {Bouquillon},
  {Bragaglia}, {Bramante}, {Breedt}, {Bressan}, {Brouillet}, {Bucciarelli},
  {Burlacu}, {Busonero}, {Butkevich}, {Buzzi}, {Caffau}, {Cancelliere},
  {C{\'a}novas}, {Cantat-Gaudin}, {Carballo}, {Carlucci}, {Carnerero},
  {Carrasco}, {Casamiquela}, {Castellani}, {Castro-Ginard}, {Castro Sampol},
  {Chaoul}, {Charlot}, {Chemin}, {Chiavassa}, {Cioni}, {Comoretto}, {Cooper},
  {Cornez}, {Cowell}, {Crifo}, {Crosta}, {Crowley}, {Dafonte}, {Dapergolas},
  {David}, {David}, {de Laverny}, {De Luise}, {De March}, {De Ridder}, {de
  Souza}, {de Teodoro}, {de Torres}, {del Peloso}, {del Pozo}, {Delbo},
  {Delgado}, {Delgado}, {Delisle}, {Di Matteo}, {Diakite}, {Diener},
  {Distefano}, {Dolding}, {Eappachen}, {Edvardsson}, {Enke}, {Esquej}, {Fabre},
  {Fabrizio}, {Faigler}, {Fedorets}, {Fernique}, {Fienga}, {Figueras},
  {Fouron}, {Fragkoudi}, {Fraile}, {Franke}, {Gai}, {Garabato},
  {Garcia-Gutierrez}, {Garc{\'\i}a-Torres}, {Garofalo}, {Gavras}, {Gerlach},
  {Geyer}, {Giacobbe}, {Gilmore}, {Girona}, {Giuffrida}, {Gomel}, {Gomez},
  {Gonzalez-Santamaria}, {Gonz{\'a}lez-Vidal}, {Granvik},
  {Guti{\'e}rrez-S{\'a}nchez}, {Guy}, {Hauser}, {Haywood}, {Helmi}, {Hidalgo},
  {Hilger}, {H{\l}adczuk}, {Hobbs}, {Holland}, {Huckle}, {Jasniewicz},
  {Jonker}, {Juaristi Campillo}, {Julbe}, {Karbevska}, {Kervella}, {Khanna},
  {Kochoska}, {Kontizas}, {Kordopatis}, {Korn}, {Kostrzewa-Rutkowska},
  {Kruszy{\'n}ska}, {Lambert}, {Lanza}, {Lasne}, {Le Campion}, {Le Fustec},
  {Lebreton}, {Lebzelter}, {Leccia}, {Leclerc}, {Lecoeur-Taibi}, {Liao},
  {Licata}, {Lindstr{\o}m}, {Lister}, {Livanou}, {Lobel}, {Madrero Pardo},
  {Managau}, {Mann}, {Marchant}, {Marconi}, {Marcos Santos}, {Marinoni},
  {Marocco}, {Marshall}, {Martin Polo}, {Mart{\'\i}n-Fleitas}, {Masip},
  {Massari}, {Mastrobuono-Battisti}, {Mazeh}, {McMillan}, {Messina},
  {Michalik}, {Millar}, {Mints}, {Molina}, {Molinaro}, {Moln{\'a}r},
  {Montegriffo}, {Mor}, {Morbidelli}, {Morel}, {Morris}, {Mulone}, {Munoz},
  {Muraveva}, {Murphy}, {Musella}, {Noval}, {Ord{\'e}novic}, {Orr{\`u}},
  {Osinde}, {Pagani}, {Pagano}, {Palaversa}, {Palicio}, {Panahi}, {Pawlak},
  {Pe{\~n}alosa Esteller}, {Penttil{\"a}}, {Piersimoni}, {Pineau}, {Plachy},
  {Plum}, {Poggio}, {Poretti}, {Poujoulet}, {Pr{\v{s}}a}, {Pulone}, {Racero},
  {Ragaini}, {Rainer}, {Raiteri}, {Rambaux}, {Ramos}, {Ramos-Lerate}, {Re
  Fiorentin}, {Regibo}, {Reyl{\'e}}, {Ripepi}, {Riva}, {Rixon}, {Robichon},
  {Robin}, {Roelens}, {Rohrbasser}, {Romero-G{\'o}mez}, {Rowell}, {Royer},
  {Rybicki}, {Sadowski}, {Sagrist{\`a} Sell{\'e}s}, {Sahlmann}, {Salgado},
  {Salguero}, {Samaras}, {Sanchez Gimenez}, {Sanna}, {Santove{\~n}a},
  {Sarasso}, {Schultheis}, {Sciacca}, {Segol}, {Segovia}, {S{\'e}gransan},
  {Semeux}, {Shahaf}, {Siddiqui}, {Siebert}, {Siltala}, {Slezak}, {Smart},
  {Solano}, {Solitro}, {Souami}, {Souchay}, {Spagna}, {Spoto}, {Steele},
  {Steidelm{\"u}ller}, {Stephenson}, {S{\"u}veges}, {Szabados}, {Szegedi-Elek},
  {Taris}, {Tauran}, {Taylor}, {Teixeira}, {Thuillot}, {Tonello}, {Torra},
  {Torra}, {Turon}, {Unger}, {Vaillant}, {van Dillen}, {Vanel}, {Vecchiato},
  {Viala}, {Vicente}, {Voutsinas}, {Weiler}, {Wevers}, {Wyrzykowski}, {Yoldas},
  {Yvard}, {Zhao}, {Zorec}, {Zucker}, {Zurbach}, \& {Zwitter}}]{Gaia21}
{Gaia Collaboration}, {Brown}, A.~G.~A., {Vallenari}, A., {et~al.} 2021, \aap,
  649, A1

\bibitem[{{Gaia Collaboration} {et~al.}(2016){Gaia Collaboration}, {Prusti},
  {de Bruijne}, {Brown}, {Vallenari}, {Babusiaux}, {Bailer-Jones}, {Bastian},
  {Biermann}, {Evans}, {Eyer}, {Jansen}, {Jordi}, {Klioner}, {Lammers},
  {Lindegren}, {Luri}, {Mignard}, {Milligan}, {Panem}, {Poinsignon},
  {Pourbaix}, {Randich}, {Sarri}, {Sartoretti}, {Siddiqui}, {Soubiran},
  {Valette}, {van Leeuwen}, {Walton}, {Aerts}, {Arenou}, {Cropper}, {Drimmel},
  {H{\o}g}, {Katz}, {Lattanzi}, {O'Mullane}, {Grebel}, {Holland}, {Huc},
  {Passot}, {Bramante}, {Cacciari}, {Casta{\~n}eda}, {Chaoul}, {Cheek}, {De
  Angeli}, {Fabricius}, {Guerra}, {Hern{\'a}ndez}, {Jean-Antoine-Piccolo},
  {Masana}, {Messineo}, {Mowlavi}, {Nienartowicz}, {Ord{\'o}{\~n}ez-Blanco},
  {Panuzzo}, {Portell}, {Richards}, {Riello}, {Seabroke}, {Tanga},
  {Th{\'e}venin}, {Torra}, {Els}, {Gracia-Abril}, {Comoretto},
  {Garcia-Reinaldos}, {Lock}, {Mercier}, {Altmann}, {Andrae}, {Astraatmadja},
  {Bellas-Velidis}, {Benson}, {Berthier}, {Blomme}, {Busso}, {Carry},
  {Cellino}, {Clementini}, {Cowell}, {Creevey}, {Cuypers}, {Davidson}, {De
  Ridder}, {de Torres}, {Delchambre}, {Dell'Oro}, {Ducourant}, {Fr{\'e}mat},
  {Garc{\'\i}a-Torres}, {Gosset}, {Halbwachs}, {Hambly}, {Harrison}, {Hauser},
  {Hestroffer}, {Hodgkin}, {Huckle}, {Hutton}, {Jasniewicz}, {Jordan},
  {Kontizas}, {Korn}, {Lanzafame}, {Manteiga}, {Moitinho}, {Muinonen},
  {Osinde}, {Pancino}, {Pauwels}, {Petit}, {Recio-Blanco}, {Robin}, {Sarro},
  {Siopis}, {Smith}, {Smith}, {Sozzetti}, {Thuillot}, {van Reeven}, {Viala},
  {Abbas}, {Abreu Aramburu}, {Accart}, {Aguado}, {Allan}, {Allasia},
  {Altavilla}, {{\'A}lvarez}, {Alves}, {Anderson}, {Andrei}, {Anglada Varela},
  {Antiche}, {Antoja}, {Ant{\'o}n}, {Arcay}, {Atzei}, {Ayache}, {Bach},
  {Baker}, {Balaguer-N{\'u}{\~n}ez}, {Barache}, {Barata}, {Barbier}, {Barblan},
  {Baroni}, {Barrado y Navascu{\'e}s}, {Barros}, {Barstow}, {Becciani},
  {Bellazzini}, {Bellei}, {Bello Garc{\'\i}a}, {Belokurov}, {Bendjoya},
  {Berihuete}, {Bianchi}, {Bienaym{\'e}}, {Billebaud}, {Blagorodnova},
  {Blanco-Cuaresma}, {Boch}, {Bombrun}, {Borrachero}, {Bouquillon}, {Bourda},
  {Bouy}, {Bragaglia}, {Breddels}, {Brouillet}, {Br{\"u}semeister},
  {Bucciarelli}, {Budnik}, {Burgess}, {Burgon}, {Burlacu}, {Busonero}, {Buzzi},
  {Caffau}, {Cambras}, {Campbell}, {Cancelliere}, {Cantat-Gaudin}, {Carlucci},
  {Carrasco}, {Castellani}, {Charlot}, {Charnas}, {Charvet}, {Chassat},
  {Chiavassa}, {Clotet}, {Cocozza}, {Collins}, {Collins}, {Costigan}, {Crifo},
  {Cross}, {Crosta}, {Crowley}, {Dafonte}, {Damerdji}, {Dapergolas}, {David},
  {David}, {De Cat}, {de Felice}, {de Laverny}, {De Luise}, {De March}, {de
  Martino}, {de Souza}, {Debosscher}, {del Pozo}, {Delbo}, {Delgado},
  {Delgado}, {di Marco}, {Di Matteo}, {Diakite}, {Distefano}, {Dolding}, {Dos
  Anjos}, {Drazinos}, {Dur{\'a}n}, {Dzigan}, {Ecale}, {Edvardsson}, {Enke},
  {Erdmann}, {Escolar}, {Espina}, {Evans}, {Eynard Bontemps}, {Fabre},
  {Fabrizio}, {Faigler}, {Falc{\~a}o}, {Farr{\`a}s Casas}, {Faye}, {Federici},
  {Fedorets}, {Fern{\'a}ndez-Hern{\'a}ndez}, {Fernique}, {Fienga}, {Figueras},
  {Filippi}, {Findeisen}, {Fonti}, {Fouesneau}, {Fraile}, {Fraser}, {Fuchs},
  {Furnell}, {Gai}, {Galleti}, {Galluccio}, {Garabato}, {Garc{\'\i}a-Sedano},
  {Gar{\'e}}, {Garofalo}, {Garralda}, {Gavras}, {Gerssen}, {Geyer}, {Gilmore},
  {Girona}, {Giuffrida}, {Gomes}, {Gonz{\'a}lez-Marcos},
  {Gonz{\'a}lez-N{\'u}{\~n}ez}, {Gonz{\'a}lez-Vidal}, {Granvik}, {Guerrier},
  {Guillout}, {Guiraud}, {G{\'u}rpide}, {Guti{\'e}rrez-S{\'a}nchez}, {Guy},
  {Haigron}, {Hatzidimitriou}, {Haywood}, {Heiter}, {Helmi}, {Hobbs},
  {Hofmann}, {Holl}, {Holland}, {Hunt}, {Hypki}, {Icardi}, {Irwin}, {Jevardat
  de Fombelle}, {Jofr{\'e}}, {Jonker}, {Jorissen}, {Julbe}, {Karampelas},
  {Kochoska}, {Kohley}, {Kolenberg}, {Kontizas}, {Koposov}, {Kordopatis},
  {Koubsky}, {Kowalczyk}, {Krone-Martins}, {Kudryashova}, {Kull}, {Bachchan},
  {Lacoste-Seris}, {Lanza}, {Lavigne}, {Le Poncin-Lafitte}, {Lebreton},
  {Lebzelter}, {Leccia}, {Leclerc}, {Lecoeur-Taibi}, {Lemaitre}, {Lenhardt},
  {Leroux}, {Liao}, {Licata}, {Lindstr{\o}m}, {Lister}, {Livanou}, {Lobel},
  {L{\"o}ffler}, {L{\'o}pez}, {Lopez-Lozano}, {Lorenz}, {Loureiro},
  {MacDonald}, {Magalh{\~a}es Fernandes}, {Managau}, {Mann}, {Mantelet},
  {Marchal}, {Marchant}, {Marconi}, {Marie}, {Marinoni}, {Marrese},
  {Marschalk{\'o}}, {Marshall}, {Mart{\'\i}n-Fleitas}, {Martino}, {Mary},
  {Matijevi{\v{c}}}, {Mazeh}, {McMillan}, {Messina}, {Mestre}, {Michalik},
  {Millar}, {Miranda}, {Molina}, {Molinaro}, {Molinaro}, {Moln{\'a}r},
  {Moniez}, {Montegriffo}, {Monteiro}, {Mor}, {Mora}, {Morbidelli}, {Morel},
  {Morgenthaler}, {Morley}, {Morris}, {Mulone}, {Muraveva}, {Musella},
  {Narbonne}, {Nelemans}, {Nicastro}, {Noval}, {Ord{\'e}novic},
  {Ordieres-Mer{\'e}}, {Osborne}, {Pagani}, {Pagano}, {Pailler}, {Palacin},
  {Palaversa}, {Parsons}, {Paulsen}, {Pecoraro}, {Pedrosa}, {Pentik{\"a}inen},
  {Pereira}, {Pichon}, {Piersimoni}, {Pineau}, {Plachy}, {Plum}, {Poujoulet},
  {Pr{\v{s}}a}, {Pulone}, {Ragaini}, {Rago}, {Rambaux}, {Ramos-Lerate},
  {Ranalli}, {Rauw}, {Read}, {Regibo}, {Renk}, {Reyl{\'e}}, {Ribeiro},
  {Rimoldini}, {Ripepi}, {Riva}, {Rixon}, {Roelens}, {Romero-G{\'o}mez},
  {Rowell}, {Royer}, {Rudolph}, {Ruiz-Dern}, {Sadowski}, {Sagrist{\`a}
  Sell{\'e}s}, {Sahlmann}, {Salgado}, {Salguero}, {Sarasso}, {Savietto},
  {Schnorhk}, {Schultheis}, {Sciacca}, {Segol}, {Segovia}, {Segransan},
  {Serpell}, {Shih}, {Smareglia}, {Smart}, {Smith}, {Solano}, {Solitro},
  {Sordo}, {Soria Nieto}, {Souchay}, {Spagna}, {Spoto}, {Stampa}, {Steele},
  {Steidelm{\"u}ller}, {Stephenson}, {Stoev}, {Suess}, {S{\"u}veges}, {Surdej},
  {Szabados}, {Szegedi-Elek}, {Tapiador}, {Taris}, {Tauran}, {Taylor},
  {Teixeira}, {Terrett}, {Tingley}, {Trager}, {Turon}, {Ulla}, {Utrilla},
  {Valentini}, {van Elteren}, {Van Hemelryck}, {van Leeuwen}, {Varadi},
  {Vecchiato}, {Veljanoski}, {Via}, {Vicente}, {Vogt}, {Voss}, {Votruba},
  {Voutsinas}, {Walmsley}, {Weiler}, {Weingrill}, {Werner}, {Wevers},
  {Whitehead}, {Wyrzykowski}, {Yoldas}, {{\v{Z}}erjal}, {Zucker}, {Zurbach},
  {Zwitter}, {Alecu}, {Allen}, {Allende Prieto}, {Amorim},
  {Anglada-Escud{\'e}}, {Arsenijevic}, {Azaz}, {Balm}, {Beck}, {Bernstein},
  {Bigot}, {Bijaoui}, {Blasco}, {Bonfigli}, {Bono}, {Boudreault}, {Bressan},
  {Brown}, {Brunet}, {Bunclark}, {Buonanno}, {Butkevich}, {Carret}, {Carrion},
  {Chemin}, {Ch{\'e}reau}, {Corcione}, {Darmigny}, {de Boer}, {de Teodoro}, {de
  Zeeuw}, {Delle Luche}, {Domingues}, {Dubath}, {Fodor}, {Fr{\'e}zouls},
  {Fries}, {Fustes}, {Fyfe}, {Gallardo}, {Gallegos}, {Gardiol}, {Gebran},
  {Gomboc}, {G{\'o}mez}, {Grux}, {Gueguen}, {Heyrovsky}, {Hoar}, {Iannicola},
  {Isasi Parache}, {Janotto}, {Joliet}, {Jonckheere}, {Keil}, {Kim},
  {Klagyivik}, {Klar}, {Knude}, {Kochukhov}, {Kolka}, {Kos}, {Kutka}, {Lainey},
  {LeBouquin}, {Liu}, {Loreggia}, {Makarov}, {Marseille}, {Martayan},
  {Martinez-Rubi}, {Massart}, {Meynadier}, {Mignot}, {Munari}, {Nguyen},
  {Nordlander}, {Ocvirk}, {O'Flaherty}, {Olias Sanz}, {Ortiz}, {Osorio},
  {Oszkiewicz}, {Ouzounis}, {Palmer}, {Park}, {Pasquato}, {Peltzer}, {Peralta},
  {P{\'e}turaud}, {Pieniluoma}, {Pigozzi}, {Poels}, {Prat}, {Prod'homme},
  {Raison}, {Rebordao}, {Risquez}, {Rocca-Volmerange}, {Rosen}, {Ruiz-Fuertes},
  {Russo}, {Sembay}, {Serraller Vizcaino}, {Short}, {Siebert}, {Silva},
  {Sinachopoulos}, {Slezak}, {Soffel}, {Sosnowska}, {Strai{\v{z}}ys}, {ter
  Linden}, {Terrell}, {Theil}, {Tiede}, {Troisi}, {Tsalmantza}, {Tur},
  {Vaccari}, {Vachier}, {Valles}, {Van Hamme}, {Veltz}, {Virtanen}, {Wallut},
  {Wichmann}, {Wilkinson}, {Ziaeepour}, \& {Zschocke}}]{Gaia16}
{Gaia Collaboration}, {Prusti}, T., {de Bruijne}, J.~H.~J., {et~al.} 2016,
  \aap, 595, A1

\bibitem[{{Garland} {et~al.}(2017){Garland}, {Dufton}, {Evans}, {Crowther},
  {Howarth}, {de Koter}, {de Mink}, {Grin}, {Langer}, {Lennon}, {McEvoy},
  {Sana}, {Schneider}, {S{\'\i}mon D{\'\i}az}, {Taylor}, {Thompson}, \&
  {Vink}}]{Garlandetal17}
{Garland}, R., {Dufton}, P.~L., {Evans}, C.~J., {et~al.} 2017, \aap, 603, A91

\bibitem[{{Garmany} \& {Stencel}(1992)}]{GaSt92}
{Garmany}, C.~D. \& {Stencel}, R.~E. 1992, \aaps, 94, 211

\bibitem[{{Garrison} \& {Kormendy}(1976)}]{GarKor76}
{Garrison}, R.~F. \& {Kormendy}, J. 1976, \pasp, 88, 865

\bibitem[{{Giddings}(1981)}]{Giddings81}
{Giddings}, J.~R. 1981, PhD thesis, (Univ. London)

\bibitem[{{Gontcharov}(2006)}]{Gontcharov06}
{Gontcharov}, G.~A. 2006, Astronomy Letters, 32, 759

\bibitem[{{Gonz{\'a}lez} {et~al.}(2019){Gonz{\'a}lez}, {Briquet}, {Przybilla},
  {Nieva}, {De Cat}, {Saesen}, {Hubrig}, {Thoul}, {P{\'a}pics}, {Palaversa},
  {Naef}, {Neveu-Van Malle}, {J{\"a}rvinen}, {Pollard}, {Kilmartin}, {Mowlavi},
  \& {Butler}}]{Gonzalezetal19}
{Gonz{\'a}lez}, J.~F., {Briquet}, M., {Przybilla}, N., {et~al.} 2019, \aap,
  626, A94

\bibitem[{{Gonz{\'a}lez} {et~al.}(2017){Gonz{\'a}lez}, {Hubrig}, {Przybilla},
  {Carroll}, {Nieva}, {Ilyin}, {J{\"a}rvinen}, {Morel}, {Sch{\"o}ller},
  {Castro}, {Barb{\'a}}, {de Koter}, {Schneider}, {Kholtygin}, {Butler},
  {Veramendi}, {Langer}, \& {BOB Collaboration}}]{Gonzalezetal17}
{Gonz{\'a}lez}, J.~F., {Hubrig}, S., {Przybilla}, N., {et~al.} 2017, \mnras,
  467, 437

\bibitem[{{Gr{\"a}fener} {et~al.}(2002){Gr{\"a}fener}, {Koesterke}, \&
  {Hamann}}]{Graefeneretal02}
{Gr{\"a}fener}, G., {Koesterke}, L., \& {Hamann}, W.~R. 2002, \aap, 387, 244

\bibitem[{{Gravity Collaboration} {et~al.}(2019){Gravity Collaboration},
  {Abuter}, {Amorim}, {Baub{\"o}ck}, {Berger}, {Bonnet}, {Brandner},
  {Cl{\'e}net}, {Coud{\'e} Du Foresto}, {de Zeeuw}, {Dexter}, {Duvert},
  {Eckart}, {Eisenhauer}, {F{\"o}rster Schreiber}, {Garcia}, {Gao}, {Gendron},
  {Genzel}, {Gerhard}, {Gillessen}, {Habibi}, {Haubois}, {Henning}, {Hippler},
  {Horrobin}, {Jim{\'e}nez-Rosales}, {Jocou}, {Kervella}, {Lacour},
  {Lapeyr{\`e}re}, {Le Bouquin}, {L{\'e}na}, {Ott}, {Paumard}, {Perraut},
  {Perrin}, {Pfuhl}, {Rabien}, {Rodriguez Coira}, {Rousset}, {Scheithauer},
  {Sternberg}, {Straub}, {Straubmeier}, {Sturm}, {Tacconi}, {Vincent}, {von
  Fellenberg}, {Waisberg}, {Widmann}, {Wieprecht}, {Wiezorrek}, {Woillez}, \&
  {Yazici}}]{GravityCollaboration19}
{Gravity Collaboration}, {Abuter}, R., {Amorim}, A., {et~al.} 2019, \aap, 625,
  L10

\bibitem[{{Gray}(2005)}]{Gray05}
{Gray}, D.~F. 2005, {The Observation and Analysis of Stellar Photospheres}, 3rd
  edn. (Cambridge: Cambridge University Press)

\bibitem[{{Griem}(1960)}]{Griem60}
{Griem}, H.~R. 1960, \apj, 132, 883

\bibitem[{{Grigsby} {et~al.}(1992){Grigsby}, {Morrison}, \&
  {Anderson}}]{Grigsbyetal92}
{Grigsby}, J.~A., {Morrison}, N.~D., \& {Anderson}, L.~S. 1992, \apjs, 78, 205

\bibitem[{{Grunhut} {et~al.}(2009){Grunhut}, {Wade}, {Marcolino}, {Petit},
  {Henrichs}, {Cohen}, {Alecian}, {Bohlender}, {Bouret}, {Kochukhov}, {Neiner},
  {St-Louis}, \& {Townsend}}]{Grunhutetal09}
{Grunhut}, J.~H., {Wade}, G.~A., {Marcolino}, W.~L.~F., {et~al.} 2009, \mnras,
  400, L94

\bibitem[{{Grunhut} {et~al.}(2017){Grunhut}, {Wade}, {Neiner}, {Oksala},
  {Petit}, {Alecian}, {Bohlender}, {Bouret}, {Henrichs}, {Hussain},
  {Kochukhov}, \& {MiMeS Collaboration}}]{Grunhutetal17}
{Grunhut}, J.~H., {Wade}, G.~A., {Neiner}, C., {et~al.} 2017, \mnras, 465, 2432

\bibitem[{{Grunhut} {et~al.}(2012){Grunhut}, {Wade}, {Sundqvist}, {ud-Doula},
  {Neiner}, {Ignace}, {Marcolino}, {Rivinius}, {Fullerton}, {Kaper},
  {Mauclaire}, {Buil}, {Garrel}, {Ribeiro}, \& {Ubaud}}]{Grunhutetal12}
{Grunhut}, J.~H., {Wade}, G.~A., {Sundqvist}, J.~O., {et~al.} 2012, \mnras,
  426, 2208

\bibitem[{{Heber}(2009)}]{Heber09}
{Heber}, U. 2009, \araa, 47, 211

\bibitem[{{Heber}(2016)}]{Heber16}
{Heber}, U. 2016, \pasp, 128, 082001

\bibitem[{{Heger} \& {Langer}(2000)}]{HeLa00}
{Heger}, A. \& {Langer}, N. 2000, \apj, 544, 1016

\bibitem[{{Hensberge} {et~al.}(2000){Hensberge}, {Pavlovski}, \&
  {Verschueren}}]{Hensbergeetal00}
{Hensberge}, H., {Pavlovski}, K., \& {Verschueren}, W. 2000, \aap, 358, 553

\bibitem[{{Herrero} {et~al.}(1992){Herrero}, {Kudritzki}, {Vilchez}, {Kunze},
  {Butler}, \& {Haser}}]{Herreroetal92}
{Herrero}, A., {Kudritzki}, R.~P., {Vilchez}, J.~M., {et~al.} 1992, \aap, 261,
  209

\bibitem[{{Herrero} {et~al.}(2002){Herrero}, {Puls}, \&
  {Najarro}}]{Herreroetal02}
{Herrero}, A., {Puls}, J., \& {Najarro}, F. 2002, \aap, 396, 949

\bibitem[{{Hillier} \& {Miller}(1998)}]{HiMi98}
{Hillier}, D.~J. \& {Miller}, D.~L. 1998, \apj, 496, 407

\bibitem[{{Hillwig} {et~al.}(2006){Hillwig}, {Gies}, {Bagnuolo}, {Huang},
  {McSwain}, \& {Wingert}}]{Hillwigetal06}
{Hillwig}, T.~C., {Gies}, D.~R., {Bagnuolo}, William~G., J., {et~al.} 2006,
  \apj, 639, 1069

\bibitem[{{Hirsch}(2009)}]{Hirsch09}
{Hirsch}, H.~A. 2009, PhD thesis, (Univ. Erlangen-N{\"u}rnberg)

\bibitem[{{Hirschi} {et~al.}(2004){Hirschi}, {Meynet}, \&
  {Maeder}}]{Hirschietal04}
{Hirschi}, R., {Meynet}, G., \& {Maeder}, A. 2004, \aap, 425, 649

\bibitem[{{Hoag} \& {Applequist}(1965)}]{HoAp65}
{Hoag}, A.~A. \& {Applequist}, N.~L. 1965, \apjs, 12, 215

\bibitem[{{Holgado} {et~al.}(2018){Holgado}, {Sim{\'o}n-D{\'\i}az},
  {Barb{\'a}}, {Puls}, {Herrero}, {Castro}, {Garcia}, {Ma{\'\i}z
  Apell{\'a}niz}, {Negueruela}, \& {Sab{\'\i}n-Sanjuli{\'a}n}}]{Holgadoetal18}
{Holgado}, G., {Sim{\'o}n-D{\'\i}az}, S., {Barb{\'a}}, R.~H., {et~al.} 2018,
  \aap, 613, A65

\bibitem[{{Holgado} {et~al.}(2022){Holgado}, {Sim{\'o}n-D{\'\i}az}, {Herrero},
  \& {Barb{\'a}}}]{Holgadoetal22}
{Holgado}, G., {Sim{\'o}n-D{\'\i}az}, S., {Herrero}, A., \& {Barb{\'a}}, R.~H.
  2022, \aap, 665, A150

\bibitem[{{Hoogerwerf} {et~al.}(2000){Hoogerwerf}, {de Bruijne}, \& {de
  Zeeuw}}]{Hoogerwerfetal00}
{Hoogerwerf}, R., {de Bruijne}, J.~H.~J., \& {de Zeeuw}, P.~T. 2000, \apjl,
  544, L133

\bibitem[{{Huang} \& {Gies}(2006)}]{HuGa06a}
{Huang}, W. \& {Gies}, D.~R. 2006, \apj, 648, 580

\bibitem[{{Hubeny} {et~al.}(1994){Hubeny}, {Hummer}, \& {Lanz}}]{Hubenyetal94}
{Hubeny}, I., {Hummer}, D.~G., \& {Lanz}, T. 1994, \aap, 282, 151

\bibitem[{{Hubeny} \& {Lanz}(1995)}]{HuLa95}
{Hubeny}, I. \& {Lanz}, T. 1995, \apj, 439, 875

\bibitem[{{Hunter} {et~al.}(2009){Hunter}, {Brott}, {Langer}, {Lennon},
  {Dufton}, {Howarth}, {Ryans}, {Trundle}, {Evans}, {de Koter}, \&
  {Smartt}}]{Hunteretal09}
{Hunter}, I., {Brott}, I., {Langer}, N., {et~al.} 2009, \aap, 496, 841

\bibitem[{{Hunter} {et~al.}(2007){Hunter}, {Dufton}, {Smartt}, {Ryans},
  {Evans}, {Lennon}, {Trundle}, {Hubeny}, \& {Lanz}}]{Hunteretal07}
{Hunter}, I., {Dufton}, P.~L., {Smartt}, S.~J., {et~al.} 2007, \aap, 466, 277

\bibitem[{{Irrgang} {et~al.}(2014){Irrgang}, {Przybilla}, {Heber}, {B{\"o}ck},
  {Hanke}, {Nieva}, \& {Butler}}]{Irrgangetal14}
{Irrgang}, A., {Przybilla}, N., {Heber}, U., {et~al.} 2014, \aap, 565, A63

\bibitem[{{J{\"a}rvinen} {et~al.}(2022){J{\"a}rvinen}, {Hubrig},
  {Sch{\"o}ller}, {Cikota}, {Ilyin}, {Hummel}, \&
  {K{\"u}ker}}]{Jaervinenetal22}
{J{\"a}rvinen}, S.~P., {Hubrig}, S., {Sch{\"o}ller}, M., {et~al.} 2022, \mnras,
  510, 4405

\bibitem[{{Kaufer} {et~al.}(1999){Kaufer}, {Stahl}, {Tubbesing},
  {N{\o}rregaard}, {Avila}, {Francois}, {Pasquini}, \&
  {Pizzella}}]{Kauferetal99}
{Kaufer}, A., {Stahl}, O., {Tubbesing}, S., {et~al.} 1999, The Messenger, 95, 8

\bibitem[{{Kharchenko} {et~al.}(2013){Kharchenko}, {Piskunov}, {Schilbach},
  {R{\"o}ser}, \& {Scholz}}]{Kharchenkoetal13}
{Kharchenko}, N.~V., {Piskunov}, A.~E., {Schilbach}, E., {R{\"o}ser}, S., \&
  {Scholz}, R.~D. 2013, \aap, 558, A53

\bibitem[{{Kimeswenger} {et~al.}(2021){Kimeswenger}, {Rainer}, {Przybilla}, \&
  {Kausch}}]{Kimeswengeretal21}
{Kimeswenger}, S., {Rainer}, M., {Przybilla}, N., \& {Kausch}, W. 2021, \aj,
  161, 66

\bibitem[{{Kun} {et~al.}(2008){Kun}, {Kiss}, \& {Balog}}]{Kunetal08}
{Kun}, M., {Kiss}, Z.~T., \& {Balog}, Z. 2008, in Handbook of Star Forming
  Regions, Volume I, ed. B.~{Reipurth} ({San Francisco: ASP}), 136

\bibitem[{{Kurucz}(1993)}]{Kurucz93}
{Kurucz}, R. 1993, CD-ROM No.~13~(Cambridge, Mass.: SAO)

\bibitem[{{Kurucz}(2005)}]{Kurucz05}
{Kurucz}, R.~L. 2005, Mem. Soc. Astron. Ital. Suppl., 8, 14

\bibitem[{{Langer}(2012)}]{Langer12}
{Langer}, N. 2012, \araa, 50, 107

\bibitem[{{Langer} \& {Kudritzki}(2014)}]{LaKu14}
{Langer}, N. \& {Kudritzki}, R.~P. 2014, \aap, 564, A52

\bibitem[{{Lanz} \& {Hubeny}(2003)}]{LaHu03}
{Lanz}, T. \& {Hubeny}, I. 2003, \apjs, 146, 417

\bibitem[{{Lindegren} {et~al.}(2021){Lindegren}, {Bastian}, {Biermann},
  {Bombrun}, {de Torres}, {Gerlach}, {Geyer}, {Hern{\'a}ndez}, {Hilger},
  {Hobbs}, {Klioner}, {Lammers}, {McMillan}, {Ramos-Lerate},
  {Steidelm{\"u}ller}, {Stephenson}, \& {van Leeuwen}}]{Lindegrenetal21}
{Lindegren}, L., {Bastian}, U., {Biermann}, M., {et~al.} 2021, \aap, 649, A4

\bibitem[{{Lorenzo} {et~al.}(2022){Lorenzo}, {Garcia}, {Najarro}, {Herrero},
  {Cervi{\~n}o}, \& {Castro}}]{Lorenzoetal22}
{Lorenzo}, M., {Garcia}, M., {Najarro}, F., {et~al.} 2022, \mnras, 516, 4164

\bibitem[{{Lucy}(2007)}]{Lucy07}
{Lucy}, L.~B. 2007, \aap, 468, 649

\bibitem[{{Lucy}(2010)}]{Lucy10}
{Lucy}, L.~B. 2010, \aap, 512, A33

\bibitem[{{Maeder} \& {Meynet}(2012)}]{MaMe12}
{Maeder}, A. \& {Meynet}, G. 2012, Reviews of Modern Physics, 84, 25

\bibitem[{{Maeder} {et~al.}(2014){Maeder}, {Przybilla}, {Nieva}, {Georgy},
  {Meynet}, {Ekstr{\"o}m}, \& {Eggenberger}}]{Maederetal14}
{Maeder}, A., {Przybilla}, N., {Nieva}, M.~F., {et~al.} 2014, \aap, 565, A39

\bibitem[{{Mahy} {et~al.}(2015){Mahy}, {Rauw}, {De Becker}, {Eenens}, \&
  {Flores}}]{Mahyetal15}
{Mahy}, L., {Rauw}, G., {De Becker}, M., {Eenens}, P., \& {Flores}, C.~A. 2015,
  \aap, 577, A23

\bibitem[{{Mahy} {et~al.}(2022){Mahy}, {Sana}, {Shenar}, {Sen}, {Langer},
  {Marchant}, {Abdul-Masih}, {Banyard}, {Bodensteiner}, {Bowman}, {Dsilva},
  {Fabry}, {Hawcroft}, {Janssens}, {Van Reeth}, \& {Eldridge}}]{Mahyetal22}
{Mahy}, L., {Sana}, H., {Shenar}, T., {et~al.} 2022, \aap, 664, A159

\bibitem[{{Manset} \& {Donati}(2003)}]{ManDon03}
{Manset}, N. \& {Donati}, J.-F. 2003, Proc. SPIE, 4843, 425

\bibitem[{{Mao} {et~al.}(2020){Mao}, {Badnell}, \& {Del Zanna}}]{Maoetal20}
{Mao}, J., {Badnell}, N.~R., \& {Del Zanna}, G. 2020, \aap, 643, A95

\bibitem[{{Marcolino} {et~al.}(2009){Marcolino}, {Bouret}, {Martins},
  {Hillier}, {Lanz}, \& {Escolano}}]{Marcolinoetal09}
{Marcolino}, W.~L.~F., {Bouret}, J.~C., {Martins}, F., {et~al.} 2009, \aap,
  498, 837

\bibitem[{{Marcolino} {et~al.}(2022){Marcolino}, {Bouret}, {Rocha-Pinto},
  {Bernini-Peron}, \& {Vink}}]{Marcolinoetal22}
{Marcolino}, W.~L.~F., {Bouret}, J.~C., {Rocha-Pinto}, H.~J., {Bernini-Peron},
  M., \& {Vink}, J.~S. 2022, \mnras, 511, 5104

\bibitem[{{Martin} {et~al.}(1993){Martin}, {Kaufman}, \&
  {Musgrove}}]{Martinetal93}
{Martin}, W.~C., {Kaufman}, V., \& {Musgrove}, A. 1993, J. Phys. Chem. Ref.
  Data, 22, 1179

\bibitem[{{Martins} {et~al.}(2015{\natexlab{a}}){Martins}, {Herv{\'e}},
  {Bouret}, {Marcolino}, {Wade}, {Neiner}, {Alecian}, {Grunhut}, \&
  {Petit}}]{Martinsetal15a}
{Martins}, F., {Herv{\'e}}, A., {Bouret}, J.~C., {et~al.} 2015{\natexlab{a}},
  \aap, 575, A34

\bibitem[{{Martins} {et~al.}(2012){Martins}, {Mahy}, {Hillier}, \&
  {Rauw}}]{Martinsetal12}
{Martins}, F., {Mahy}, L., {Hillier}, D.~J., \& {Rauw}, G. 2012, \aap, 538, A39

\bibitem[{{Martins} {et~al.}(2004){Martins}, {Schaerer}, {Hillier}, \&
  {Heydari-Malayeri}}]{Martinsetal04}
{Martins}, F., {Schaerer}, D., {Hillier}, D.~J., \& {Heydari-Malayeri}, M.
  2004, \aap, 420, 1087

\bibitem[{{Martins} {et~al.}(2005){Martins}, {Schaerer}, {Hillier},
  {Meynadier}, {Heydari-Malayeri}, \& {Walborn}}]{Martinsetal05}
{Martins}, F., {Schaerer}, D., {Hillier}, D.~J., {et~al.} 2005, \aap, 441, 735

\bibitem[{{Martins} {et~al.}(2015{\natexlab{b}}){Martins},
  {Sim{\'o}n-D{\'\i}az}, {Palacios}, {Howarth}, {Georgy}, {Walborn}, {Bouret},
  \& {Barb{\'a}}}]{Martinsetal15b}
{Martins}, F., {Sim{\'o}n-D{\'\i}az}, S., {Palacios}, A., {et~al.}
  2015{\natexlab{b}}, \aap, 578, A109

\bibitem[{{Mason} {et~al.}(2001){Mason}, {Wycoff}, {Hartkopf}, {Douglass}, \&
  {Worley}}]{Masonetal01}
{Mason}, B.~D., {Wycoff}, G.~L., {Hartkopf}, W.~I., {Douglass}, G.~G., \&
  {Worley}, C.~E. 2001, \aj, 122, 3466

\bibitem[{{Massey} {et~al.}(1995){Massey}, {Johnson}, \&
  {Degioia-Eastwood}}]{Masseyetal95}
{Massey}, P., {Johnson}, K.~E., \& {Degioia-Eastwood}, K. 1995, \apj, 454, 151

\bibitem[{{Massey} {et~al.}(2013){Massey}, {Neugent}, {Hillier}, \&
  {Puls}}]{Masseyetal13}
{Massey}, P., {Neugent}, K.~F., {Hillier}, D.~J., \& {Puls}, J. 2013, \apj,
  768, 6

\bibitem[{{McLeod} {et~al.}(2015){McLeod}, {Dale}, {Ginsburg}, {Ercolano},
  {Gritschneder}, {Ramsay}, \& {Testi}}]{McLeodetal15}
{McLeod}, A.~F., {Dale}, J.~E., {Ginsburg}, A., {et~al.} 2015, \mnras, 450,
  1057

\bibitem[{{Megeath} {et~al.}(2008){Megeath}, {Townsley}, {Oey}, \&
  {Tieftrunk}}]{Megeathetal08}
{Megeath}, S.~T., {Townsley}, L.~K., {Oey}, M.~S., \& {Tieftrunk}, A.~R. 2008,
  in Handbook of Star Forming Regions, Volume I, ed. B.~{Reipurth} ({San
  Francisco: ASP}), 264

\bibitem[{{Mermilliod}(1982)}]{Mermilliod82}
{Mermilliod}, J.~C. 1982, \aap, 109, 37

\bibitem[{{Mermilliod}(1997)}]{Mermilliod97}
{Mermilliod}, J.~C. 1997, VizieR Online Data Catalog, 2168

\bibitem[{{Meynet} {et~al.}(2011){Meynet}, {Eggenberger}, \&
  {Maeder}}]{Meynetetal11}
{Meynet}, G., {Eggenberger}, P., \& {Maeder}, A. 2011, \aap, 525, L11

\bibitem[{{Meynet} \& {Maeder}(2000)}]{MeMa00}
{Meynet}, G. \& {Maeder}, A. 2000, \aap, 361, 101

\bibitem[{{Mokiem} {et~al.}(2005){Mokiem}, {de Koter}, {Puls}, {Herrero},
  {Najarro}, \& {Villamariz}}]{Mokiemetal05}
{Mokiem}, M.~R., {de Koter}, A., {Puls}, J., {et~al.} 2005, \aap, 441, 711

\bibitem[{{Moore}(1993)}]{Moore93}
{Moore}, C.~E. 1993, {Tables of Spectra of Hydrogen, Carbon, Nitrogen, and
  Oxygen Atoms and Ions} ({Boca Raton, FL: CRC Press})

\bibitem[{{Morel} \& {Magnenat}(1978)}]{MoMa78}
{Morel}, M. \& {Magnenat}, P. 1978, \aaps, 34, 477

\bibitem[{{Morel} {et~al.}(2022){Morel}, {Blaz{\`e}re}, {Semaan}, {Gosset},
  {Zorec}, {Fr{\'e}mat}, {Blomme}, {Daflon}, {Lobel}, {Nieva}, {Przybilla},
  {Gebran}, {Herrero}, {Mahy}, {Santos}, {Tautvai{\v{s}}ien{\.{e}}}, {Gilmore},
  {Randich}, {Alfaro}, {Bergemann}, {Carraro}, {Damiani}, {Franciosini},
  {Morbidelli}, {Pancino}, {Worley}, \& {Zaggia}}]{Moreletal22}
{Morel}, T., {Blaz{\`e}re}, A., {Semaan}, T., {et~al.} 2022, \aap, 665, A108

\bibitem[{{Morel} \& {Butler}(2008)}]{MoBu08}
{Morel}, T. \& {Butler}, K. 2008, \aap, 487, 307

\bibitem[{{Morel} {et~al.}(2015){Morel}, {Castro}, {Fossati}, {Hubrig},
  {Langer}, {Przybilla}, {Sch{\"o}ller}, {Carroll}, {Ilyin}, {Irrgang},
  {Oskinova}, {Schneider}, {D{\'\i}az}, {Briquet}, {Gonz{\'a}lez},
  {Kharchenko}, {Nieva}, {Scholz}, {de Koter}, {Hamann}, {Herrero}, {Ma{\'\i}z
  Apell{\'a}niz}, {Sana}, {Arlt}, {Barb{\'a}}, {Dufton}, {Kholtygin}, {Mathys},
  {Piskunov}, {Reisenegger}, {Spruit}, \& {Yoon}}]{Moreletal15}
{Morel}, T., {Castro}, N., {Fossati}, L., {et~al.} 2015, in New Windows on
  Massive Stars, ed. G.~{Meynet}, C.~{Georgy}, J.~{Groh}, \& P.~{Stee}, Vol.
  307, 342

\bibitem[{{Najarro} {et~al.}(2006){Najarro}, {Hillier}, {Puls}, {Lanz}, \&
  {Martins}}]{Najarroetal06}
{Najarro}, F., {Hillier}, D.~J., {Puls}, J., {Lanz}, T., \& {Martins}, F. 2006,
  \aap, 456, 659

\bibitem[{Nelder \& Mead(1965)}]{NeMe65}
Nelder, J.~A. \& Mead, R. 1965, Computer Journal, 7, 308

\bibitem[{{Nieva} \& {Przybilla}(2006)}]{NiPr06}
{Nieva}, M.~F. \& {Przybilla}, N. 2006, ApJ, 639, L39

\bibitem[{{Nieva} \& {Przybilla}(2007)}]{NiPr07}
{Nieva}, M.~F. \& {Przybilla}, N. 2007, \aap, 467, 295

\bibitem[{{Nieva} \& {Przybilla}(2008)}]{NiPr08}
{Nieva}, M.~F. \& {Przybilla}, N. 2008, \aap, 481, 199

\bibitem[{{Nieva} \& {Przybilla}(2012)}]{NiPr12}
{Nieva}, M.~F. \& {Przybilla}, N. 2012, \aap, 539, A143

\bibitem[{{Nieva} \& {Przybilla}(2014)}]{NiPr14}
{Nieva}, M.~F. \& {Przybilla}, N. 2014, \aap, 566, A7

\bibitem[{{Nieva} \& {Sim{\'o}n-D{\'\i}az}(2011)}]{NiSi11}
{Nieva}, M.~F. \& {Sim{\'o}n-D{\'\i}az}, S. 2011, \aap, 532, A2

\bibitem[{{Odenkirchen} \& {Brosche}(1992)}]{OdBr92}
{Odenkirchen}, M. \& {Brosche}, P. 1992, Astronomische Nachrichten, 313, 69

\bibitem[{{Pfeiffer} {et~al.}(1998){Pfeiffer}, {Frank}, {Baumueller},
  {Fuhrmann}, \& {Gehren}}]{Pfeifferetal98}
{Pfeiffer}, M.~J., {Frank}, C., {Baumueller}, K., {Fuhrmann}, K., \& {Gehren},
  T. 1998, \aaps, 130, 381

\bibitem[{{Pourbaix} {et~al.}(2004){Pourbaix}, {Tokovinin}, {Batten}, {Fekel},
  {Hartkopf}, {Levato}, {Morrell}, {Torres}, \& {Udry}}]{Pourbaixetal04}
{Pourbaix}, D., {Tokovinin}, A.~A., {Batten}, A.~H., {et~al.} 2004, \aap, 424,
  727

\bibitem[{{Poveda} {et~al.}(1967){Poveda}, {Ruiz}, \& {Allen}}]{Povedaetal67}
{Poveda}, A., {Ruiz}, J., \& {Allen}, C. 1967, Bol. Obs. Tonantzintla Tacubaya,
  4, 86

\bibitem[{{Przybilla}(2005)}]{Przybilla05}
{Przybilla}, N. 2005, \aap, 443, 293

\bibitem[{{Przybilla} \& {Butler}(2001)}]{PrBu01}
{Przybilla}, N. \& {Butler}, K. 2001, \aap, 379, 955

\bibitem[{{Przybilla} \& {Butler}(2004)}]{PrBu04}
{Przybilla}, N. \& {Butler}, K. 2004, \apj, 609, 1181

\bibitem[{{Przybilla} {et~al.}(2001{\natexlab{a}}){Przybilla}, {Butler},
  {Becker}, \& {Kudritzki}}]{Przybillaetal01a}
{Przybilla}, N., {Butler}, K., {Becker}, S.~R., \& {Kudritzki}, R.~P.
  2001{\natexlab{a}}, \aap, 369, 1009

\bibitem[{{Przybilla} {et~al.}(2006{\natexlab{a}}){Przybilla}, {Butler},
  {Becker}, \& {Kudritzki}}]{Przybillaetal06a}
{Przybilla}, N., {Butler}, K., {Becker}, S.~R., \& {Kudritzki}, R.~P.
  2006{\natexlab{a}}, \aap, 445, 1099

\bibitem[{{Przybilla} {et~al.}(2000){Przybilla}, {Butler}, {Becker},
  {Kudritzki}, \& {Venn}}]{Przybillaetal00}
{Przybilla}, N., {Butler}, K., {Becker}, S.~R., {Kudritzki}, R.~P., \& {Venn},
  K.~A. 2000, \aap, 359, 1085

\bibitem[{{Przybilla} {et~al.}(2001{\natexlab{b}}){Przybilla}, {Butler}, \&
  {Kudritzki}}]{Przybillaetal01b}
{Przybilla}, N., {Butler}, K., \& {Kudritzki}, R.~P. 2001{\natexlab{b}}, \aap,
  379, 936

\bibitem[{{Przybilla} {et~al.}(2010){Przybilla}, {Firnstein}, {Nieva},
  {Meynet}, \& {Maeder}}]{Przybillaetal10}
{Przybilla}, N., {Firnstein}, M., {Nieva}, M.~F., {Meynet}, G., \& {Maeder}, A.
  2010, \aap, 517, A38

\bibitem[{{Przybilla} {et~al.}(2008){Przybilla}, {Nieva}, \&
  {Butler}}]{Przybillaetal08}
{Przybilla}, N., {Nieva}, M.~F., \& {Butler}, K. 2008, \apjl, 688, L103

\bibitem[{{Przybilla} {et~al.}(2006{\natexlab{b}}){Przybilla}, {Nieva}, \&
  {Edelmann}}]{Przybillaetal06b}
{Przybilla}, N., {Nieva}, M.~F., \& {Edelmann}, H. 2006{\natexlab{b}}, Baltic
  Astronomy, 15, 107

\bibitem[{{Przybilla} {et~al.}(2013){Przybilla}, {Nieva}, {Irrgang}, \&
  {Butler}}]{Przybillaetal13}
{Przybilla}, N., {Nieva}, M.~F., {Irrgang}, A., \& {Butler}, K. 2013, EAS
  Publ.~Ser., 63, 13

\bibitem[{{Puls} {et~al.}(1996){Puls}, {Kudritzki}, {Herrero}, {Pauldrach},
  {Haser}, {Lennon}, {Gabler}, {Voels}, {Vilchez}, {Wachter}, \&
  {Feldmeier}}]{Pulsetal96}
{Puls}, J., {Kudritzki}, R.~P., {Herrero}, A., {et~al.} 1996, \aap, 305, 171

\bibitem[{{Puls} {et~al.}(2020){Puls}, {Najarro}, {Sundqvist}, \&
  {Sen}}]{Pulsetal20}
{Puls}, J., {Najarro}, F., {Sundqvist}, J.~O., \& {Sen}, K. 2020, \aap, 642,
  A172

\bibitem[{{Puls} {et~al.}(2005){Puls}, {Urbaneja}, {Venero}, {Repolust},
  {Springmann}, {Jokuthy}, \& {Mokiem}}]{Pulsetal05}
{Puls}, J., {Urbaneja}, M.~A., {Venero}, R., {et~al.} 2005, \aap, 435, 669

\bibitem[{{Puls} {et~al.}(2008){Puls}, {Vink}, \& {Najarro}}]{Pulsetal08}
{Puls}, J., {Vink}, J.~S., \& {Najarro}, F. 2008, \aapr, 16, 209

\bibitem[{{Quintana} \& {Wright}(2021)}]{QuWr21}
{Quintana}, A.~L. \& {Wright}, N.~J. 2021, \mnras, 508, 2370

\bibitem[{{Rauw} \& {Naz{\'e}}(2016)}]{RaNa16}
{Rauw}, G. \& {Naz{\'e}}, Y. 2016, \aap, 594, A82

\bibitem[{{Reid} {et~al.}(2019){Reid}, {Menten}, {Brunthaler}, {Zheng}, {Dame},
  {Xu}, {Li}, {Sakai}, {Wu}, {Immer}, {Zhang}, {Sanna}, {Moscadelli}, {Rygl},
  {Bartkiewicz}, {Hu}, {Quiroga-Nu{\~n}ez}, \& {van Langevelde}}]{Reidetal19}
{Reid}, M.~J., {Menten}, K.~M., {Brunthaler}, A., {et~al.} 2019, \apj, 885, 131

\bibitem[{{Rodrigo} \& {Solano}(2020)}]{Rodrigo20}
{Rodrigo}, C. \& {Solano}, E. 2020, Contributions to the XIV.0 Scientific
  Meeting of the Spanish Astronomical Society

\bibitem[{{Rodrigo} {et~al.}(2012){Rodrigo}, {Solano}, \& {Bayo}}]{Rodrigo12}
{Rodrigo}, C., {Solano}, E., \& {Bayo}, A. 2012, IVOA Working Draft

\bibitem[{{Rom{\'a}n-Z{\'u}{\~n}iga} \& {Lada}(2008)}]{RoLa08}
{Rom{\'a}n-Z{\'u}{\~n}iga}, C.~G. \& {Lada}, E.~A. 2008, in Handbook of Star
  Forming Regions, Volume I, ed. B.~{Reipurth} ({San Francisco: ASP}), 928

\bibitem[{{Rybicki} \& {Hummer}(1991)}]{RyHu91}
{Rybicki}, G.~B. \& {Hummer}, D.~G. 1991, \aap, 245, 171

\bibitem[{{Sana} {et~al.}(2012){Sana}, {de Mink}, {de Koter}, {Langer},
  {Evans}, {Gieles}, {Gosset}, {Izzard}, {Le Bouquin}, \&
  {Schneider}}]{Sanaetal12}
{Sana}, H., {de Mink}, S.~E., {de Koter}, A., {et~al.} 2012, Science, 337, 444

\bibitem[{{Sana} {et~al.}(2014){Sana}, {Le Bouquin}, {Lacour}, {Berger},
  {Duvert}, {Gauchet}, {Norris}, {Olofsson}, {Pickel}, {Zins}, {Absil}, {de
  Koter}, {Kratter}, {Schnurr}, \& {Zinnecker}}]{Sanaetal14}
{Sana}, H., {Le Bouquin}, J.~B., {Lacour}, S., {et~al.} 2014, \apjs, 215, 15

\bibitem[{{Santolaya-Rey} {et~al.}(1997){Santolaya-Rey}, {Puls}, \&
  {Herrero}}]{SantolayaReyetal97}
{Santolaya-Rey}, A.~E., {Puls}, J., \& {Herrero}, A. 1997, \aap, 323, 488

\bibitem[{{Sch{\"o}ller} {et~al.}(2017){Sch{\"o}ller}, {Hubrig}, {Fossati},
  {Carroll}, {Briquet}, {Oskinova}, {J{\"a}rvinen}, {Ilyin}, {Castro}, {Morel},
  {Langer}, {Przybilla}, {Nieva}, {Kholtygin}, {Sana}, {Herrero}, {Barb{\'a}},
  {de Koter}, \& {BOB Collaboration}}]{Schoelleretal17}
{Sch{\"o}ller}, M., {Hubrig}, S., {Fossati}, L., {et~al.} 2017, \aap, 599, A66

\bibitem[{{Sch\"onberner} {et~al.}(1988){Sch\"onberner}, {Herrero}, {Becker},
  {Eber}, {Butler}, {Kudritzki}, \& {Simon}}]{Schoenberneretal88}
{Sch\"onberner}, D., {Herrero}, A., {Becker}, S., {et~al.} 1988, \aap, 197, 209

\bibitem[{{Sch\"oning} \& {Butler}(1989{\natexlab{a}})}]{SchBu89a}
{Sch\"oning}, T. \& {Butler}, K. 1989{\natexlab{a}}, \aaps, 78, 51

\bibitem[{{Sch\"oning} \& {Butler}(1989{\natexlab{b}})}]{SchBu89b}
{Sch\"oning}, T. \& {Butler}, K. 1989{\natexlab{b}}, \aap, 219, 326

\bibitem[{{Seaton}(1962)}]{Seaton62}
{Seaton}, M.~J. 1962, in Atomic and Molecular Processes, ed. D.~R. {Bates}, 375

\bibitem[{{Seaton} {et~al.}(1994){Seaton}, {Yan}, {Mihalas}, \&
  {Pradhan}}]{Seatonetal94}
{Seaton}, M.~J., {Yan}, Y., {Mihalas}, D., \& {Pradhan}, A.~K. 1994, \mnras,
  266, 805

\bibitem[{{Sen} {et~al.}(2022){Sen}, {Langer}, {Marchant}, {Menon}, {de Mink},
  {Schootemeijer}, {Sch{\"u}rmann}, {Mahy}, {Hastings}, {Nathaniel}, {Sana},
  {Wang}, \& {Xu}}]{Senetal22}
{Sen}, K., {Langer}, N., {Marchant}, P., {et~al.} 2022, \aap, 659, A98

\bibitem[{{Shi} \& {Hu}(1999)}]{ShiHu99}
{Shi}, H.~M. \& {Hu}, J.~Y. 1999, \aaps, 136, 313

\bibitem[{{Shultz} {et~al.}(2019){Shultz}, {Le Bouquin}, {Rivinius}, {Wade},
  {Kochukhov}, {Alecian}, {Petit}, {Pfuhl}, {Karl}, {Gao}, {Grellmann}, {Lin},
  {Garcia}, {Lacour}, {MiMeS Collaboration}, \& {BinaMIcS
  Collaboration}}]{Shultzetal19}
{Shultz}, M., {Le Bouquin}, J.~B., {Rivinius}, T., {et~al.} 2019, \mnras, 482,
  3950

\bibitem[{{Sim{\'o}n-D{\'\i}az} {et~al.}(2011){Sim{\'o}n-D{\'\i}az},
  {Garc{\'\i}a-Rojas}, {Esteban}, {Stasi{\'n}ska}, {L{\'o}pez-S{\'a}nchez}, \&
  {Morisset}}]{SimonDiazetal11}
{Sim{\'o}n-D{\'\i}az}, S., {Garc{\'\i}a-Rojas}, J., {Esteban}, C., {et~al.}
  2011, \aap, 530, A57

\bibitem[{{Sim{\'o}n-D{\'\i}az} {et~al.}(2017){Sim{\'o}n-D{\'\i}az}, {Godart},
  {Castro}, {Herrero}, {Aerts}, {Puls}, {Telting}, \&
  {Grassitelli}}]{SimonDiazetal17}
{Sim{\'o}n-D{\'\i}az}, S., {Godart}, M., {Castro}, N., {et~al.} 2017, \aap,
  597, A22

\bibitem[{{Sim{\'o}n-D{\'\i}az} \& {Herrero}(2014)}]{SiHe14}
{Sim{\'o}n-D{\'\i}az}, S. \& {Herrero}, A. 2014, \aap, 562, A135

\bibitem[{{Sim{\'o}n-D{\'\i}az} {et~al.}(2006){Sim{\'o}n-D{\'\i}az}, {Herrero},
  {Esteban}, \& {Najarro}}]{SimonDiazetal06}
{Sim{\'o}n-D{\'\i}az}, S., {Herrero}, A., {Esteban}, C., \& {Najarro}, F. 2006,
  \aap, 448, 351

\bibitem[{{Sota} {et~al.}(2014){Sota}, {Ma{\'\i}z Apell{\'a}niz}, {Morrell},
  {Barb{\'a}}, {Walborn}, {Gamen}, {Arias}, \& {Alfaro}}]{Sotaetal14}
{Sota}, A., {Ma{\'\i}z Apell{\'a}niz}, J., {Morrell}, N.~I., {et~al.} 2014,
  \apjs, 211, 10

\bibitem[{{Sota} {et~al.}(2011){Sota}, {Ma{\'\i}z Apell{\'a}niz}, {Walborn},
  {Alfaro}, {Barb{\'a}}, {Morrell}, {Gamen}, \& {Arias}}]{Sotaetal11}
{Sota}, A., {Ma{\'\i}z Apell{\'a}niz}, J., {Walborn}, N.~R., {et~al.} 2011,
  \apjs, 193, 24

\bibitem[{{Stoop} {et~al.}(2022){Stoop}, {Kaper}, {de Koter}, {Guo}, {Lamers},
  \& {Rieder}}]{Stoopetal22}
{Stoop}, M., {Kaper}, L., {de Koter}, A., {et~al.} 2022, arXiv e-prints,
  arXiv:2207.08452

\bibitem[{{Sundqvist} {et~al.}(2019){Sundqvist}, {Bj{\"o}rklund}, {Puls}, \&
  {Najarro}}]{Sundqvistetal19}
{Sundqvist}, J.~O., {Bj{\"o}rklund}, R., {Puls}, J., \& {Najarro}, F. 2019,
  \aap, 632, A126

\bibitem[{{Sung} {et~al.}(2017){Sung}, {Bessell}, {Chun}, {Yi}, {Naz{\'e}},
  {Lim}, {Karimov}, {Rauw}, {Park}, \& {Hur}}]{Sungetal17}
{Sung}, H., {Bessell}, M.~S., {Chun}, M.-Y., {et~al.} 2017, \apjs, 230, 3

\bibitem[{{Tayal}(2007)}]{Tayal07}
{Tayal}, S.~S. 2007, \apjs, 171, 331

\bibitem[{{Tayal} \& {Zatsarinny}(2017)}]{TaZa17}
{Tayal}, S.~S. \& {Zatsarinny}, O. 2017, \apj, 850, 147

\bibitem[{{Thompson} {et~al.}(1995){Thompson}, {Nandy}, {Jamar}, {Monfils},
  {Houziaux L.}, {Carnochan}, \& {Wilson}}]{Thompsonetal95}
{Thompson}, G.~I., {Nandy}, K., {Jamar}, C., {et~al.} 1995, VizieR Online Data
  Catalog, II/59B

\bibitem[{{Tremblay} \& {Bergeron}(2009)}]{TrBe09}
{Tremblay}, P.~E. \& {Bergeron}, P. 2009, \apj, 696, 1755

\bibitem[{{ud-Doula} {et~al.}(2009){ud-Doula}, {Owocki}, \&
  {Townsend}}]{udDoulaetal09}
{ud-Doula}, A., {Owocki}, S.~P., \& {Townsend}, R. H.~D. 2009, \mnras, 392,
  1022

\bibitem[{{van Leeuwen}(2007)}]{vanLeeuwen07a}
{van Leeuwen}, F. 2007, \aap, 474, 653

\bibitem[{{van Regemorter}(1962)}]{vanRegemorter62}
{van Regemorter}, H. 1962, \apj, 136, 906

\bibitem[{{Vanbeveren} {et~al.}(1998){Vanbeveren}, {De Loore}, \& {Van
  Rensbergen}}]{Vanbeverenetal98}
{Vanbeveren}, D., {De Loore}, C., \& {Van Rensbergen}, W. 1998, \aapr, 9, 63

\bibitem[{{Villamariz} \& {Herrero}(2000)}]{ViHe00}
{Villamariz}, M.~R. \& {Herrero}, A. 2000, \aap, 357, 597

\bibitem[{{Villamariz} \& {Herrero}(2005)}]{ViHe05}
{Villamariz}, M.~R. \& {Herrero}, A. 2005, \aap, 442, 263

\bibitem[{{Villamariz} {et~al.}(2002){Villamariz}, {Herrero}, {Becker}, \&
  {Butler}}]{Villamarizetal02}
{Villamariz}, M.~R., {Herrero}, A., {Becker}, S.~R., \& {Butler}, K. 2002,
  \aap, 388, 940

\bibitem[{{Vink} {et~al.}(2000){Vink}, {de Koter}, \& {Lamers}}]{Vinketal00}
{Vink}, J.~S., {de Koter}, A., \& {Lamers}, H.~J.~G.~L.~M. 2000, \aap, 362, 295

\bibitem[{{Vink} {et~al.}(2001){Vink}, {de Koter}, \& {Lamers}}]{Vinketal01}
{Vink}, J.~S., {de Koter}, A., \& {Lamers}, H.~J.~G.~L.~M. 2001, \aap, 369, 574

\bibitem[{{Vink} \& {Sander}(2021)}]{ViSa21}
{Vink}, J.~S. \& {Sander}, A. A.~C. 2021, \mnras, 504, 2051

\bibitem[{{Voels} {et~al.}(1989){Voels}, {Bohannan}, {Abbott}, \&
  {Hummer}}]{Voelsetal89}
{Voels}, S.~A., {Bohannan}, B., {Abbott}, D.~C., \& {Hummer}, D.~G. 1989, \apj,
  340, 1073

\bibitem[{{Wade} {et~al.}(2016){Wade}, {Neiner}, {Alecian}, {Grunhut}, {Petit},
  {Batz}, {Bohlender}, {Cohen}, {Henrichs}, {Kochukhov}, {Landstreet},
  {Manset}, {Martins}, {Mathis}, {Oksala}, {Owocki}, {Rivinius}, {Shultz},
  {Sundqvist}, {Townsend}, {ud-Doula}, {Bouret}, {Braithwaite}, {Briquet},
  {Carciofi}, {David-Uraz}, {Folsom}, {Fullerton}, {Leroy}, {Marcolino},
  {Moffat}, {Naz{\'e}}, {Louis}, {Auri{\`e}re}, {Bagnulo}, {Bailey},
  {Barb{\'a}}, {Blaz{\`e}re}, {B{\"o}hm}, {Catala}, {Donati}, {Ferrario},
  {Harrington}, {Howarth}, {Ignace}, {Kaper}, {L{\"u}ftinger}, {Prinja},
  {Vink}, {Weiss}, \& {Yakunin}}]{Wadeetal16}
{Wade}, G.~A., {Neiner}, C., {Alecian}, E., {et~al.} 2016, \mnras, 456, 2

\bibitem[{{Walborn} {et~al.}(2011){Walborn}, {Ma{\'\i}z Apell{\'a}niz}, {Sota},
  {Alfaro}, {Morrell}, {Barb{\'a}}, {Arias}, \& {Gamen}}]{Walbornetal11}
{Walborn}, N.~R., {Ma{\'\i}z Apell{\'a}niz}, J., {Sota}, A., {et~al.} 2011,
  \aj, 142, 150

\bibitem[{{Wellstein} {et~al.}(2001){Wellstein}, {Langer}, \&
  {Braun}}]{Wellsteinetal01}
{Wellstein}, S., {Langer}, N., \& {Braun}, H. 2001, \aap, 369, 939

\bibitem[{{Wesselius} {et~al.}(1982){Wesselius}, {van Duinen}, {de Jonge},
  {Aalders}, {Luinge}, \& {Wildeman}}]{Wesseliusetal82}
{Wesselius}, P.~R., {van Duinen}, R.~J., {de Jonge}, A.~R.~W., {et~al.} 1982,
  \aaps, 49, 427

\bibitem[{{We{\ss}mayer} {et~al.}(2022){We{\ss}mayer}, {Przybilla}, \&
  {Butler}}]{Wessmayeretal22}
{We{\ss}mayer}, D., {Przybilla}, N., \& {Butler}, K. 2022, \aap, 668, A92

\bibitem[{{Wiese} {et~al.}(1996){Wiese}, {Fuhr}, \& {Deters}}]{Wieseetal96}
{Wiese}, W.~L., {Fuhr}, J.~R., \& {Deters}, T.~M. 1996, J. Phys. \& Chem. Ref.
  Data., Monograph 7 (Melville, NY: AIP Press)

\bibitem[{{Xiang} {et~al.}(2022){Xiang}, {Rix}, {Ting}, {Kudritzki}, {Conroy},
  {Zari}, {Shi}, {Przybilla}, {Ramirez-Tannus}, {Tkachenko}, {Gebruers}, \&
  {Liu}}]{Xiangetal22}
{Xiang}, M., {Rix}, H.-W., {Ting}, Y.-S., {et~al.} 2022, \aap, 662, A66

\bibitem[{{Xu} {et~al.}(2021){Xu}, {Hou}, {Bian}, {Hao}, {Liu}, {Li}, \&
  {Li}}]{Xuetal21}
{Xu}, Y., {Hou}, L.~G., {Bian}, S.~B., {et~al.} 2021, \aap, 645, L8

\bibitem[{{Yu} {et~al.}(2021){Yu}, {Wang}, {Cui}, {Li}, {Liu}, {Zhang}, {Tian},
  {Huo}, {Ju}, {Liu}, {Wen}, \& {Feng}}]{Yuetal21}
{Yu}, Y., {Wang}, H.-F., {Cui}, W.-Y., {et~al.} 2021, \apj, 922, 80

\bibitem[{{Zorec} {et~al.}(2017){Zorec}, {Rieutord}, {Espinosa Lara},
  {Fr{\'e}mat}, {Domiciano de Souza}, \& {Royer}}]{Zorecetal17}
{Zorec}, J., {Rieutord}, M., {Espinosa Lara}, F., {et~al.} 2017, \aap, 606, A32

\end{thebibliography}

\appendix

\section{The distance to 10 Lac}\label{appendix:A}

The distance determination to the important MK standard star 10~Lac (HD~214680) deserves a dedicated discussion. The distance is
unfortunately not
tightly constrained because the star was too distant for the Hipparcos mission to measure a reliable parallax (both in the original and the
revised data reduction), and at $V$\,=\,4.88\,mag it is bright for the \textit{Gaia} mission \citep{Gaia16}, in both data releases, DR2 and EDR3.
Table~\ref{tab:appendix1} summarises the available parallax $\varpi$ and proper motion measurements $\mu_\alpha$ and $\mu_\delta$ (in right ascension $\alpha$ and declination $\delta$, respectively) from the Hipparcos and \textit{Gaia}
missions, as well as the resulting distances, and the spectroscopic distance from the present work. The parallax and proper motion
measurements disagree and a resolution of the discrepancies is beyond the scope of the present work. This possibly points to the existence
of unconsidered systematic effects.

\begin{figure}[ht]
   \centering
   \includegraphics[width=.96\hsize]{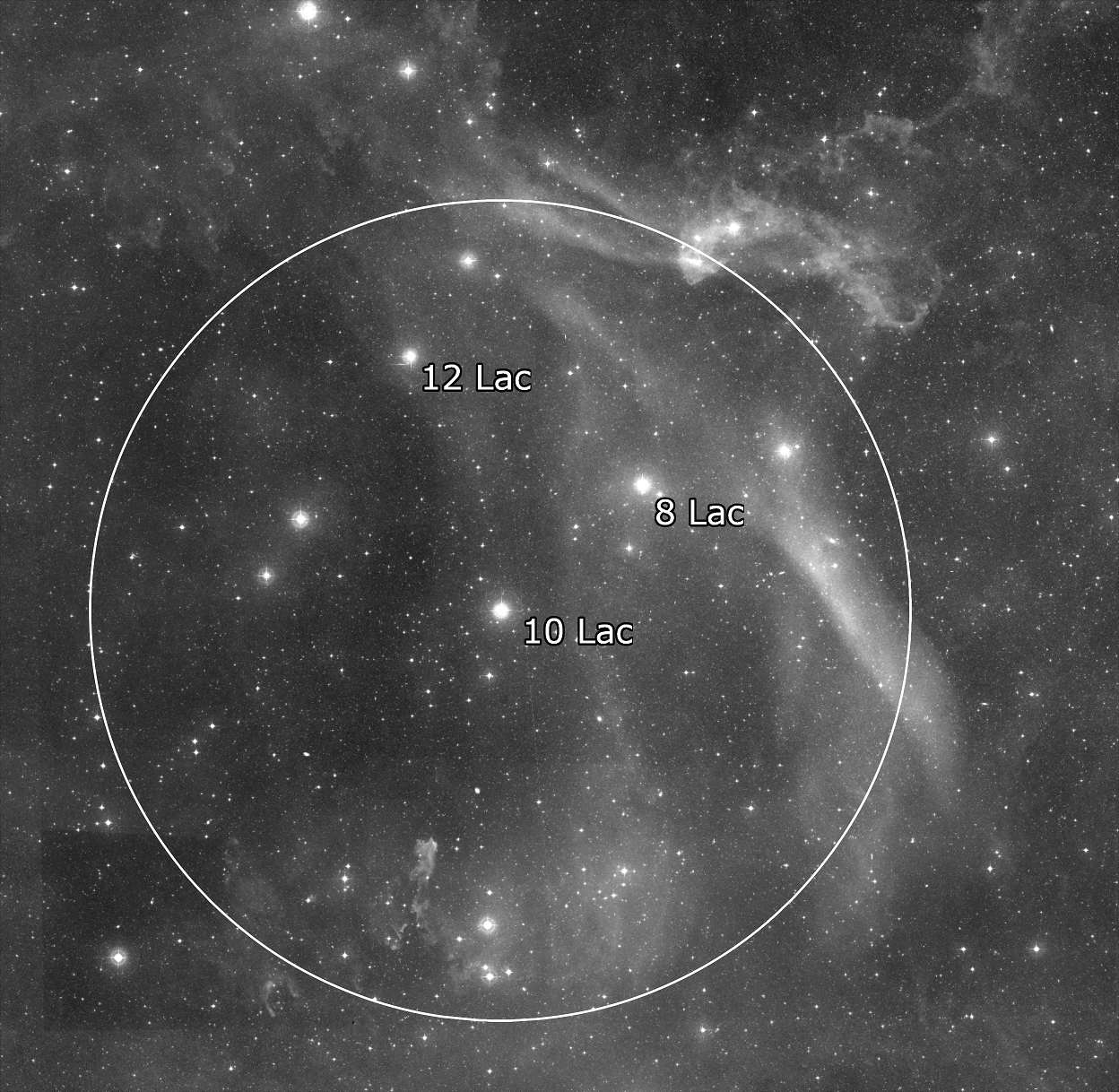}
      \caption{Image of the Lac OB1 association, adopted from the Digital Sky Survey in the red band.
      The central extent of the Lac OB1b association is encircled, a few bright stars are labelled.}
         \label{fig:10Laccomplex}
\end{figure}

Only the \citet{vanLeeuwen07a} value is compatible with our spectroscopic distance, which may be by chance, so an independent
constraint would be desirable. The star 10~Lac is the ionising source of the 10~Lac nebula complex (consisting in particular of the
\ion{H}{ii} region \object{Sh2-126}) that coincides with the Lac OB1b association \citep{ChLe08} (see our Fig.~\ref{fig:10Laccomplex}). Two
bright early B-type stars are identified, while two other Lac OB1 stars lie outside of the field shown.
An indirect distance estimation can be achieved by analysis of the parallaxes of the fainter Lac OB1b members, for which \textit{Gaia} should have
delivered more reliable data.

\begin{figure}[ht]
   \centering
   \includegraphics[width=.99\hsize]{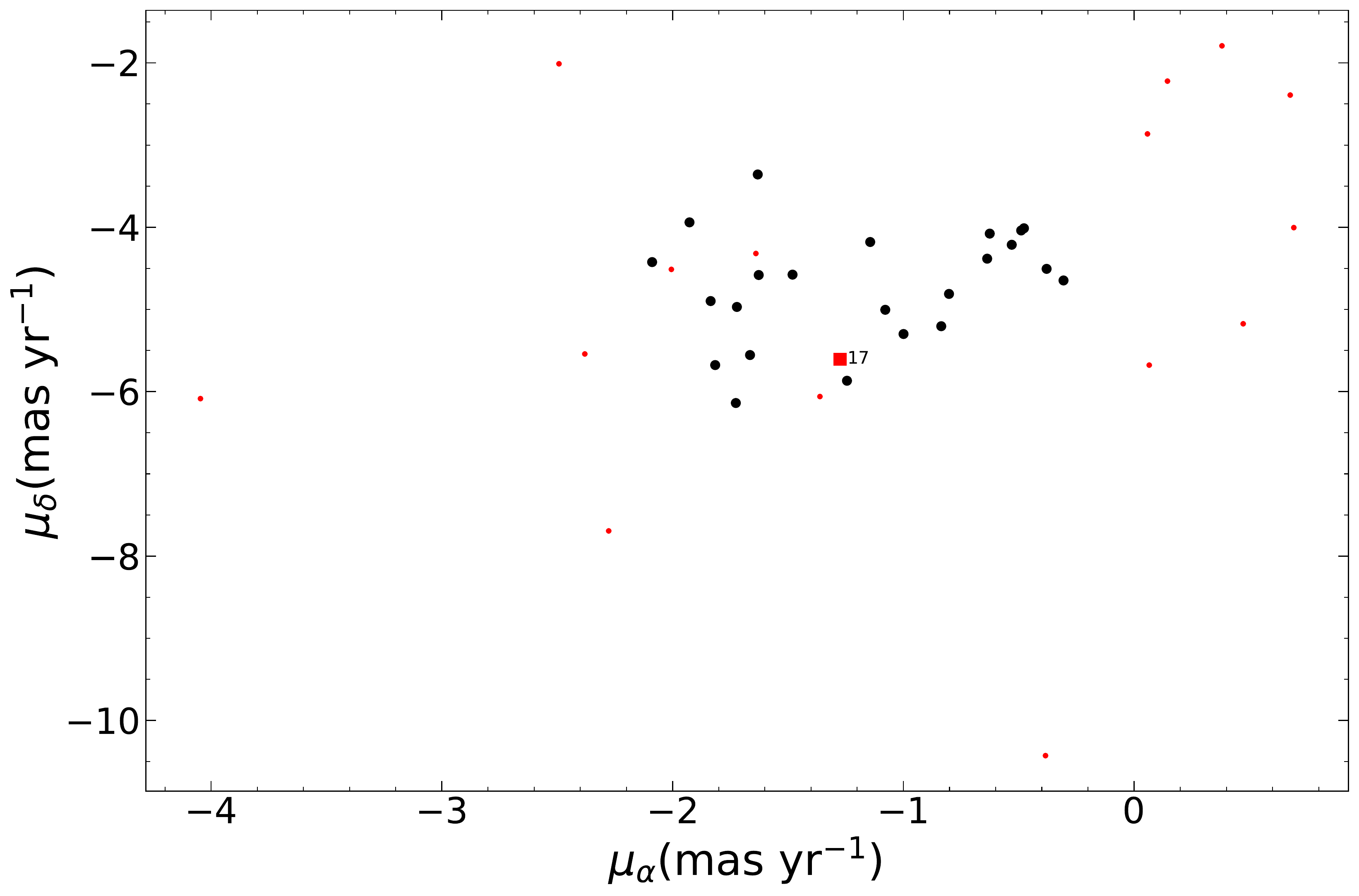}
      \caption{\textit{Gaia} EDR3 proper motions of stars in the Lac OB1b association. Larger black dots indicate stars retained, small red dots indicate stars removed for the distance estimation of the association based on their brightness, proper motion or RUWE. The sample star 10~Lac (ID\#17) is marked by the square.}
         \label{figure:lac_ob1b}
\end{figure}

We therefore selected Lac OB1b association member stars of spectral type B (and 10~Lac) according to \citet{ChLe08}.\  A summary of \textit{Gaia}
EDR3 measurements for these is provided Table~\ref{table:lac_ob1b}, where the name, parallax, proper motion components, the $G$ magnitude and the RUWE are tabulated.\ We then applied the following selection criteria. Stars brighter than the sixth $G$ magnitude were excluded, as were
stars with unusual proper motions (see Fig.~\ref{figure:lac_ob1b}) and those with a large RUWE, above a
value of 1.20. The distance to the Lac OB1b association was calculated from the inverted parallaxes of the remaining 23 objects to
$d_{\mathrm{Lac OB1b}}$\,=\,542$_{-52}^{+65}$\,pc (1$\sigma$ standard deviation), or $d_{\mathrm{Lac OB1b}}$\,=\,542$_{-11}^{+14}$\,pc (standard error of the mean), which agrees well with our spectroscopic distance for 10~Lac.

\begin{table*}[!ht]
\centering
\caption{Astrometric data and derived distances for 10~Lac.}\label{tab:appendix1}
 \small
 \begin{tabular}{lllll}
    \hline
    \hline
    $\varpi$\,(mas) & $\mu_\alpha$\,(mas\,yr$^{-1}$) & $\mu_\delta$\,(mas\,yr$^{-1}$) & $d$\,(pc) & Reference\\
    \hline
    3.08$\pm$0.62     & $-$0.29$\pm$0.66   & $-$5.70$\pm$0.52   & 325$^{+82}_{-55}$          & Hipparcos, \citet{ESA97}\\
    1.89$\pm$0.22     & $-$0.32$\pm$0.25   & $-$5.46$\pm$0.19   & 529$^{+70}_{-55}$          & Hipparcos, \citet{vanLeeuwen07a}\\
    2.7876$\pm$0.2321 & $-$1.437$\pm$0.296 & $-$4.803$\pm$0.368 & 391.31$^{+32.58}_{-27.57}$ & \textit{Gaia} DR2, \citet{Gaia18}\\
    2.1920$\pm$0.1314 & $-$1.274$\pm$0.118 & $-$5.605$\pm$0.120 & 456.20$^{+29.10}_{-25.80}$ & \textit{Gaia} EDR3, \citet{Gaia21}\\
    ...               & ...                & ...                & 552$\pm$32                 & spectroscopic, this work\\
    ...               & ...                & ...                & 542$^{+65}_{-52}$ & \textit{Gaia} EDR3 distance to Lac OB1b, this work\\
    \hline
\end{tabular}
\end{table*}

\begin{table*}[!ht]
    \centering
    \caption{\textit{Gaia} EDR3 parallaxes and photometric and astrometric measurements for stars in the Lac OB1b association.}\label{table:lac_ob1b}
    \small
    \begin{tabular}{l r r r r cr}
    \hline
    \hline
    Name & $\varpi (\si{\mas})$ & {$\mu_{\alpha} (\si{\mas\per\year})$} & {$\mu_{\delta} (\si{\mas\per\year})$} & {$G (\si{\mag})$} & RUWE\\
    \hline
    6 Lac & $2.8367 \pm 0.1336$ & $-2.380 \pm 0.098$ & $-5.541 \pm 0.124$ & $4.4775$ & $1.65$\\
    10 Lac & $2.1920 \pm 0.1314$ & $-1.274 \pm 0.118$ & $-5.605 \pm 0.120$ & $4.8255$ & $1.21$\\
    12 Lac & $2.5877 \pm 0.1327$ & $-2.005 \pm 0.150$ & $-4.512 \pm 0.117$ & $5.2102$ & $1.21$\\
    16 Lac & $2.2387 \pm 0.0736$ & $-1.639 \pm 0.044$ & $-4.319 \pm 0.064$ & $5.5638$ & $1.26$\\
    \textbf{HD\,212978} & $2.0135 \pm 0.0515$ & $-1.244 \pm 0.031$ & $-5.867 \pm 0.045$ & $6.1241$ & $1.09$\\
    HD\,217101 & $2.2918 \pm 0.0423$ & $0.065 \pm 0.048$ & $-5.677 \pm 0.047$ & $6.1392$ & $0.90$\\
    \textbf{HD\,215191} & $2.1623 \pm 0.0519$ & $-0.999 \pm 0.045$ & $-5.299 \pm 0.041$ & $6.3920$ & $1.11$\\
    \textbf{8 Lac} & $2.1711 \pm 0.0598$ & $-1.721 \pm 0.069$ & $-4.969 \pm 0.060$ & $6.4409$ & $1.19$\\
    \textbf{HD\,214263} & $1.9604 \pm 0.0403$ & $-1.726 \pm 0.038$ & $-6.138 \pm 0.036$ & $6.8144$ & $1.01$\\
    \textbf{HD\,213976} & $2.0603 \pm 0.0446$ & $-1.631 \pm 0.027$ & $-3.357 \pm 0.041$ & $7.0124$ & $1.08$\\
    \textbf{HD\,214432} & $1.9787 \pm 0.0382$ & $-1.835 \pm 0.043$ & $-4.898 \pm 0.039$ & $7.5643$ & $0.95$\\
    {HD\,216684} & $1.6846 \pm 0.0319$ & $-0.530 \pm 0.024$ & $-4.213 \pm 0.031$ & $7.7390$ & $1.11$\\
    \textbf{V423 Lac} & $1.6817 \pm 0.0280$ & $-0.477 \pm 0.021$ & $-4.012 \pm 0.026$ & $7.9560$ & $1.20$\\
    \textbf{HD\,212668} & $1.8088 \pm 0.0329$ & $-1.078 \pm 0.029$ & $-5.004 \pm 0.031$ & $8.0774$ & $1.06$\\
    HD\,213484 & $1.5504 \pm 0.0268$ & $0.381 \pm 0.023$ & $-1.793 \pm 0.026$ & $8.2749$ & $1.17$\\
    \textbf{HD\,214243} & $2.0388 \pm 0.0390$ & $-2.089 \pm 0.029$ & $-4.424 \pm 0.037$ & $8.2813$ & $1.08$\\
    \textbf{HD\,214098} & $1.6315 \pm 0.0347$ & $-0.306 \pm 0.026$ & $-4.647 \pm 0.029$ & $8.3035$ & $1.08$\\
    HD\,212153 & $0.4510 \pm 0.6390$ & $-4.046 \pm 0.498$ & $-6.085 \pm 0.537$ & $8.6457$ & $9.10$\\
    \textbf{HD\,216815} & $1.6617 \pm 0.0276$ & $-0.489 \pm 0.017$ & $-4.039 \pm 0.022$ & $8.3336$ & $1.16$\\
    \textbf{HD\,215211} & $1.8734 \pm 0.0311$ & $-0.836 \pm 0.034$ & $-5.204 \pm 0.029$ & $8.6368$ & $1.11$\\
    \textbf{HD\,216537} & $1.6559 \pm 0.0233$ & $-0.626 \pm 0.016$ & $-4.077 \pm 0.020$ & $8.7603$ & $1.13$\\
    HD\,213390 & $1.3726 \pm 0.0167$ & $0.145 \pm 0.014$ & $-2.222 \pm 0.017$ & $8.8383$ & $0.96$\\
    HD\,213800 & $1.6625 \pm 0.0229$ & $0.676 \pm 0.016$ & $-2.392 \pm 0.021$ & $9.1629$ & $1.06$\\
    \textbf{BD\,+42 4370} & $1.6166 \pm 0.0175$ & $-0.802 \pm 0.015$ & $-4.811 \pm 0.017$ & $9.2809$ & $0.97$\\
    \textbf{HD\,214977} & $1.7629 \pm 0.0204$ & $-1.627 \pm 0.011$ & $-4.581 \pm 0.019$ & $9.3238$ & $1.06$\\
    BD\,+40 4771 & $1.1461 \pm 0.0229$ & $0.692 \pm 0.018$ & $-4.004 \pm 0.021$ & $9.4695$ & $1.16$\\
    HD\,212732 & $1.6620 \pm 0.0253$ & $-2.277 \pm 0.022$ & $-7.694 \pm 0.026$ & $9.4676$ & $1.33$\\
    BD\,+38 4883 & $2.0092 \pm 0.0239$ & $0.473 \pm 0.024$ & $-5.174 \pm 0.027$ & $9.4627$ & $1.03$\\
    \textbf{BD\,+38 4834} & $2.0849 \pm 0.0208$ & $-1.815 \pm 0.021$ & $-5.676 \pm 0.021$ & $9.5604$ & $1.17$\\
    BD\,+40 4831 & $2.6886 \pm 0.0144$ & $0.058 \pm 0.010$ & $-2.864 \pm 0.014$ & $9.5632$ & $0.94$\\
    \textbf{HD\,214179} & $1.5980 \pm 0.0232$ & $-0.637 \pm 0.018$ & $-4.382 \pm 0.020$ & $9.6340$ & $1.10$\\
    BD\,+36 4896 & $1.8517 \pm 0.0571$ & $-1.362 \pm 0.051$ & $-6.059 \pm 0.047$ & $9.6878$ & $3.51$\\
    \textbf{BD\,+43 4205} & $1.5696 \pm 0.0198$ & $-0.379 \pm 0.017$ & $-4.506 \pm 0.018$ & $9.8823$ & $1.12$\\
    BD\,+39 4917 & $2.4690 \pm 0.0915$ & $-0.384 \pm 0.094$ & $-10.428 \pm 0.080$ & $9.8811$ & $4.61$\\
    \textbf{BD\,+39 4920} & $1.7941 \pm 0.0201$ & $-1.480 \pm 0.018$ & $-4.576 \pm 0.018$ & $9.9312$ & $1.02$\\
    \textbf{BD\,+40 4852} & $1.6312 \pm 0.0236$ & $-1.144 \pm 0.017$ & $-4.179 \pm 0.021$ & $9.9281$ & $1.16$\\
    BD\,+42 4429 & $0.8706 \pm 0.0221$ & $-2.492 \pm 0.018$ & $-2.011 \pm 0.022$ & $9.9108$ & $1.17$\\
    \textbf{HD\,215271} & $2.1022 \pm 0.0177$ & $-1.927 \pm 0.013$ & $-3.940 \pm 0.017$ & $10.4557$ & $1.09$\\
    \textbf{BD\,+36 4868} & $1.9325 \pm 0.0182$ & $-1.664 \pm 0.015$ & $-5.554 \pm 0.016$ & $10.4764$ & $1.07$\\
\hline
    \end{tabular}
    \tablefoot{
    The names of objects used for the distance estimate of the association are in bold. We rejected the other objects based on their RUWE factor, high brightness or discrepant proper motions.}
\end{table*}

\section{Spectral fit for 10 Lac}\label{appendix:B}
An example for the fit quality of the global synthetic spectrum with observations is given in Fig.~\ref{fig:10Lac_fit} for the O9\,V MK
standard star 10~Lac. While the full blue wavelength range is covered, only selected wavelengths are considered beyond that, because the
density of diagnostic lines is largely reduced there in the late O-type stars. Overall, excellent agreement is found. One exception is the
region around $\sim$3925\,{\AA}. In contrast to the \ion{He}{ii} Pickering transitions longwards of 4000\,{\AA}, \ion{He}{ii}
$\lambda$3923.49\,{\AA} was not covered by the Stark broadening tables for \ion{He}{ii} from the detailed quantum-mechanical calculations
by \citet{SchBu89a,SchBu89b}. Instead, the \citet{Griem60} theory for Stark-broadened hydrogenic lines as implemented by \citet{AuMi72}
was employed, as for the shallow \ion{He}{ii} lines originating from the $n$\,=\,5 level, for example \ion{He}{ii} $\lambda\lambda$6683.209\,{\AA}
and 6527.107\,{\AA}. A few residuals from the data reduction are also present, as witnessed by a few emission spikes in the last panel of
Fig.~\ref{fig:10Lac_fit}.

Several observed features are not covered by the synthetic spectrum. These mostly stem from \ion{N}{iii}.  The available nitrogen model atom is insufficient to model  the observed lines reliably (this will be improved in future work). Also,
some chemical species, such as sulphur or iron, are completely absent as we lack reliable model atoms. However, they produce only relatively
weak features. In addition, some narrow absorption lines, such as the Ca H+K lines, the Na D lines, and a \ion{K}{i}
resonance line,\footnote{Only the \ion{K}{i} $\lambda$7698.9\,{\AA} line is clearly visible, while the other fine-structure component
\ion{K}{i} $\lambda$7664.9\,{\AA} overlaps with a saturated telluric O$_2$ line \citep[see e.g.][]{Kimeswengeretal21}, which depends on the
radial velocity of the target star.} are of interstellar origin. Diffuse interstellar bands (DIBs; increasingly important in
more reddened stars) and the many telluric absorption features due to O$_2$ (concentrated in the A-, B- and $\gamma$-bands of O$_2$) and H$_2$O bands (e.g. the weak sharp features near
H$\alpha$ or the forest of strong lines beyond $\sim$8900\,{\AA} in Fig.~\ref{fig:10Lac_fit}) that are spread over the red wavelength range are also omitted.

A number of photospheric emission lines are present in the spectrum of 10~Lac and in the other sample stars. They become more numerous towards the red part
of the spectrum, where non-LTE effects are amplified \citep[see e.g.][]{PrBu04}. They occur between high-lying energy levels. Two examples of the fit of the global synthetic
spectra to an \ion{O}{iii} and a \ion{Si}{iii} emission line in four sample stars are given in Fig.~\ref{fig:emission}. This gives additional confidence in the
quality of the model atoms as a good fit is achieved without the
necessity of fine adjustments of the selected atomic data. We note the stronger
contamination of 10~Lac (HD~214680) by telluric H$_2$O in the $\sim$8100\,{\AA} range in the spectrum taken at Calar Alto in Spain as compared with the
higher and drier Mauna Kea on Hawaii (spectra of the other three stars were taken at the CFHT).

\begin{figure*}
    \centering
    \includegraphics[width=0.985\hsize]{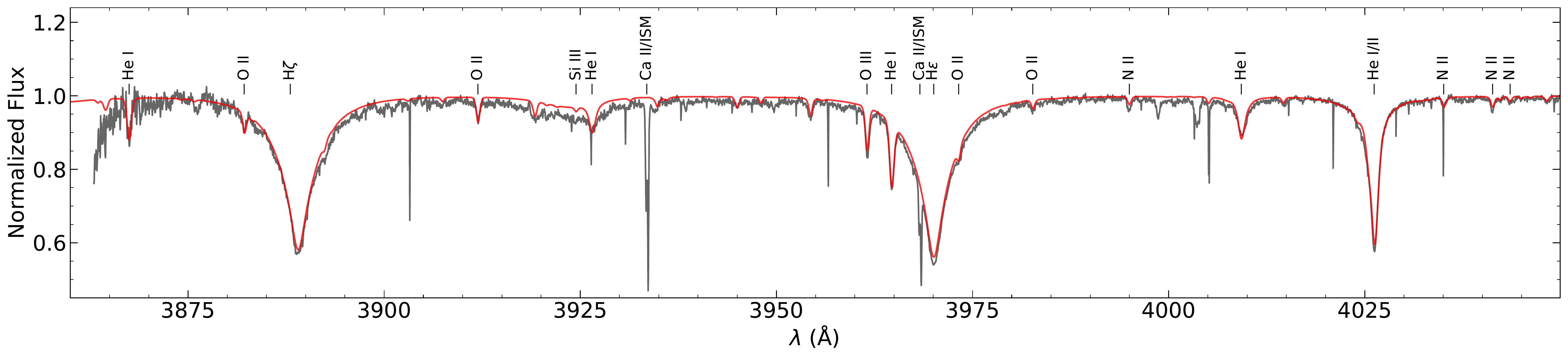}
    \includegraphics[width=0.985\hsize]{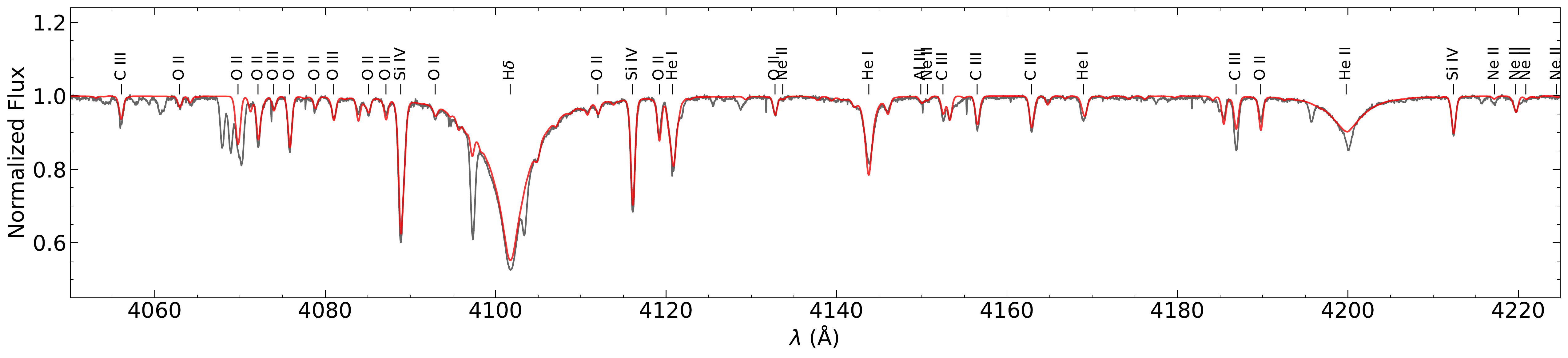}
    \includegraphics[width=0.985\hsize]{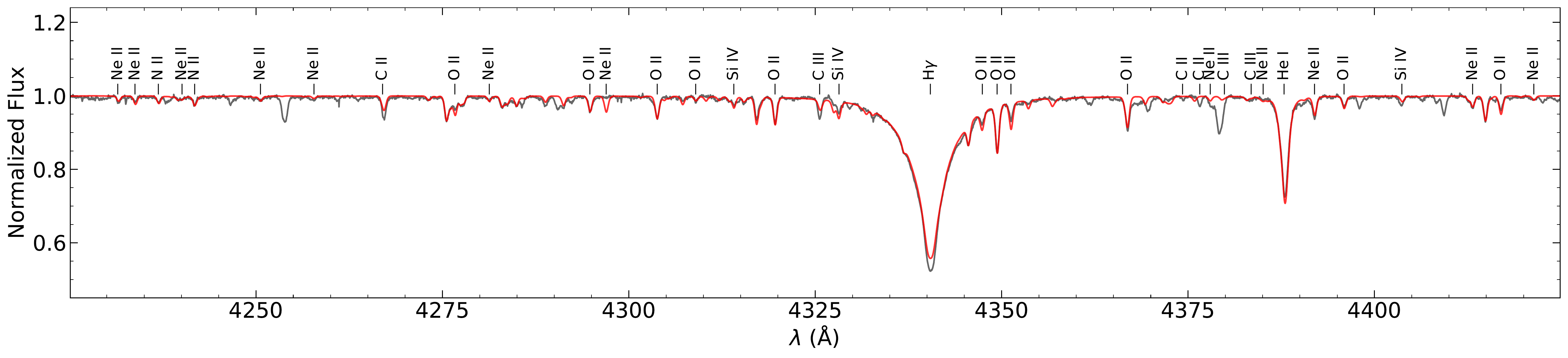}
    \includegraphics[width=0.985\hsize]{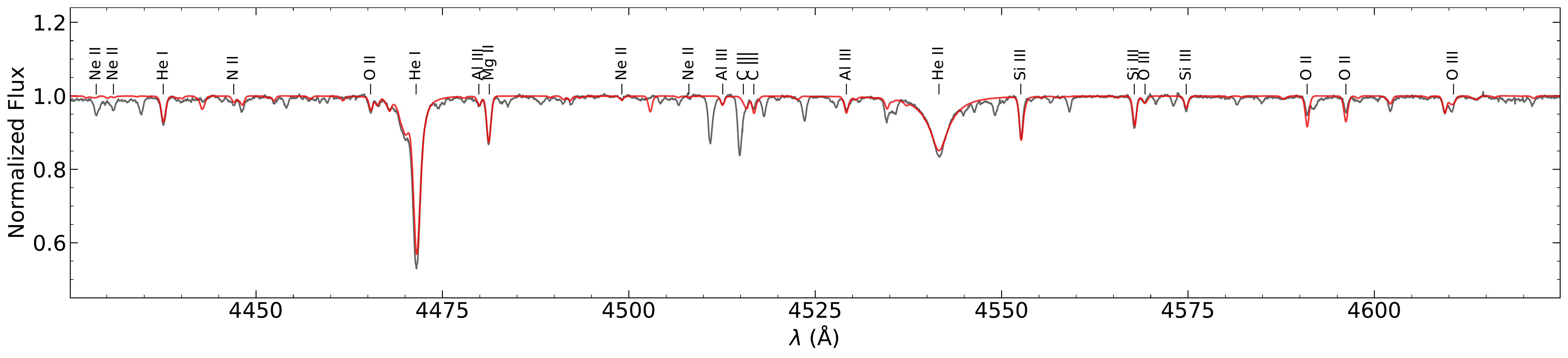}
    \includegraphics[width=0.985\hsize]{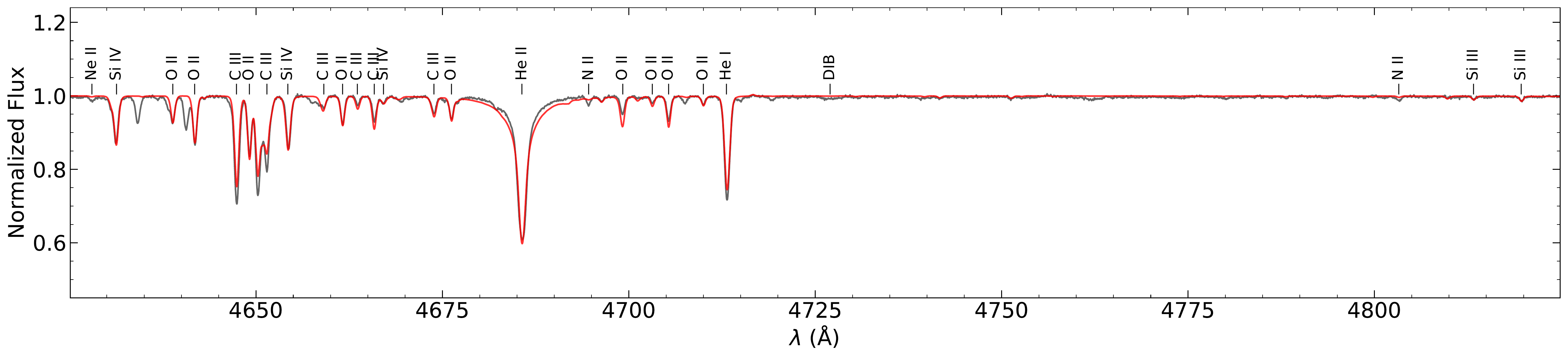}
    \caption{Comparison between the global synthetic (red) and observed spectrum (black line) for the O9\,V standard star 10~Lac (HD~214680) in the spectral range $\lambda\lambda$3900 to 4800\,{\AA}. Many of the stronger diagnostic lines are identified.}
    \label{fig:10Lac_fit}
\end{figure*}

\addtocounter{figure}{-1}
\begin{figure*}
    \centering
    \includegraphics[width=0.985\hsize]{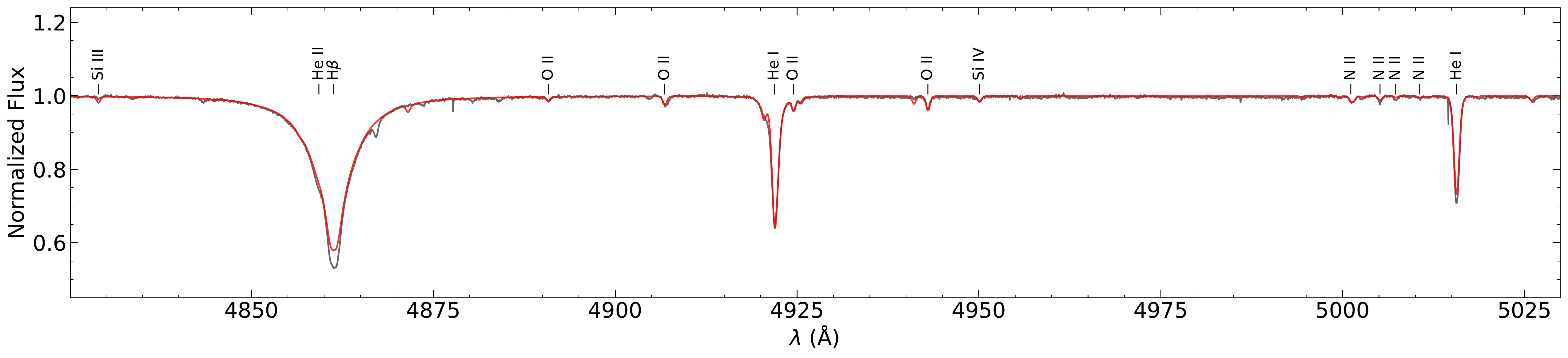}
    \includegraphics[width=0.985\hsize]{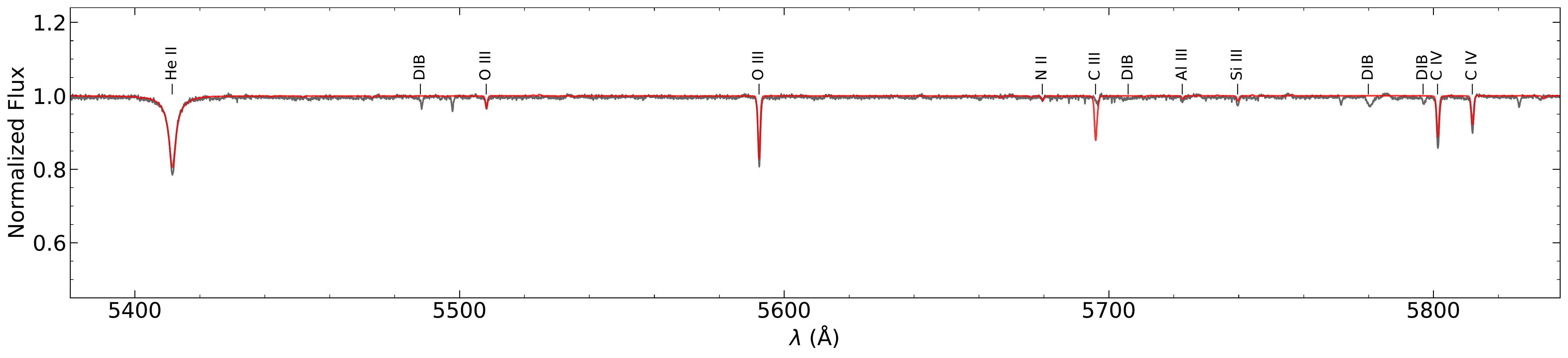}
    \includegraphics[width=0.985\hsize]{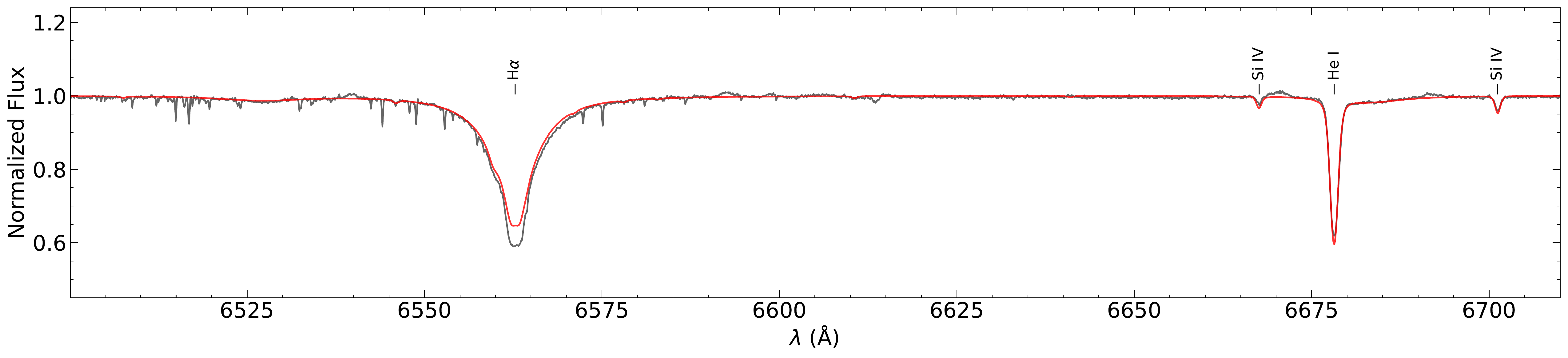}
    \includegraphics[width=0.985\hsize]{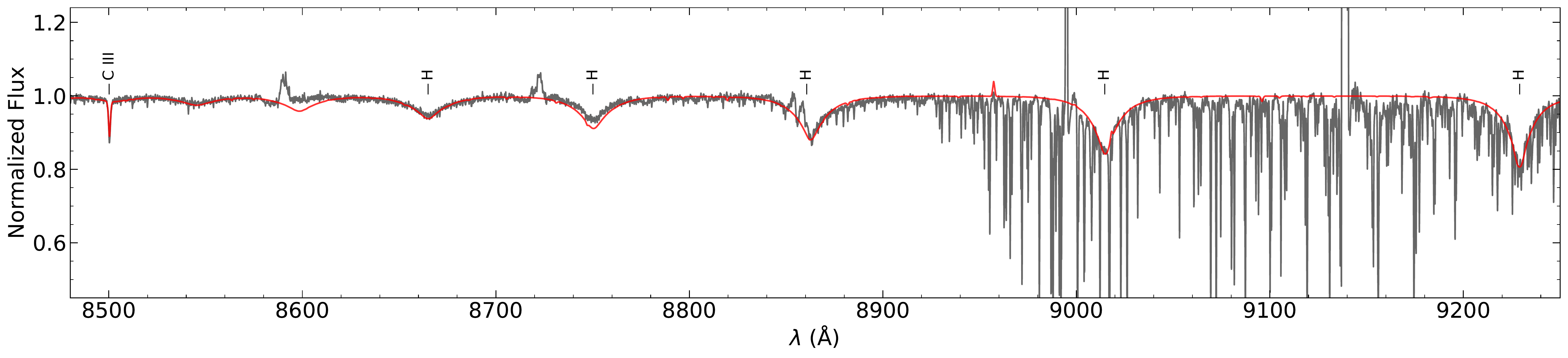}
    \caption{{\bf (cont.)} Selected wavelength ranges between $\sim$4800 and 9200\,{\AA}.}
\end{figure*}
\begin{figure*}
    \centering
    \includegraphics[width=0.94\hsize]{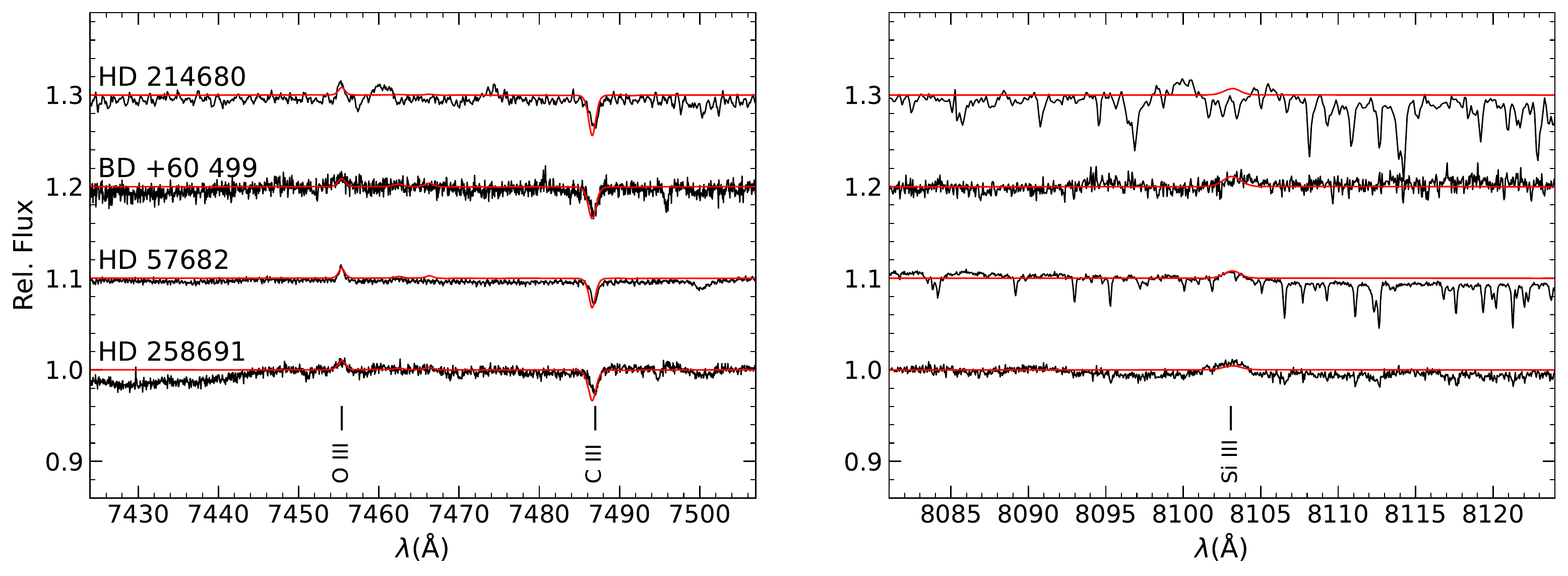}
    \caption{Same as Fig.~\ref{fig:10Lac_fit}, but for examples of \ion{O}{iii} ({\em left panel}) and \ion{Si}{iii} emission lines ({\em right panel}),
    for 10~Lac and three other sample stars.}\label{fig:emission}
\end{figure*}

\section{The distance to HD~46202, HD~258691, and the Rosette Nebula cluster NGC~2244}\label{appendix:C}
The stars HD~46202 and HD~258691 belong to the seven O stars of the open cluster NGC~2244. These seven stars are the main sources of ionising photons that excite the Rosette Nebula. In addition, 24 early B-type stars of spectral types B0 to B3 are contained in NGC~2244 {\citep[see Table~2 of][]{RoLa08}}. We adopt these 31 stars as basis for the estimation of a cluster distance to NGC~2244 in analogy to Appendix~\ref{appendix:A}, with the data
\begin{figure*}[ht]
   \centering
   \includegraphics[width=.86\hsize]{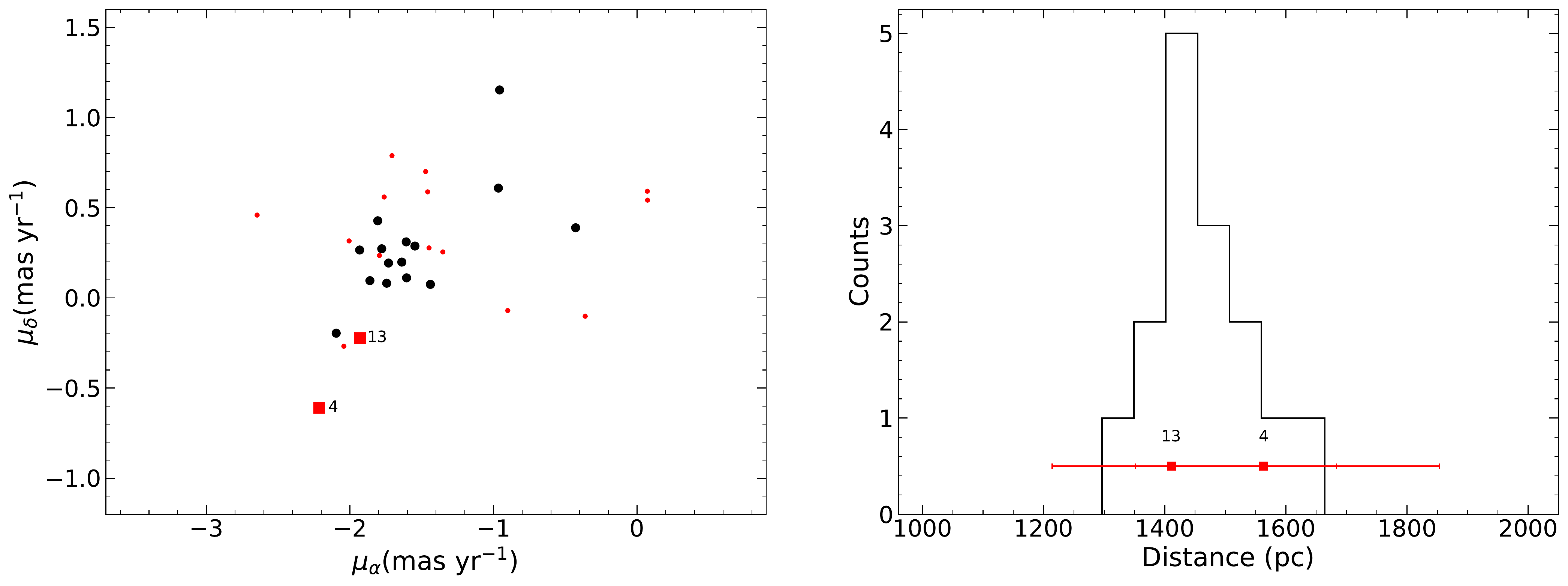}
      \caption{Proper motion and distances of OB stars in NGC~2244. {\em Left panel}: \textit{Gaia} EDR3 proper motions of OB stars in the NGC~2244 cluster. Data marked with small red dots are subject to selection criteria and are excluded from the histogram in the right panel. {\em Right panel}: Distribution of the \textit{Gaia} EDR3 distances, computed from the inversion of the parallaxes. \textit{Gaia} EDR3 distances to the stars HD~46202 (ID\#13) and HD~258691 (ID\#4) are marked by the squares.}\label{fig:NGC2244}
\end{figure*}
\begin{table*}[!ht]
    \centering
    \caption{\textit{Gaia} EDR3 parallaxes and photometric and astrometric measurements for OB stars in the Rosette nebula cluster NGC~2244.}\label{table:ngc_2244}
    \begin{tabular}{l r r r r cr}
    \hline
    \hline
    Name & $\varpi (\si{\mas})$ & {$\mu_{\alpha} (\si{\mas\per\year})$} & {$\mu_{\delta} (\si{\mas\per\year})$} & {$G (\si{\mag})$} & RUWE\\
    \hline
    \textbf{HD\,46150} & $0.6431 \pm 0.0340$ & $-2.096 \pm 0.031$ & $-0.196 \pm 0.030$ & 6.7138 & 1.04 \\
    \textbf{HD\,46223} & $0.6968 \pm 0.0264$ & $-1.605 \pm 0.026$ & $0.111 \pm 0.023$ & 7.1998 & 1.06 \\
    \textbf{HD\,46149} & $0.6385 \pm 0.0323$ & $-1.608 \pm 0.032$ & $0.310 \pm 0.028$ & 7.5508 & 1.06 \\
    \textbf{HD\,46484} & $0.7025 \pm 0.0279$ & $-0.957 \pm 0.028$ & $1.153 \pm 0.025$ & 7.6406 & 1.11 \\
    HD\,46106 & $0.6468 \pm 0.0388$ & $-2.006 \pm 0.035$ & $0.316 \pm 0.029$ & 7.8970 & 0.81 \\
    HD\,46202 & $0.7087 \pm 0.1148$ & $-1.929 \pm 0.110$ & $-0.223 \pm 0.110$ & 8.1580 & 4.98 \\
    \textbf{HD\,46485} & $0.7713 \pm 0.0236$ & $-1.639 \pm 0.027$ & $0.198 \pm 0.024$ & 8.1657 & 0.90 \\
    \textbf{HD\,46056} & $0.6744 \pm 0.0265$ & $-1.861 \pm 0.026$ & $0.095 \pm 0.024$ & 8.1836 & 0.94 \\
    HD\,259135 & $0.6499 \pm 0.0396$ & $-1.762 \pm 0.038$ & $0.559 \pm 0.034$ & 8.5275 & 1.31 \\
    HD\,259012 & $1.3861 \pm 0.1309$ & $-2.647 \pm 0.140$ & $0.459 \pm 0.124$ & 9.2978 & 6.00 \\
    \textbf{HD\,259105} & $0.7132 \pm 0.0244$ & $-1.744 \pm 0.022$ & $0.081 \pm 0.021$ & 9.3355 & 1.16 \\
    HD\,258691 & $0.6396 \pm 0.1001$ & $-2.215 \pm 0.100$ & $-0.609 \pm 0.090$ & 9.5245 & 6.97 \\
    \textbf{GSC\,00154-02337} & $0.6991 \pm 0.0220$ & $-1.547 \pm 0.020$ & $0.288 \pm 0.018$ & 9.6817 & 0.97 \\
    \textbf{GSC\,00154-02504} & $0.7163 \pm 0.0303$ & $-1.932 \pm 0.033$ & $0.265 \pm 0.032$ & 10.5832 & 0.99 \\
    HD\,259172 & $0.7800 \pm 0.0430$ & $-1.707 \pm 0.041$ & $0.789 \pm 0.035$ & 10.6365 & 2.11 \\
    HD\,259300 & $0.6971 \pm 0.0326$ & $-1.795 \pm 0.030$ & $0.235 \pm 0.029$ & 10.6983 & 2.06 \\
    NGC\,2244 193 & $0.5579 \pm 0.1402$ & $-1.353 \pm 0.120$ & $0.255 \pm 0.107$ & 10.7182 & 5.58 \\
    HD\,259238 & $0.6460 \pm 0.0278$ & $-1.458 \pm 0.026$ & $0.588 \pm 0.024$ & 11.0311 & 1.34 \\
    \textbf{GSC\,00154-01247} & $0.6486 \pm 0.0218$ & $-0.966 \pm 0.025$ & $0.609 \pm 0.021$ & 11.0388 & 1.04 \\
    GSC\,00154-01007 & $0.6727 \pm 0.0193$ & $-1.473 \pm 0.019$ & $0.700 \pm 0.017$ & 11.1379 & 0.84 \\
    \textbf{GSC\,00154-02187} & $0.6883 \pm 0.0186$ & $-0.428 \pm 0.017$ & $0.389 \pm 0.016$ & 11.1721 & 0.90 \\
    HD\,259268 & $0.4636 \pm 0.1233$ & $-1.449 \pm 0.091$ & $0.277 \pm 0.090$ & 11.1782 & 3.50 \\
    \textbf{GSC\,00154-01016} & $0.6706 \pm 0.0213$ & $-1.440 \pm 0.020$ & $0.075 \pm 0.020$ & 11.3139 & 0.87 \\
    GSC\,00154-00234 & $0.4683 \pm 0.0200$ & $-0.901 \pm 0.020$ & $-0.071 \pm 0.017$ & 11.4294 & 1.15 \\
    \textbf{GSC\,00154-02141} & $0.7172 \pm 0.0199$ & $-1.779 \pm 0.026$ & $0.272 \pm 0.022$ & 11.5531 & 1.11 \\
    \textbf{GSC\,00154-01753} & $0.6008 \pm 0.0226$ & $-1.731 \pm 0.021$ & $0.193 \pm 0.020$ & 11.8841 & 1.17 \\
    \textbf{GSC\,00154-02247} & $0.6644 \pm 0.0155$ & $-1.806 \pm 0.016$ & $0.427 \pm 0.014$ & 12.7080 & 1.00 \\
    GSC\,00154-02164 & $0.6654 \pm 0.0231$ & $-2.042 \pm 0.023$ & $-0.268 \pm 0.021$ & 12.7863 & 1.24 \\
    ALS\,15339 & $0.2269 \pm 0.0264$ & $-0.361 \pm 0.027$ & $-0.102 \pm 0.024$ & 14.5999 & 1.09 \\
    ALS\,15335 & $0.1942 \pm 0.0247$ & $0.073 \pm 0.026$ & $0.542 \pm 0.023$ & 14.6297 & 1.02 \\
    ALS\,15336 & $0.1959 \pm 0.0324$ & $0.072 \pm 0.029$ & $0.591 \pm 0.027$ & 15.0479 & 0.90 \\
\hline
    \end{tabular}
    \tablefoot{
    The names of objects used for the distance estimate of the cluster are in bold. We rejected the other objects based on their RUWE factor, discrepant distance or discrepant proper motions.}
\end{table*}
collected in Table \ref{table:ngc_2244}. The same selection criteria were applied to the \textit{Gaia} EDR3 data. We note that with $G$ magnitudes between $\sim$7 and 15\,mag the stars are ideal to deliver high-quality parallaxes from EDR3. The distribution of all OB-type member stars of NGC~2244 in proper motion space is visualised in the left panel of Fig.~\ref{fig:NGC2244}, with data from stars that were omitted marked by small red dots. The two sample stars {-- HD~46202 (ID\#13) and HD~258691 (ID\#4) --} are among those with the most negative proper motions, both in right ascension and in declination. The distribution of \textit{Gaia} EDR3 distances from the 15 remaining cluster stars (from the inverted parallaxes) are shown in the right panel. We estimate a mean inverted parallax distance of $d_{\mathrm{NGC~2244}}$\,=\,1464$_{-84}^{+94}$\,pc (1$\sigma$ standard deviation) to the cluster ($d_{\mathrm{NGC~2244}}$\,=\,1464$_{-22}^{+24}$\,pc, standard error of the mean). Both sample stars agree nicely with the cluster distance estimate in terms of \textit{Gaia} EDR3 distance. However, HD~46202 falls short in terms of spectroscopic distance (possibly because of its binarity), while the spectroscopic distance to HD~258691 matches well.

\addtocounter{section}{1}
\addtocounter{figure}{-1}
\begin{figure*}[ht]
   \centering
   \includegraphics[width=.89\hsize]{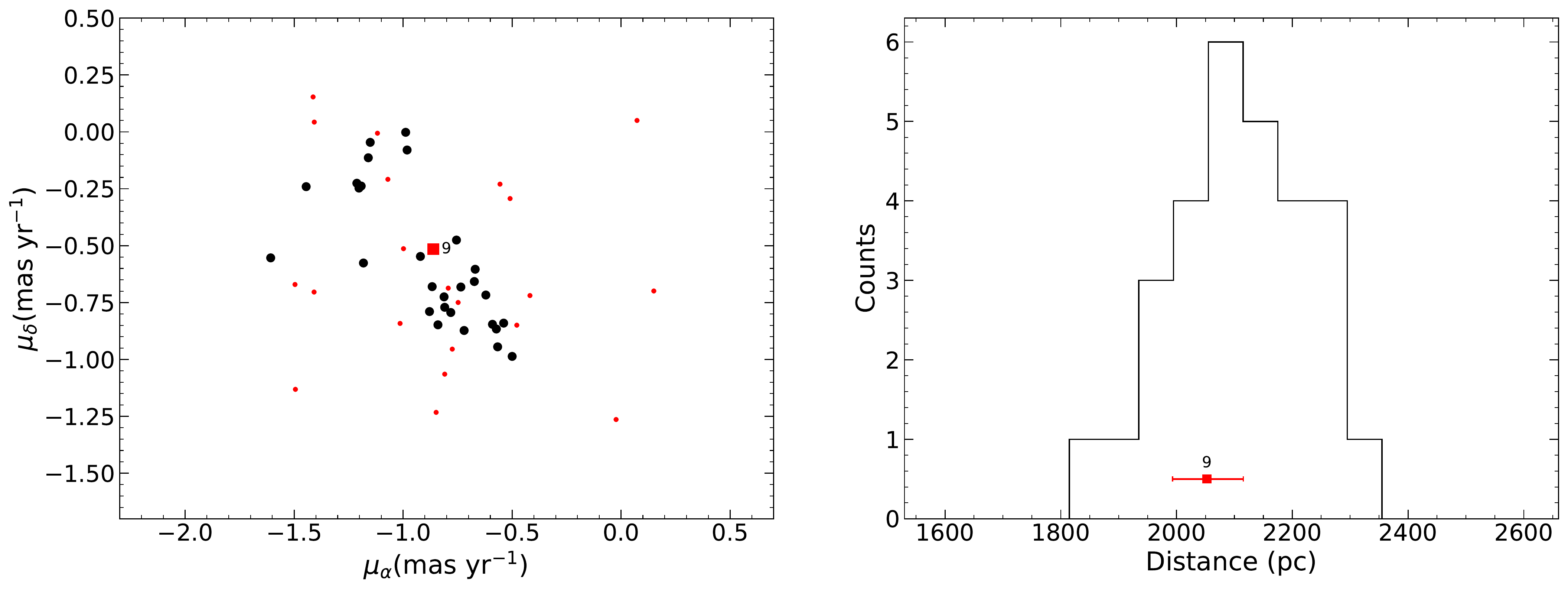}
      \caption{Same as Fig.~\ref{fig:NGC2244}, but for OB stars in the
      cluster IC~1805.}
         \label{fig:Collinder26}
\end{figure*}

\addtocounter{section}{1}
\addtocounter{figure}{-1}
\begin{figure*}[ht]
   \centering
   \includegraphics[width=.89\hsize]{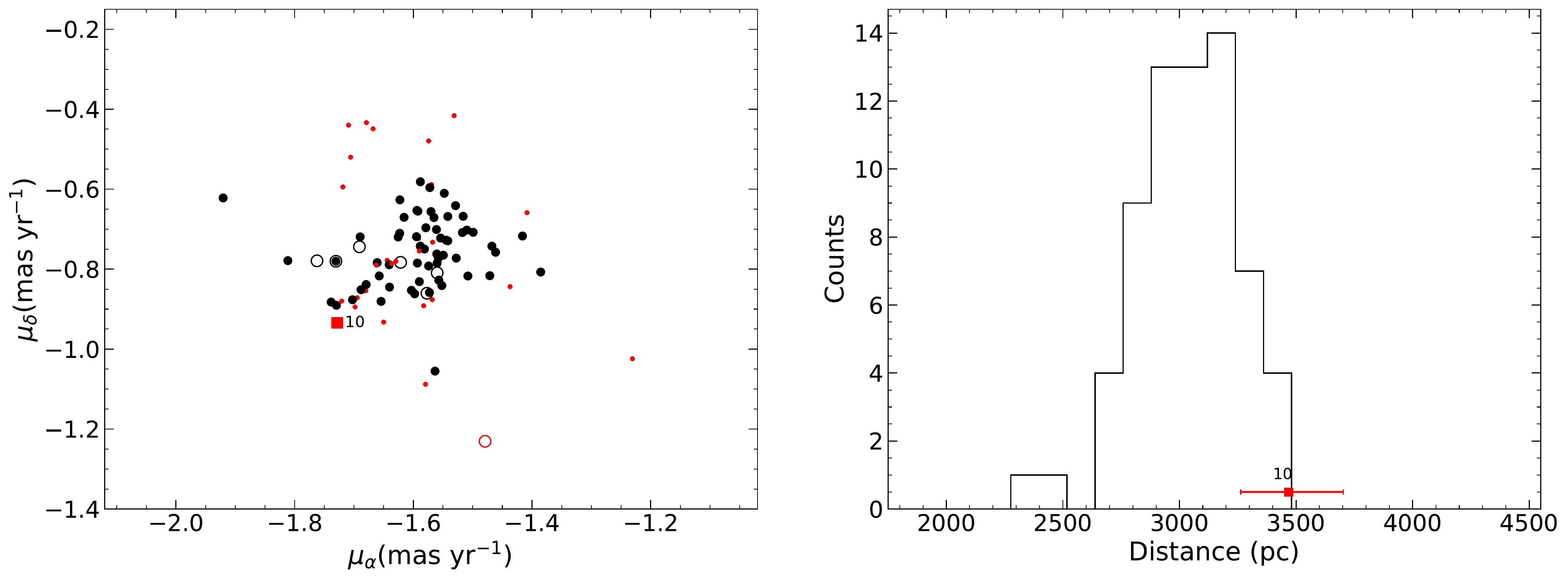}
      \caption{Proper motion and distances of stars in NGC~457. {\em Left panel}: \textit{Gaia} EDR3 proper motion of stars in the open cluster NGC~457. Stars marked
      with small red dots or red circles are subject to selection criteria and are excluded from the histogram in the right panel. Massive stars are marked by open circles.
      In addition to the massive stars, we included stars with a membership probability $\geq 70\%$ according to the
      WEBDA database. {\em Right panel}: Distribution of the \textit{Gaia} DR3 distances, computed from the inversion of the parallaxes.  The sample star
      BD~+57~247 (ID\#10) is marked by a square.}
         \label{fig:NGC457}
\end{figure*}
\addtocounter{section}{-2}

\section{The distance to BD+60~499 and the Heart Nebula cluster IC~1805}\label{appendix:D}

The Heart Nebula cluster IC~1805 (Melotte~15, Collinder~26) is the central
cluster of the association Cas~OB6 in the Perseus spiral arm, with about forty OB star members \citep{Masseyetal95,ShiHu99}. The cluster and the surrounding star-forming region W4, and their connection to the close-by star-forming regions W3 and W5, are discussed in detail by \citet{Megeathetal08} and \citet{Sungetal17}. Distances to Cas~OB6, including IC~1805, in the range of 2.2 to 2.35\,kpc were usually employed in the past \citep[see][]{Megeathetal08}, with more recent work indicating 2.4$\pm$0.2\,kpc for IC~1805 \citep{Sungetal17}.

An initial sample of 51 stars of spectral types B3 and earlier with cluster membership probability $\geq 70\%$ according to \citet{ShiHu99} and including the four brightest O stars of the cluster \citep{Masseyetal95} was assembled
for the distance estimation to IC~1805 (see Table~\ref{table:collinder_26}).
Proper motions for these 51 stars and a histogram of the distance distribution of the 29 cluster members, after applying the same selection criteria as for NGC~2244, are shown in Fig.~\ref{fig:Collinder26}. A distance value of $d_{\mathrm{IC~1805}}$\,=\,2108$_{-115}^{+129}$\,pc (1$\sigma$ standard deviation) or $d_{\mathrm{IC~1805}}$\,=\,2108$_{-21}^{+24}$\,pc (standard error of the mean) is derived, slightly shorter than the commonly assumed distance values. Both the spectroscopic and the \textit{Gaia} EDR3-based distance
to BD~+60~499 agree with the cluster distance within the 1$\sigma$-uncertainties.
The star's proper motions also fit nicely into the proper motion distribution of the cluster stars.
We note that the two bright O-star members of IC~1805, HD~15558 and BD~+60~497,
that were dropped from the cluster average calculations are known binaries \citep{Hillwigetal06,DeBeckeretal06}; some other stars from Table~\ref{table:collinder_26} are probably binaries as well.
\begin{table*}[!ht]
    \centering
    \caption{\textit{Gaia} EDR3 parallaxes and photometric and astrometric measurements for OB stars in the open cluster IC~1805.}\label{table:collinder_26}
    \small
    \begin{tabular}{l r r r r cr}
    \hline
    \hline
    Name & $\varpi (\si{\mas})$ & {$\mu_{\alpha} (\si{\mas\per\year})$} & {$\mu_{\delta} (\si{\mas\per\year})$} & {$G (\si{\mag})$} & RUWE\\
    \hline
HD~15558            & $0.5764 \pm 0.0662$ &     $-0.478 \pm 0.063$ & $-0.849 \pm     0.064$ &  7.7869 & 3.06 \\
\textbf{HD~15570}   & $0.4424 \pm 0.0226$ & $-0.720 \pm 0.021$ & $-0.873 \pm     0.023$ &  7.8909 & 0.95 \\
\textbf{HD~15629}   & $0.4848 \pm 0.0178$ & $-0.673 \pm 0.016$ & $-0.658 \pm     0.021$ &  8.2985 & 0.91 \\
BD +60 497          & $0.3852 \pm 0.0200$ & $-1.496 \pm 0.019$ & $-0.671 \pm     0.024$ &  8.6066 & 1.19 \\
\textbf{BD +60 513} & $0.4734 \pm 0.0153$ & $-0.539 \pm 0.014$ & $-0.840 \pm 0.017$ & 9.2564 & 0.90 \\
\textbf{BD +60 501} & $0.5060 \pm 0.0134$ & $-0.867 \pm 0.013$ & $-0.680 \pm 0.014$ & 9.4690 & 0.96 \\
\textbf{BD +60 498} & $0.4515 \pm 0.0192$ & $-1.192 \pm 0.017$ & $-0.238 \pm 0.021$ & 9.7939 & 1.01 \\
\textbf{BD +60 499} & $0.4872 \pm 0.0145$ & $-0.861 \pm 0.013$ & $-0.516 \pm 0.015$ & 10.1307 & 0.97 \\
IC 1805 113 & $0.4477 \pm 0.0470$ & $-1.413 \pm 0.047$ & $0.154 \pm 0.043$ & 10.4683 & 3.15 \\
BD +60 496 & $0.4706 \pm 0.0198$ & $-0.848 \pm 0.018$ & $-1.233 \pm 0.023$ & 10.4816 & 1.21 \\
IC 1805 161 & $0.7636 \pm 0.0695$ & $-0.809 \pm 0.063$ & $-1.065 \pm 0.075$ & 10.7771 & 5.00 \\
IC 1805 259 & $0.2844 \pm 0.3464$ & $-1.494 \pm 0.328$ & $-1.131 \pm 0.343$ & 10.8137 & 19.63 \\
IC 1805 21 & $0.5048 \pm 0.0158$ & $-1.408 \pm 0.013$ & $-0.704 \pm 0.015$ & 10.8269 & 1.25 \\
IC 1805 211 & $0.3872 \pm 0.1163$ & $-0.023 \pm 0.113$ & $-1.264 \pm 0.122$ & 10.9194 & 7.32 \\
IC 1805 136 & $0.6399 \pm 0.1187$ & $-1.118 \pm 0.115$ & $-0.006 \pm 0.125$ & 10.9250 & 7.85 \\
\textbf{BD +60 506} & $0.4614 \pm 0.0194$ & $-0.781 \pm 0.019$ & $-0.794 \pm 0.019$ & 10.9932 & 1.18 \\
IC 1805 288 & $0.4002 \pm 0.0179$ & $-0.418 \pm 0.017$ & $-0.719 \pm 0.020$ & 10.9983 & 1.07 \\
\textbf{IC 1805 149} & $0.4445 \pm 0.0178$ & $-0.590 \pm 0.017$ & $-0.846 \pm 0.019$ & 11.0848 & 1.03 \\
IC 1805 23 & $0.5544 \pm 0.0506$ & $0.150 \pm 0.045$ & $-0.699 \pm 0.054$ & 11.1545 & 3.39 \\
IC 1805 143 & $0.4324 \pm 0.0236$ & $-0.814 \pm 0.023$ & $-0.717 \pm 0.024$ & 11.2542 & 1.33 \\
IC 1805 111 & $0.4274 \pm 0.0217$ & $-0.998 \pm 0.021$ & $-0.513 \pm 0.027$ & 11.3151 & 1.24 \\
\textbf{IC 1805 121} & $0.4405 \pm 0.0195$ & $-0.879 \pm 0.019$ & $-0.789 \pm 0.020$ & 11.4125 & 1.10 \\
\textbf{IC 1805 260} & $0.4517 \pm 0.0217$ & $-0.621 \pm 0.020$ & $-0.717 \pm 0.024$ & 11.4156 & 1.07 \\
IC 1805 174 & $0.4858 \pm 0.0234$ & $-0.748 \pm 0.023$ & $-0.750 \pm 0.024$ & 11.4523 & 1.21 \\
IC 1805 169 & $0.4852 \pm 0.0216$ & $-1.014 \pm 0.021$ & $-0.842 \pm 0.023$ & 11.4875 & 1.25 \\
IC 1805 166 & $0.4102 \pm 0.0170$ & $-0.793 \pm 0.017$ & $-0.687 \pm 0.017$ & 11.8302 & 1.14 \\
\textbf{IC 1805 163} & $0.4726 \pm 0.0129$ & $-0.840 \pm 0.012$ & $-0.848 \pm 0.014$ & 12.0228 & 0.98 \\
IC 1805 262 & $0.5504 \pm 0.0162$ & $-0.509 \pm 0.015$ & $-0.293 \pm 0.016$ & 12.0467 & 1.25 \\
\textbf{IC 1805 72} & $0.4246 \pm 0.0138$ & $-1.202 \pm 0.013$ & $-0.247 \pm 0.014$ & 12.0523 & 1.00 \\
\textbf{IC 1805 167} & $0.4930 \pm 0.0117$ & $-0.755 \pm 0.011$ & $-0.475 \pm 0.013$ & 12.1096 & 0.87 \\
\textbf{IC 1805 82} & $0.4576 \pm 0.0110$ & $-1.160 \pm 0.010$ & $-0.114 \pm 0.013$ & 12.1524 & 0.91 \\
\textbf{IC 1805 69} & $0.5029 \pm 0.0161$ & $-1.212 \pm 0.014$ & $-0.226 \pm 0.016$ & 12.1665 & 1.17 \\
\textbf{IC 1805 62} & $0.4891 \pm 0.0128$ & $-1.607 \pm 0.011$ & $-0.554 \pm 0.013$ & 12.2608 & 1.03 \\
IC 1805 49 & $0.5230 \pm 0.0154$ & $-1.070 \pm 0.014$ & $-0.208 \pm 0.015$ & 12.2991 & 1.26 \\
IC 1805 18 & $0.4090 \pm 0.0104$ & $-0.556 \pm 0.009$ & $-0.230 \pm 0.010$ & 12.3769 & 0.86 \\
IC 1805 252 & $0.5635 \pm 0.0111$ & $0.073 \pm 0.010$ & $0.050 \pm 0.011$ & 12.3856 & 0.90 \\
\textbf{IC 1805 188} & $0.4745 \pm 0.0134$ & $-0.921 \pm 0.013$ & $-0.547 \pm 0.016$ & 12.4787 & 0.93 \\
\textbf{IC 1805 53} & $0.5510 \pm 0.0145$ & $-1.445 \pm 0.012$ & $-0.241 \pm 0.014$ & 12.5451 & 1.03 \\
\textbf{IC 1805 277} & $0.4556 \pm 0.0127$ & $-0.988 \pm 0.011$ & $-0.002 \pm 0.014$ & 12.5471 & 0.88 \\
\textbf{IC 1805 165} & $0.4729 \pm 0.0115$ & $-0.812 \pm 0.010$ & $-0.725 \pm 0.011$ & 12.6282 & 0.90 \\
IC 1805 152 & $0.5371 \pm 0.0356$ & $-0.774 \pm 0.033$ & $-0.955 \pm 0.036$ & 12.6705 & 2.64 \\
\textbf{IC 1805 180} & $0.4784 \pm 0.0138$ & $-0.809 \pm 0.013$ & $-0.771 \pm 0.014$ & 12.7325 & 0.92 \\
\textbf{IC 1805 276} & $0.4819 \pm 0.0150$ & $-0.669 \pm 0.014$ & $-0.604 \pm 0.016$ & 12.7507 & 0.97 \\
\textbf{IC 1805 191} & $0.4438 \pm 0.0127$ & $-1.151 \pm 0.012$ & $-0.046 \pm 0.014$ & 12.7637 & 0.95 \\
\textbf{IC 1805 52} & $0.5221 \pm 0.0130$ & $-1.182 \pm 0.011$ & $-0.576 \pm 0.012$ & 12.7866 & 0.98 \\
\textbf{IC 1805 332} & $0.5062 \pm 0.0159$ & $-0.982 \pm 0.016$ & $-0.080 \pm 0.017$ & 12.8183 & 1.05 \\
\textbf{IC 1805 175} & $0.4638 \pm 0.0167$ & $-0.566 \pm 0.015$ & $-0.945 \pm 0.016$ & 12.8432 & 1.03 \\
IC 1805 68 & $0.4145 \pm 0.0183$ & $-1.408 \pm 0.016$ & $0.043 \pm 0.019$ & 12.9683 & 1.09 \\
\textbf{IC 1805 212} & $0.4931 \pm 0.0192$ & $-0.572 \pm 0.020$ & $-0.866 \pm 0.022$ & 12.9863 & 1.10 \\
\textbf{IC 1805 157} & $0.4598 \pm 0.0126$ & $-0.735 \pm 0.012$ & $-0.682 \pm 0.013$ & 13.2631 & 0.98 \\
\textbf{IC 1805 225} & $0.4705 \pm 0.0145$ & $-0.500 \pm 0.014$ & $-0.987 \pm 0.016$ & 13.4357 & 0.94 \\
\hline
    \end{tabular}
    \tablefoot{
    The names of objects used for the distance estimate of the cluster are in bold. We rejected the other objects based on their RUWE factor, discrepant distance or discrepant proper motions.}
\end{table*}

\section{The distance to BD~+57~247 and the Owl cluster NGC~457}\label{appendix:E}
The open cluster NGC~457 in the Perseus spiral arm is different with respect to all other clusters harbouring O stars from our sample, by lacking active star formation in its vicinity and nebular emission. On the other hand, it possibly harbours three evolved stars, the mid B-type supergiant HD~7902, the yellow supergiant HD~7927 ($\phi$~Cas) and the red supergiant HD~236697. This is because of its supposedly much older age of about 24\,Myr \citep{Kharchenkoetal13,Diasetal21}. Both these data compilations give also similar distances to NGC~457, of 2400\,pc \citep{Kharchenkoetal13} and 2540$\pm$133\,pc \citep[thought their \textit{Gaia} DR2-based cluster mean parallax of 0.318$\pm$0.047\,mas implies a longer distance]{Diasetal21}. The cluster is centred in proper motion space on $\mu_\alpha$\,=\,$-$1.58\,mas\,yr$^{-1}$ and $\mu_\delta$\,=\,$-$0.76\,mas\,yr$^{-1}$ from an
analysis of 458 cluster members \citep{Flynnetal22}, based on \textit{Gaia} EDR3 data.

\begin{table*}[!ht]
    \centering
    \caption{\textit{Gaia} EDR3 parallaxes and photometric and astrometric measurements for massive stars in the open cluster NGC~457.}\label{table:ngc_457}
    \small
    \begin{tabular}{l r r r r cr}
    \hline
    \hline
    Name & $\varpi (\si{\mas})$ & {$\mu_{\alpha} (\si{\mas\per\year})$} & {$\mu_{\delta} (\si{\mas\per\year})$} & {$G (\si{\mag})$} & RUWE\\
    \hline
HD 7927             & $0.2142 \pm 0.0838$ & $-1.479 \pm 0.056$ & $-1.231 \pm 0.068$ & 4.7361 & 0.92 \\
\textbf{HD 7902}    & $0.3679 \pm 0.0169$ & $-1.560 \pm 0.013$ & $-0.809 \pm 0.016$ & 6.8426 & 0.92 \\
\textbf{HD 236697}   & $0.3102 \pm 0.0182$ & $-1.730 \pm 0.012$ & $-0.780 \pm 0.015$ & 7.5056 & 0.87 \\
\textbf{BD +57 243} & $0.3335 \pm 0.0122$ & $-1.577 \pm 0.009$ & $-0.860 \pm 0.011$ & 9.2586 & 0.97 \\
\textbf{BD +57 252} & $0.3179 \pm 0.0134$ & $-1.621 \pm 0.009$ & $-0.783 \pm 0.011$ & 9.4132 & 0.88 \\
\textbf{HD 236689} & $0.3501 \pm 0.0147$ & $-1.762 \pm 0.010$ & $-0.779 \pm 0.012$ & 9.4658 & 0.95 \\
\textbf{BD +57 249} & $0.3402 \pm 0.0133$ & $-1.691 \pm 0.010$ & $-0.744 \pm 0.011$ & 9.7400 & 0.90 \\
BD +57 247 & $0.2883 \pm 0.0182$ & $-1.728 \pm 0.013$ & $-0.935 \pm 0.017$ & 9.8529 & 0.94 \\
\hline
    \end{tabular}
\tablefoot{
    The names of objects used for the distance estimate of the cluster are in bold. We rejected HD~7927 and BD~+57~247 based on the parallax.}
\end{table*}

The five OB stars plus possibly three supergiants (see Table~\ref{table:ngc_457}) are a too small number of stars to pursue a discussion as for the other clusters here. We therefore added all NGC~457 stars with a membership probability $\geq 70\%$ according to the WEBDA database for our distance investigation (these stars are not included in Table~\ref{table:ngc_457}). The same selection criteria as used before were employed.
The proper motion distribution of this test star ensemble is displayed in the left panel of Fig.~\ref{fig:NGC457}, matching nicely the above stated central value. A histogram of the distance distribution of the stars that passed the selection procedure are shown in the right panel of Fig.~\ref{fig:NGC457}. Based on this, we estimate a mean inverted parallax distance of
$d_{\mathrm{NGC~457}}$\,=\,3026$_{-223}^{+262}$\,pc (1$\sigma$ standard deviation) to NGC~457, or $d_{\mathrm{NGC~457}}$\,=\,3026$_{-27}^{+32}$\,pc (standard error of the mean). This is significantly farther away than previously assumed in the literature.

Our sample star BD~+57~247 overlaps both in \textit{Gaia} EDR3 and spectroscopic distance with the farthest bin of the distance distribution of NGC~457. From a statistical viewpoint it may therefore be a cluster member, as its proper motions also  match those of other cluster stars. However, its $G$ magnitude is 0.6\,mag fainter than that of the brightest main-sequence star of the cluster,  the B0\,IV star BD~+57~243. The other three massive main-sequence stars, of spectral types B1 to B1.5, are also brighter, though they should be fainter by more than a magnitude if located at the same distance. The difference is so large that it cannot be resolved by differential extinction (the dust emission in the WISE bands appears rather smooth towards the cluster) or binarity (the maximum magnitude difference would be 0.75\,mag for two equally bright components). A blue straggler scenario for the star \citep{Mermilliod82} can therefore also be excluded, as it should be brighter than the cluster main-sequence B stars. As a consequence, BD~+57~247 is probably lying about 400\,pc farther away than NGC~457. This is further supported by stronger absorption in the interstellar Na\,D lines in BD~+57~247 when compared to the cluster member supergiant HD~7902 \citep[and see below]{Wessmayeretal22}, and in particular in the unsaturated interstellar \ion{K}{i} resonance line.

Finally, we want to comment on the cluster age. An age of $\sim$24\,Myr as given in the literature {\citep{Kharchenkoetal13,Diasetal21}} is incompatible with any of the massive member stars on the main sequence. Even the least massive of spectral type B1.5\,V would have reached termination of core hydrogen-burning after this time. NGC~457 must therefore be younger and the presence of the supergiants helps enormously in the age determination. In fact, a high-resolution spectrum of the B6\,Ia star HD~7902 has recently been analysed using the same analysis methodology as employed here \citep{Wessmayeretal22}. A spectroscopic distance of 2900$\pm$280\,pc confirms it to be a cluster member. It had a ZAMS mass of $\sim$20\,$M_\odot$ {and is with an age of $\sim$9 to 10\,Myr way younger then the literature cluster age.} Such an age would also be compatible with the luminosity class IV of the B0 star and of luminosity class V of the B1 and B1.5 stars. We do not consider the EDR3 parallax for the F0\,Ia supergiant $\phi$ Cas to be reliable as the star is bright for the \textit{Gaia} mission, similar to the case of 10~Lac. Nonetheless, its enormous luminosity ($B.C.$ is close to zero) may be compatible with cluster membership and cluster age if the star was initially somewhat more massive than HD~7902 and has already advanced well beyond the red supergiant stage (see Fig.~1 of \citealt{Hirschietal04} for an exemplary evolution track). The red supergiant HD~236697 appears to be a cluster member from its parallax and kinematics, but it would require substantial extinction to raise its luminosity to fit the $\sim$9 to 10\,Myr cluster age. Such extinction could be provided by dust formation in its circumstellar envelope, which is not uncommon among massive red supergiants.

\section{The distance to HD~206183, HD~207538, and the cluster IC~1396}\label{appendix:F}
The cluster \object{IC 1396} (Trumpler 37, Collinder 439) in the Cep OB2 association in the Local spiral arm is one of the youngest among the relatively nearby open clusters showing active star formation \citep[for an overview, see][]{Kunetal08,Errmannetal13}. Its age is in the range of 3 to 5\,Myr and a distance of about 800 and 870\,pc, respectively, in the direction of Galactic rotation was advocated by the two sources.

\begin{figure*}[ht]
   \centering
   \includegraphics[width=.9\hsize]{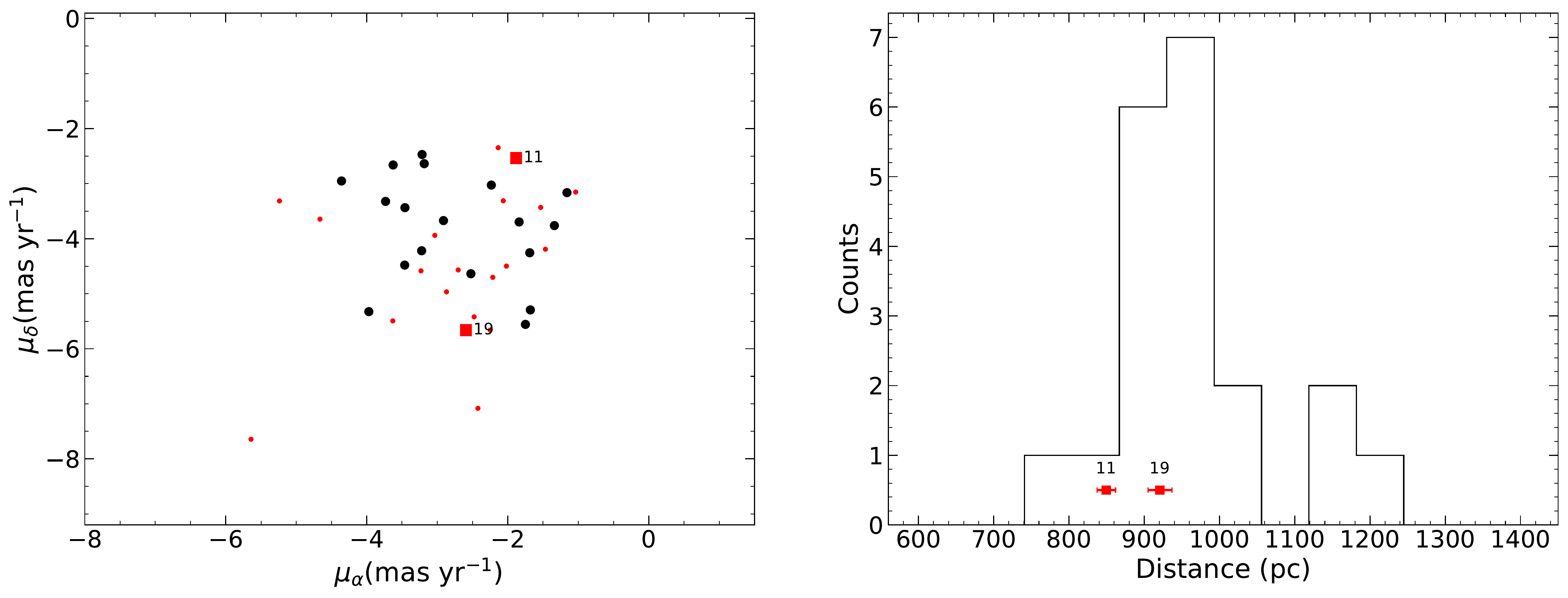}
      \caption{Same as Fig.~\ref{fig:NGC2244}, but for OB stars in the
      cluster IC~1396.}
\label{fig:IC1396}
\end{figure*}

We used an analogous procedure and selection criteria to identify member stars as discussed before. A starting sample of 38 OB stars of spectral type B3 and earlier was based on the data of \citet{GarKor76} (see our Table \ref{table:ic_1396}). The proper motion distribution of these stars is visualised in the left panel of Fig.~\ref{fig:IC1396} and a histogram of the distance distribution of the 20 stars that passed the selection procedure are shown in the right panel. From the
mean inverted parallax we estimate a distance to IC~1396 of
$d_{\mathrm{IC~1396}}$\,=\,954$_{-95}^{+119}$\,pc (1$\sigma$ standard deviation) or
$d_{\mathrm{IC~1396}}$\,=\,954$_{-21}^{+27}$\,pc (standard error of the mean),
slightly farther away than suggested in the pre-\textit{Gaia} era. The \textit{Gaia} EDR3 and spectroscopic distances of both sample stars in IC~1393, HD~206183 and HD~207538 agree with this cluster distance within the mutual 1$\sigma$ uncertainties.
Also, the stars' proper motions fit into the range spanned by the other cluster members.

\begin{table*}[!ht]
    \centering
    \caption{\textit{Gaia} EDR3 parallaxes and photometric and astrometric measurements for OB stars in the cluster IC~1396.}\label{table:ic_1396}
    \small
    \begin{tabular}{l r r r r cr}
    \hline
    \hline
    Name & $\varpi (\si{\mas})$ & {$\mu_{\alpha} (\si{\mas\per\year})$} & {$\mu_{\delta} (\si{\mas\per\year})$} & {$G (\si{\mag})$} & RUWE\\
    \hline
HD 206267 & $1.3604 \pm 0.2188$ & $-3.631 \pm 0.224$ & $-5.493 \pm 0.281$ & 5.6124 & 5.07 \\
\textbf{HD 206773} & $1.0958 \pm 0.0191$ & $-1.688 \pm 0.024$ & $-4.255 \pm 0.026$ & 6.7327 & 0.98 \\
\textbf{HD 205196} & $0.9713 \pm 0.0155$ & $-3.733 \pm 0.021$ & $-3.322 \pm 0.018$ & 7.1805 & 0.96 \\
\textbf{HD 207538} & $1.1773 \pm 0.0170$ & $-1.882 \pm 0.020$ & $-2.535 \pm 0.019$ & 7.2152 & 0.95 \\
\textbf{HD 206183} & $1.0862 \pm 0.0188$ & $-2.591 \pm 0.022$ & $-5.662 \pm 0.022$ & 7.3983 & 0.89 \\
HD 204827 & $1.0765 \pm 0.0988$ & $-3.231 \pm 0.132$ & $-4.584 \pm 0.128$ & 7.7181 & 6.67 \\
HD 205329 & $0.3658 \pm 0.4385$ & $-2.063 \pm 0.527$ & $-3.311 \pm 0.446$ & 7.9544 & 24.84 \\
\textbf{HD 239729} & $1.0704 \pm 0.0195$ & $-2.523 \pm 0.023$ & $-4.636 \pm 0.021$ & 8.2288 & 1.07 \\
HD 205794 & $1.0843 \pm 0.0144$ & $-1.466 \pm 0.017$ & $-4.191 \pm 0.016$ & 8.2861 & 0.84 \\
HD 239712 & $1.4020 \pm 0.0156$ & $-5.642 \pm 0.019$ & $-7.643 \pm 0.018$ & 8.3937 & 1.03 \\
HD 239738 & $1.0478 \pm 0.0323$ & $-2.136 \pm 0.040$ & $-2.347 \pm 0.038$ & 8.5064 & 1.59 \\
\textbf{HD 205948} & $1.0679 \pm 0.0162$ & $-3.222 \pm 0.021$ & $-4.219 \pm 0.019$ & 8.5433 & 0.87 \\
HD 206081 & $1.1328 \pm 0.0407$ & $-3.035 \pm 0.058$ & $-3.940 \pm 0.046$ & 8.5563 & 2.23 \\
\textbf{HD 239748} & $1.0778 \pm 0.0221$ & $-1.160 \pm 0.031$ & $-3.164 \pm 0.027$ & 8.6936 & 1.01 \\
HD 239689 & $1.1303 \pm 0.0133$ & $-2.018 \pm 0.018$ & $-4.498 \pm 0.014$ & 8.7965 & 0.82 \\
\textbf{HD 239745} & $1.0706 \pm 0.0163$ & $-1.680 \pm 0.018$ & $-5.294 \pm 0.015$ & 8.8278 & 0.87 \\
V* V427 Cep & $0.8514 \pm 0.0123$ & $-2.703 \pm 0.017$ & $-4.568 \pm 0.014$ & 8.9443 & 0.85 \\
HD 239724 & $0.4645 \pm 0.0124$ & $-4.664 \pm 0.014$ & $-3.644 \pm 0.013$ & 8.9996 & 0.81 \\
\textbf{HD 239731} & $1.0486 \pm 0.0133$ & $-2.233 \pm 0.016$ & $-3.026 \pm 0.015$ & 9.0408 & 1.08 \\
V* AI Cep & $0.5235 \pm 0.0141$ & $-5.238 \pm 0.017$ & $-3.314 \pm 0.015$ & 9.0573 & 1.06 \\
HD 239725 & $1.0922 \pm 0.0454$ & $-2.248 \pm 0.050$ & $-5.660 \pm 0.052$ & 9.0685 & 2.75 \\
\textbf{HD 239703} & $0.9766 \pm 0.0121$ & $-2.912 \pm 0.017$ & $-3.671 \pm 0.016$ & 9.1017 & 0.99 \\
\textbf{HD 239683} & $1.0857 \pm 0.0128$ & $-4.358 \pm 0.015$ & $-2.952 \pm 0.016$ & 9.2378 & 0.98 \\
\textbf{HD 239722} & $1.3496 \pm 0.0106$ & $-3.971 \pm 0.013$ & $-5.325 \pm 0.012$ & 9.3433 & 0.88 \\
\textbf{HD 239742} & $1.0242 \pm 0.0186$ & $-1.749 \pm 0.021$ & $-5.557 \pm 0.020$ & 9.3456 & 1.14 \\
\textbf{HD 239710} & $1.0686 \pm 0.0134$ & $-3.462 \pm 0.017$ & $-4.480 \pm 0.015$ & 9.3663 & 0.96 \\
BD +57 2395 & $1.0964 \pm 0.0195$ & $-1.037 \pm 0.023$ & $-3.152 \pm 0.024$ & 9.4444 & 1.44 \\
\textbf{HD 239693} & $1.0975 \pm 0.0156$ & $-1.839 \pm 0.020$ & $-3.696 \pm 0.017$ & 9.5120 & 1.19 \\
\textbf{BD +57 2376} & $0.8582 \pm 0.0129$ & $-3.459 \pm 0.017$ & $-3.436 \pm 0.015$ & 9.6299 & 0.97 \\
\textbf{HD 239746} & $0.8483 \pm 0.0112$ & $-3.215 \pm 0.015$ & $-2.471 \pm 0.012$ & 9.7440 & 0.90 \\
V* V738 Cep & $1.1851 \pm 0.0088$ & $-2.423 \pm 0.011$ & $-7.081 \pm 0.012$ & 9.7550 & 0.81 \\
\textbf{BD +58 2292} & $0.8033 \pm 0.0162$ & $-3.185 \pm 0.018$ & $-2.638 \pm 0.018$ & 9.7551 & 0.96 \\
\textbf{BD +58 2294} & $1.0619 \pm 0.0128$ & $-1.338 \pm 0.015$ & $-3.762 \pm 0.013$ & 9.8482 & 0.96 \\
BD +56 2596 & $1.1322 \pm 0.0296$ & $-2.870 \pm 0.039$ & $-4.965 \pm 0.031$ & 9.9765 & 2.04 \\
BD +56 2622 & $1.0595 \pm 0.0118$ & $-2.213 \pm 0.014$ & $-4.702 \pm 0.013$ & 10.0479 & 0.79 \\
\textbf{BD +57 2358} & $1.1290 \pm 0.0127$ & $-3.626 \pm 0.016$ & $-2.660 \pm 0.014$ & 10.0703 & 0.99 \\
BD +57 2395B & $1.0984 \pm 0.0192$ & $-1.533 \pm 0.023$ & $-3.432 \pm 0.023$ & 10.1156 & 1.46 \\
Cl Trumpler 37 430 & $1.0707 \pm 0.0185$ & $-2.477 \pm 0.026$ & $-5.419 \pm 0.019$ & 10.2841 & 0.85 \\

\hline
    \end{tabular}
    \tablefoot{
    The names of objects used for the distance estimate of the cluster are in bold. We rejected the other objects based on their RUWE factor, discrepant distance or discrepant proper motions.}
\end{table*}

\end{document}